\documentclass[a4paper,USenglish,numberwithinsect,cleveref]{lipics-v2021}

\hideLIPIcs

\usepackage{url}
\usepackage{fancyvrb}
\usepackage{mdwlist}  % Miscellaneous list-related commands
\usepackage{xspace}   % Smart spacing
\usepackage{bookmark}
\usepackage{comment}
\usepackage{footmisc} % symbol footnote

\usepackage{amsmath}
\usepackage{amssymb}
\usepackage{bm}       % Bold symbols in maths mode
\usepackage{tikz} % copy from esop2019 distributive
\usetikzlibrary{matrix} % copy from esop2019 distributive
\usetikzlibrary{arrows,automata} % copy from esop2019 distributive
\usetikzlibrary{positioning} % copy from esop2019 distributive
\usetikzlibrary{shapes.geometric} % copy from esop2019 distributive

\usepackage{mathtools}  % For "::=" ( \Coloneqq )

\usepackage{savesym} % copy from esop2019 distributive
\savesymbol{checkmark} % copy from esop2019 distributive
\usepackage{dingbat} % copy from esop2019 distributive
\usepackage{longtable} % copy from esop2019 distributive
\usepackage{subcaption} % copy from esop2019 distributive

\usepackage{xcolor}
\usepackage{array}

\usepackage{sourcecodepro}

\definecolor{arsenic}{rgb}{0.23, 0.27, 0.29}
\usepackage{listings}
 \usepackage{styles/mathpartir}  % by Didier Rémy (http://gallium.inria.fr/~remy/latex/mathpartir.html

\usepackage[implicitLineBreakHack]{styles/ottalt}  % http://users.eecs.northwestern.edu/~jesse/code/latex/

\usepackage{cite}

\usepackage{graphicx}

\usepackage{appendix}

\usepackage{tablefootnote}

\newcommand{\hlmath}[2][gray!40]{\colorbox{#1}{$\displaystyle#2$}}

\DeclareMathSymbol{\mlq}{\mathord}{operators}{``}
\DeclareMathSymbol{\mrq}{\mathord}{operators}{`'}

\newcommand{\concatOp}{+\kern-1.3ex+\kern0.8ex}  % http://tex.stackexchange.com/a/4195/73122

\newcommand\mynote[3]{\textcolor{#2}{#1: #3}}

\ifdefined\ispure
\renewcommand\mynote[3]{} % hide all comments
\fi

\newcommand{\tikzcircle}[2][black,fill=black]{\tikz[baseline=-0.5ex]\draw[#1,radius=#2] (0,0) circle ;}

\newcommand{\emptycircle}{\tikzcircle[black,fill=white]{2pt}\xspace}
\newcommand{\fullcircle}{\tikzcircle{2pt}\xspace}

\newcommand{\ottdrule}[4][]{{\displaystyle\frac{\begin{array}{l}#2\end{array}}{#3}\quad\ottdrulename{#4}}}
\newcommand{\ottusedrule}[1]{\[#1\]}
\newcommand{\ottpremise}[1]{ #1 \\}
\newenvironment{ottdefnblock}[3][]{ \framebox{\mbox{#2}} \quad #3 \\[0pt]}{}

\newcommand{\ottnt}[1]{\mathit{#1}}
\newcommand{\ottmv}[1]{\mathit{#1}}

\newcommand{\ottsym}[1]{#1}
\newcommand{\ottcom}[1]{\text{#1}}
\newcommand{\ottdrulename}[1]{\textsc{#1}}

\newcommand{\ottafterlastrule}{\\}

\newcommand{\ottdruleSXXvar}[1]{\ottdrule[#1]{%
\ottpremise{\vdash  \Delta}%
\ottpremise{\Delta  \vdash  \ottmv{X}}%
}{
 \Delta   \vdash   \ottmv{X}  \leq  \ottmv{X} }{%
{\ottdrulename{S\_var}}{}%
}}

\newcommand{\ottdruleSXXint}[1]{\ottdrule[#1]{%
\ottpremise{\vdash  \Delta}%
}{
 \Delta   \vdash    \mathsf{Int}   \leq   \mathsf{Int}  }{%
{\ottdrulename{S\_int}}{}%
}}

\newcommand{\ottdruleSXXtop}[1]{\ottdrule[#1]{%
\ottpremise{\Delta  \vdash  \mathit{A}}%
\ottpremise{ \Delta   \vdash  \rceil B^\circ \lceil }%
}{
 \Delta   \vdash   \mathit{A}  \leq  B^\circ }{%
{\ottdrulename{S\_top}}{}%
}}

\newcommand{\ottdruleSXXbot}[1]{\ottdrule[#1]{%
\ottpremise{\vdash  \Delta}%
\ottpremise{\Delta  \vdash  A^\circ}%
}{
 \Delta   \vdash    \mathsf{Bot}   \leq  A^\circ }{%
{\ottdrulename{S\_bot}}{}%
}}

\newcommand{\ottdruleSXXandl}[1]{\ottdrule[#1]{%
\ottpremise{\Delta  \vdash  \mathit{B}}%
\ottpremise{ \Delta   \vdash   \mathit{A}  \leq  C^\circ }%
}{
 \Delta   \vdash   \mathit{A}  \, \& \,  \mathit{B}  \leq  C^\circ }{%
{\ottdrulename{S\_andl}}{}%
}}

\newcommand{\ottdruleSXXandr}[1]{\ottdrule[#1]{%
\ottpremise{\Delta  \vdash  \mathit{A}}%
\ottpremise{ \Delta   \vdash   \mathit{B}  \leq  C^\circ }%
}{
 \Delta   \vdash   \mathit{A}  \, \& \,  \mathit{B}  \leq  C^\circ }{%
{\ottdrulename{S\_andr}}{}%
}}

\newcommand{\ottdruleSXXarrow}[1]{\ottdrule[#1]{%
\ottpremise{ \Delta   \vdash   \mathit{A}_{{\mathrm{2}}}  \leq  \mathit{A}_{{\mathrm{1}}} }%
\ottpremise{ \Delta   \vdash   \mathit{B}_{{\mathrm{1}}}  \leq  B^\circ_{{\mathrm{2}}} }%
}{
 \Delta   \vdash   \mathit{A}_{{\mathrm{1}}}  \rightarrow  \mathit{B}_{{\mathrm{1}}}  \leq  \mathit{A}_{{\mathrm{2}}}  \rightarrow  B^\circ_{{\mathrm{2}}} }{%
{\ottdrulename{S\_arrow}}{}%
}}

\newcommand{\ottdruleSXXall}[1]{\ottdrule[#1]{%
\ottpremise{ \Delta   \vdash   \mathit{B}_{{\mathrm{1}}}  \leq  \mathit{A}_{{\mathrm{1}}} }%
\ottpremise{ \Delta  ,  \ottmv{X}  *  \mathit{B}_{{\mathrm{1}}}   \vdash   \mathit{A}_{{\mathrm{2}}}  \leq  B^\circ_{{\mathrm{2}}} }%
}{
 \Delta   \vdash    \forall   \ottmv{X} * \mathit{A}_{{\mathrm{1}}} .\, \mathit{A}_{{\mathrm{2}}}   \leq   \forall   \ottmv{X} * \mathit{B}_{{\mathrm{1}}} .\, B^\circ_{{\mathrm{2}}}  }{%
{\ottdrulename{S\_all}}{}%
}}

\newcommand{\ottdruleSXXrcd}[1]{\ottdrule[#1]{%
\ottpremise{ \Delta   \vdash   \mathit{A}  \leq  B^\circ }%
}{
 \Delta   \vdash   \ottsym{\{}  \ottmv{l}  \!\vcentcolon\!  \mathit{A}  \ottsym{\}}  \leq  \ottsym{\{}  \ottmv{l}  \!\vcentcolon\!  B^\circ  \ottsym{\}} }{%
{\ottdrulename{S\_rcd}}{}%
}}

\newcommand{\ottdruleSXXand}[1]{\ottdrule[#1]{%
\ottpremise{ \hlmath{  \mathit{B}_{{\mathrm{1}}}  \mathbin{\lhd}  \mathit{B}  \mathbin{\rhd}  \mathit{B}_{{\mathrm{2}}}  } }%
\ottpremise{  \Delta   \vdash   \mathit{A}  \leq  \mathit{B}_{{\mathrm{1}}}   \and   \Delta   \vdash   \mathit{A}  \leq  \mathit{B}_{{\mathrm{2}}}  }%
}{
 \Delta   \vdash   \mathit{A}  \leq  \mathit{B} }{%
{\ottdrulename{S\_and}}{}%
}}

\newcommand{\ottdefnalgoXXsub}[1]{\begin{ottdefnblock}[#1]{$ \Delta   \vdash   \mathit{A}  \leq  \mathit{B} $}{\ottcom{Algorithmic Subtyping}}
\ottusedrule{\ottdruleSXXvar{}}
\ottusedrule{\ottdruleSXXint{}}
\ottusedrule{\ottdruleSXXtop{}}
\ottusedrule{\ottdruleSXXbot{}}
\ottusedrule{\ottdruleSXXandl{}}
\ottusedrule{\ottdruleSXXandr{}}
\ottusedrule{\ottdruleSXXarrow{}}
\ottusedrule{\ottdruleSXXall{}}
\ottusedrule{\ottdruleSXXrcd{}}
\ottusedrule{\ottdruleSXXand{}}
\end{ottdefnblock}}

\newcommand{\ottdefnsAlgorithmicSubtyping}{
\ottdefnalgoXXsub{}}

\newcommand{\ottdruleDaxXXintArrow}[1]{\ottdrule[#1]{%
}{
 \mathsf{Int}   *_{ax}  \mathit{A}_{{\mathrm{1}}}  \rightarrow  \mathit{A}_{{\mathrm{2}}}}{%
{\ottdrulename{Dax\_intArrow}}{}%
}}

\newcommand{\ottdruleDaxXXintRcd}[1]{\ottdrule[#1]{%
}{
 \mathsf{Int}   *_{ax}  \ottsym{\{}  \ottmv{l}  \!\vcentcolon\!  \mathit{A}  \ottsym{\}}}{%
{\ottdrulename{Dax\_intRcd}}{}%
}}

\newcommand{\ottdruleDaxXXintAll}[1]{\ottdrule[#1]{%
}{
 \mathsf{Int}   *_{ax}   \forall   \ottmv{X} * \mathit{A} .\, \mathit{B} }{%
{\ottdrulename{Dax\_intAll}}{}%
}}

\newcommand{\ottdruleDaxXXarrowRcd}[1]{\ottdrule[#1]{%
}{
\mathit{A}_{{\mathrm{1}}}  \rightarrow  \mathit{A}_{{\mathrm{2}}}  *_{ax}  \ottsym{\{}  \ottmv{l}  \!\vcentcolon\!  \mathit{A}  \ottsym{\}}}{%
{\ottdrulename{Dax\_arrowRcd}}{}%
}}

\newcommand{\ottdruleDaxXXarrowAll}[1]{\ottdrule[#1]{%
}{
\mathit{A}_{{\mathrm{1}}}  \rightarrow  \mathit{A}_{{\mathrm{2}}}  *_{ax}   \forall   \ottmv{X} * \mathit{A} .\, \mathit{B} }{%
{\ottdrulename{Dax\_arrowAll}}{}%
}}

\newcommand{\ottdruleDaxXXrcdAll}[1]{\ottdrule[#1]{%
}{
\ottsym{\{}  \ottmv{l}  \!\vcentcolon\!  \mathit{C}  \ottsym{\}}  *_{ax}   \forall   \ottmv{X} * \mathit{A} .\, \mathit{B} }{%
{\ottdrulename{Dax\_rcdAll}}{}%
}}

\newcommand{\ottdruleDaxXXarrowInt}[1]{\ottdrule[#1]{%
}{
\mathit{A}_{{\mathrm{1}}}  \rightarrow  \mathit{A}_{{\mathrm{2}}}  *_{ax}   \mathsf{Int} }{%
{\ottdrulename{Dax\_arrowInt}}{}%
}}

\newcommand{\ottdruleDaxXXrcdInt}[1]{\ottdrule[#1]{%
}{
\ottsym{\{}  \ottmv{l}  \!\vcentcolon\!  \mathit{A}  \ottsym{\}}  *_{ax}   \mathsf{Int} }{%
{\ottdrulename{Dax\_rcdInt}}{}%
}}

\newcommand{\ottdruleDaxXXallInt}[1]{\ottdrule[#1]{%
}{
 \forall   \ottmv{X} * \mathit{A} .\, \mathit{B}   *_{ax}   \mathsf{Int} }{%
{\ottdrulename{Dax\_allInt}}{}%
}}

\newcommand{\ottdruleDaxXXrcdArrow}[1]{\ottdrule[#1]{%
}{
\ottsym{\{}  \ottmv{l}  \!\vcentcolon\!  \mathit{A}  \ottsym{\}}  *_{ax}  \mathit{A}_{{\mathrm{1}}}  \rightarrow  \mathit{A}_{{\mathrm{2}}}}{%
{\ottdrulename{Dax\_rcdArrow}}{}%
}}

\newcommand{\ottdruleDaxXXallArrow}[1]{\ottdrule[#1]{%
}{
 \forall   \ottmv{X} * \mathit{A} .\, \mathit{B}   *_{ax}  \mathit{A}_{{\mathrm{1}}}  \rightarrow  \mathit{A}_{{\mathrm{2}}}}{%
{\ottdrulename{Dax\_allArrow}}{}%
}}

\newcommand{\ottdruleDaxXXallRcd}[1]{\ottdrule[#1]{%
}{
 \forall   \ottmv{X} * \mathit{A} .\, \mathit{B}   *_{ax}  \ottsym{\{}  \ottmv{l}  \!\vcentcolon\!  \mathit{C}  \ottsym{\}}}{%
{\ottdrulename{Dax\_allRcd}}{}%
}}

\newcommand{\ottdruleDaxXXrcdNeq}[1]{\ottdrule[#1]{%
\ottpremise{ \ottmv{l_{{\mathrm{1}}}}  \neq  \ottmv{l_{{\mathrm{2}}}} }%
}{
\ottsym{\{}  \ottmv{l_{{\mathrm{1}}}}  \!\vcentcolon\!  \mathit{A}  \ottsym{\}}  *_{ax}  \ottsym{\{}  \ottmv{l_{{\mathrm{2}}}}  \!\vcentcolon\!  \mathit{B}  \ottsym{\}}}{%
{\ottdrulename{Dax\_rcdNeq}}{}%
}}

\newcommand{\ottdefndisjointXXaxiom}[1]{\begin{ottdefnblock}[#1]{$\mathit{A}  *_{ax}  \mathit{B}$}{\ottcom{Disjointness Axioms}}
\ottusedrule{\ottdruleDaxXXintArrow{}}
\ottusedrule{\ottdruleDaxXXintRcd{}}
\ottusedrule{\ottdruleDaxXXintAll{}}
\ottusedrule{\ottdruleDaxXXarrowRcd{}}
\ottusedrule{\ottdruleDaxXXarrowAll{}}
\ottusedrule{\ottdruleDaxXXrcdAll{}}
\ottusedrule{\ottdruleDaxXXarrowInt{}}
\ottusedrule{\ottdruleDaxXXrcdInt{}}
\ottusedrule{\ottdruleDaxXXallInt{}}
\ottusedrule{\ottdruleDaxXXrcdArrow{}}
\ottusedrule{\ottdruleDaxXXallArrow{}}
\ottusedrule{\ottdruleDaxXXallRcd{}}
\ottusedrule{\ottdruleDaxXXrcdNeq{}}
\end{ottdefnblock}}

\newcommand{\ottdefnsDisjointnessAxiom}{
\ottdefndisjointXXaxiom{}}

\newcommand{\ottdruleDXXax}[1]{\ottdrule[#1]{%
\ottpremise{   \vdash  \Delta  \and  \Delta  \vdash  \mathit{A}  ,  \mathit{B}    \and  \mathit{A}  *_{ax}  \mathit{B} }%
}{
\Delta  \vdash  \mathit{A}  *  \mathit{B}}{%
{\ottdrulename{D\_ax}}{}%
}}

\newcommand{\ottdruleDXXtopl}[1]{\ottdrule[#1]{%
\ottpremise{\Delta  \vdash  \mathit{B}}%
\ottpremise{ \Delta   \vdash  \rceil \mathit{A} \lceil }%
}{
\Delta  \vdash  \mathit{A}  *  \mathit{B}}{%
{\ottdrulename{D\_topl}}{}%
}}

\newcommand{\ottdruleDXXtopr}[1]{\ottdrule[#1]{%
\ottpremise{\Delta  \vdash  \mathit{A}}%
\ottpremise{ \Delta   \vdash  \rceil \mathit{B} \lceil }%
}{
\Delta  \vdash  \mathit{A}  *  \mathit{B}}{%
{\ottdrulename{D\_topr}}{}%
}}

\newcommand{\ottdruleDXXarrow}[1]{\ottdrule[#1]{%
\ottpremise{ \Delta  \vdash  \mathit{A}_{{\mathrm{1}}}  ,  \mathit{A}_{{\mathrm{2}}}  \ottlinebreakhack }%
\ottpremise{\Delta  \vdash  \mathit{B}_{{\mathrm{1}}}  *  \mathit{B}_{{\mathrm{2}}}}%
}{
\Delta  \vdash  \mathit{A}_{{\mathrm{1}}}  \rightarrow  \mathit{B}_{{\mathrm{1}}}  *  \mathit{A}_{{\mathrm{2}}}  \rightarrow  \mathit{B}_{{\mathrm{2}}}}{%
{\ottdrulename{D\_arrow}}{}%
}}

\newcommand{\ottdruleDXXrcdEq}[1]{\ottdrule[#1]{%
\ottpremise{\Delta  \vdash  \mathit{A}  *  \mathit{B}}%
}{
\Delta  \vdash  \ottsym{\{}  \ottmv{l}  \!\vcentcolon\!  \mathit{A}  \ottsym{\}}  *  \ottsym{\{}  \ottmv{l}  \!\vcentcolon\!  \mathit{B}  \ottsym{\}}}{%
{\ottdrulename{D\_rcdEq}}{}%
}}

\newcommand{\ottdruleDXXall}[1]{\ottdrule[#1]{%
\ottpremise{\Delta  \vdash  \mathit{A}_{{\mathrm{1}}}  ,  \mathit{A}_{{\mathrm{2}}}}%
\ottpremise{\Delta  ,  \ottmv{X}  *  \mathit{A}_{{\mathrm{1}}}  \, \& \,  \mathit{A}_{{\mathrm{2}}}  \vdash  \mathit{B}_{{\mathrm{1}}}  *  \mathit{B}_{{\mathrm{2}}}}%
}{
\Delta  \vdash   \forall   \ottmv{X} * \mathit{A}_{{\mathrm{1}}} .\, \mathit{B}_{{\mathrm{1}}}   *   \forall   \ottmv{X} * \mathit{A}_{{\mathrm{2}}} .\, \mathit{B}_{{\mathrm{2}}} }{%
{\ottdrulename{D\_all}}{}%
}}

\newcommand{\ottdruleDXXvarl}[1]{\ottdrule[#1]{%
\ottpremise{  \ottmv{X} * \mathit{A} \in \Delta   \ottlinebreakhack }%
\ottpremise{ \Delta   \vdash   \mathit{A}  \leq  \mathit{B} }%
}{
\Delta  \vdash  \ottmv{X}  *  \mathit{B}}{%
{\ottdrulename{D\_varl}}{}%
}}

\newcommand{\ottdruleDXXvarr}[1]{\ottdrule[#1]{%
\ottpremise{  \ottmv{X} * \mathit{A} \in \Delta   \ottlinebreakhack }%
\ottpremise{ \Delta   \vdash   \mathit{A}  \leq  \mathit{B} }%
}{
\Delta  \vdash  \mathit{B}  *  \ottmv{X}}{%
{\ottdrulename{D\_varr}}{}%
}}

\newcommand{\ottdruleDXXandl}[1]{\ottdrule[#1]{%
\ottpremise{ \mathit{A}_{{\mathrm{1}}}  \mathbin{\lhd}  \mathit{A}  \mathbin{\rhd}  \mathit{A}_{{\mathrm{2}}} }%
\ottpremise{\Delta  \vdash  \mathit{A}_{{\mathrm{1}}}  *  \mathit{B}}%
\ottpremise{\Delta  \vdash  \mathit{A}_{{\mathrm{2}}}  *  \mathit{B}}%
}{
\Delta  \vdash  \mathit{A}  *  \mathit{B}}{%
{\ottdrulename{D\_andl}}{}%
}}

\newcommand{\ottdruleDXXandr}[1]{\ottdrule[#1]{%
\ottpremise{ \mathit{B}_{{\mathrm{1}}}  \mathbin{\lhd}  \mathit{B}  \mathbin{\rhd}  \mathit{B}_{{\mathrm{2}}} }%
\ottpremise{\Delta  \vdash  \mathit{A}  *  \mathit{B}_{{\mathrm{1}}}}%
\ottpremise{\Delta  \vdash  \mathit{A}  *  \mathit{B}_{{\mathrm{2}}}}%
}{
\Delta  \vdash  \mathit{A}  *  \mathit{B}}{%
{\ottdrulename{D\_andr}}{}%
}}

\newcommand{\ottdefndisjoint}[1]{\begin{ottdefnblock}[#1]{$\Delta  \vdash  \mathit{A}  *  \mathit{B}$}{\ottcom{Type Disjointness}}
\ottusedrule{\ottdruleDXXax{}}
\ottusedrule{\ottdruleDXXtopl{}}
\ottusedrule{\ottdruleDXXtopr{}}
\ottusedrule{\ottdruleDXXarrow{}}
\ottusedrule{\ottdruleDXXrcdEq{}}
\ottusedrule{\ottdruleDXXall{}}
\ottusedrule{\ottdruleDXXvarl{}}
\ottusedrule{\ottdruleDXXvarr{}}
\ottusedrule{\ottdruleDXXandl{}}
\ottusedrule{\ottdruleDXXandr{}}
\end{ottdefnblock}}

\newcommand{\ottdefnsTypeDisjointness}{
\ottdefndisjoint{}}

\newcommand{\ottdruleISXXrefl}[1]{\ottdrule[#1]{%
}{
 \mathit{A}  \lesssim  \mathit{A} }{%
{\ottdrulename{IS\_refl}}{}%
}}

\newcommand{\ottdruleISXXand}[1]{\ottdrule[#1]{%
\ottpremise{  \mathit{B}_{{\mathrm{1}}}  \mathbin{\lhd}  \mathit{B}  \mathbin{\rhd}  \mathit{B}_{{\mathrm{2}}}   \and     \mathit{A}_{{\mathrm{1}}}  \lesssim  \mathit{B}_{{\mathrm{1}}}   \and   \mathit{A}_{{\mathrm{2}}}  \lesssim  \mathit{B}_{{\mathrm{2}}}    }%
}{
 \mathit{A}_{{\mathrm{1}}}  \, \& \,  \mathit{A}_{{\mathrm{2}}}  \lesssim  \mathit{B} }{%
{\ottdrulename{IS\_and}}{}%
}}

\newcommand{\ottdefnsubsub}[1]{\begin{ottdefnblock}[#1]{$ \mathit{A}  \lesssim  \mathit{B} $}{\ottcom{Isomorphic Subtyping}}
\ottusedrule{\ottdruleISXXrefl{}}
\ottusedrule{\ottdruleISXXand{}}
\end{ottdefnblock}}

\newcommand{\ottdefnsIsomorphicSubtyping}{
\ottdefnsubsub{}}

\newcommand{\ottdruleADXXandArrow}[1]{\ottdrule[#1]{%
\ottpremise{  \mathit{A}_{{\mathrm{1}}}  \rhd  \mathit{B}_{{\mathrm{1}}}  \rightarrow  \mathit{C}_{{\mathrm{1}}}   \and   \mathit{A}_{{\mathrm{2}}}  \rhd  \mathit{B}_{{\mathrm{2}}}  \rightarrow  \mathit{C}_{{\mathrm{2}}}  }%
}{
 \mathit{A}_{{\mathrm{1}}}  \, \& \,  \mathit{A}_{{\mathrm{2}}}  \rhd  \mathit{B}_{{\mathrm{1}}}  \, \& \,  \mathit{B}_{{\mathrm{2}}}  \rightarrow  \mathit{C}_{{\mathrm{1}}}  \, \& \,  \mathit{C}_{{\mathrm{2}}} }{%
{\ottdrulename{AD\_andArrow}}{}%
}}

\newcommand{\ottdruleADXXandRcd}[1]{\ottdrule[#1]{%
\ottpremise{  \mathit{A}_{{\mathrm{1}}}  \rhd  \ottsym{\{}  \ottmv{l}  \!\vcentcolon\!  \mathit{B}_{{\mathrm{1}}}  \ottsym{\}}   \and   \mathit{A}_{{\mathrm{2}}}  \rhd  \ottsym{\{}  \ottmv{l}  \!\vcentcolon\!  \mathit{B}_{{\mathrm{2}}}  \ottsym{\}}  }%
}{
 \mathit{A}_{{\mathrm{1}}}  \, \& \,  \mathit{A}_{{\mathrm{2}}}  \rhd  \ottsym{\{}  \ottmv{l}  \!\vcentcolon\!  \mathit{B}_{{\mathrm{1}}}  \, \& \,  \mathit{B}_{{\mathrm{2}}}  \ottsym{\}} }{%
{\ottdrulename{AD\_andRcd}}{}%
}}

\newcommand{\ottdruleADXXandAll}[1]{\ottdrule[#1]{%
\ottpremise{  \mathit{A}_{{\mathrm{1}}}  \rhd   \forall   \ottmv{X} * \mathit{B}_{{\mathrm{1}}} .\, \mathit{C}_{{\mathrm{1}}}    \and   \mathit{A}_{{\mathrm{2}}}  \rhd   \forall   \ottmv{X} * \mathit{B}_{{\mathrm{2}}} .\, \mathit{C}_{{\mathrm{2}}}   }%
}{
 \mathit{A}_{{\mathrm{1}}}  \, \& \,  \mathit{A}_{{\mathrm{2}}}  \rhd   \forall   \ottmv{X} * \mathit{B}_{{\mathrm{1}}}  \, \& \,  \mathit{B}_{{\mathrm{2}}} .\, \ottsym{(}  \mathit{C}_{{\mathrm{1}}}  \, \& \,  \mathit{C}_{{\mathrm{2}}}  \ottsym{)}  }{%
{\ottdrulename{AD\_andAll}}{}%
}}

\newcommand{\ottdruleADXXrefl}[1]{\ottdrule[#1]{%
}{
 \mathit{A}  \rhd  \mathit{A} }{%
{\ottdrulename{AD\_refl}}{}%
}}

\newcommand{\ottdefnappDist}[1]{\begin{ottdefnblock}[#1]{$ \mathit{A}  \rhd  \mathit{B} $}{\ottcom{Applicative Distribution}}
\ottusedrule{\ottdruleADXXandArrow{}}
\ottusedrule{\ottdruleADXXandRcd{}}
\ottusedrule{\ottdruleADXXandAll{}}
\ottusedrule{\ottdruleADXXrefl{}}
\end{ottdefnblock}}

\newcommand{\ottdefnsApplicativeDistribution}{
\ottdefnappDist{}}

\newcommand{\ottdrulePAppXXabs}[1]{\ottdrule[#1]{%
\ottpremise{ \mathit{B}  \rhd  \mathit{C}_{{\mathrm{1}}}  \rightarrow  \mathit{C}_{{\mathrm{2}}} }%
\ottpremise{ \mathit{e}_{{\mathrm{2}}} \,\rightsquigarrow_{ \mathit{A} }\, \mathit{u} }%
}{
 \ottsym{(}   \lambda  \ottmv{x} \!\vcentcolon\! \mathit{A} .\, \mathit{e}_{{\mathrm{1}}}   \ottsym{)}  \!\vcentcolon\!  \mathit{B}  \bullet  \mathit{e}_{{\mathrm{2}}} ~ \hookrightarrow ~ \ottsym{(}  \mathit{e}_{{\mathrm{1}}}  \ottsym{[}  \ottmv{x}  \mapsto  \mathit{u}  \ottsym{]}  \ottsym{)}  \!\vcentcolon\!  \mathit{C}_{{\mathrm{2}}} }{%
{\ottdrulename{PApp\_abs}}{}%
}}

\newcommand{\ottdrulePAppXXtabs}[1]{\ottdrule[#1]{%
\ottpremise{ \mathit{A}  \rhd   \forall   \ottmv{X} * \mathit{B}_{{\mathrm{1}}} .\, \mathit{B}_{{\mathrm{2}}}  }%
}{
 \ottsym{(}   \Lambda  \ottmv{X} .\, \mathit{e}   \ottsym{)}  \!\vcentcolon\!  \mathit{A}  \bullet  \mathit{C} ~ \hookrightarrow ~ \ottsym{(}  \mathit{e}  \ottsym{[}  \ottmv{X}  \mapsto  \mathit{C}  \ottsym{]}  \ottsym{)}  \!\vcentcolon\!  \ottsym{(}  \mathit{B}_{{\mathrm{2}}}  \ottsym{[}  \ottmv{X}  \mapsto  \mathit{C}  \ottsym{]}  \ottsym{)} }{%
{\ottdrulename{PApp\_tabs}}{}%
}}

\newcommand{\ottdrulePAppXXproj}[1]{\ottdrule[#1]{%
\ottpremise{ \mathit{A}  \rhd  \ottsym{\{}  \ottmv{l}  \!\vcentcolon\!  \mathit{B}  \ottsym{\}} }%
}{
  \{  \ottmv{l} \ottsym{=} \mathit{e}  \}   \!\vcentcolon\!  \mathit{A}  \bullet  \ottsym{\{}  \ottmv{l}  \ottsym{\}} ~ \hookrightarrow ~ \mathit{e}  \!\vcentcolon\!  \mathit{B} }{%
{\ottdrulename{PApp\_proj}}{}%
}}

\newcommand{\ottdrulePAppXXmerge}[1]{\ottdrule[#1]{%
\ottpremise{  \mathit{v}_{{\mathrm{1}}}  \bullet  \ottnt{arg} ~ \hookrightarrow ~ \mathit{u}_{{\mathrm{1}}}   \and   \mathit{v}_{{\mathrm{2}}}  \bullet  \ottnt{arg} ~ \hookrightarrow ~ \mathit{u}_{{\mathrm{2}}}  }%
}{
 \mathit{v}_{{\mathrm{1}}}  \,,,\,  \mathit{v}_{{\mathrm{2}}}  \bullet  \ottnt{arg} ~ \hookrightarrow ~ \mathit{u}_{{\mathrm{1}}}  \,,,\,  \mathit{u}_{{\mathrm{2}}} }{%
{\ottdrulename{PApp\_merge}}{}%
}}

\newcommand{\ottdefnpapp}[1]{\begin{ottdefnblock}[#1]{$ \mathit{v}  \bullet  \ottnt{arg} ~ \hookrightarrow ~ \mathit{u} $}{\ottcom{Parallel Application}}
\ottusedrule{\ottdrulePAppXXabs{}}
\ottusedrule{\ottdrulePAppXXtabs{}}
\ottusedrule{\ottdrulePAppXXproj{}}
\ottusedrule{\ottdrulePAppXXmerge{}}
\end{ottdefnblock}}

\newcommand{\ottdefnsParallelApplication}{
\ottdefnpapp{}}

\newcommand{\ottdrulePTXXtop}[1]{\ottdrule[#1]{%
}{
 \top  :   \mathsf{Top}  }{%
{\ottdrulename{PT\_top}}{}%
}}

\newcommand{\ottdrulePTXXint}[1]{\ottdrule[#1]{%
}{
 \ottmv{i}  :   \mathsf{Int}  }{%
{\ottdrulename{PT\_int}}{}%
}}

\newcommand{\ottdrulePTXXanno}[1]{\ottdrule[#1]{%
}{
 \ottsym{(}  \mathit{e}  \!\vcentcolon\!  \mathit{A}  \ottsym{)}  :  \mathit{A} }{%
{\ottdrulename{PT\_anno}}{}%
}}

\newcommand{\ottdrulePTXXmerge}[1]{\ottdrule[#1]{%
\ottpremise{ \mathit{u}_{{\mathrm{1}}}  :  \mathit{A} }%
\ottpremise{ \mathit{u}_{{\mathrm{2}}}  :  \mathit{B} }%
}{
 \ottsym{(}  \mathit{u}_{{\mathrm{1}}}  \,,,\,  \mathit{u}_{{\mathrm{2}}}  \ottsym{)}  :  \ottsym{(}  \mathit{A}  \, \& \,  \mathit{B}  \ottsym{)} }{%
{\ottdrulename{PT\_merge}}{}%
}}

\newcommand{\ottdefnpType}[1]{\begin{ottdefnblock}[#1]{$ \mathit{u}  :  \mathit{A} $}{\ottcom{Principal Type of Pre-Values}}
\ottusedrule{\ottdrulePTXXtop{}}
\ottusedrule{\ottdrulePTXXint{}}
\ottusedrule{\ottdrulePTXXanno{}}
\ottusedrule{\ottdrulePTXXmerge{}}
\end{ottdefnblock}}

\newcommand{\ottdefnsPrincipalType}{
\ottdefnpType{}}

\newcommand{\ottdruleCXXlit}[1]{\ottdrule[#1]{%
}{
\ottmv{i}  \approx  \ottmv{i}}{%
{\ottdrulename{C\_lit}}{}%
}}

\newcommand{\ottdruleCXXanno}[1]{\ottdrule[#1]{%
}{
\mathit{e}  \!\vcentcolon\!  \mathit{A}  \approx  \mathit{e}  \!\vcentcolon\!  \mathit{B}}{%
{\ottdrulename{C\_anno}}{}%
}}

\newcommand{\ottdruleCXXdisjoint}[1]{\ottdrule[#1]{%
\ottpremise{ \cdot   \vdash  \mathit{A}  *  \mathit{B}}%
\ottpremise{  \mathit{u}_{{\mathrm{1}}}  :  \mathit{A}   \and   \mathit{u}_{{\mathrm{2}}}  :  \mathit{B}  }%
}{
\mathit{u}_{{\mathrm{1}}}  \approx  \mathit{u}_{{\mathrm{2}}}}{%
{\ottdrulename{C\_disjoint}}{}%
}}

\newcommand{\ottdruleCXXmergel}[1]{\ottdrule[#1]{%
\ottpremise{\mathit{u}_{{\mathrm{1}}}  \approx  \mathit{u}}%
\ottpremise{\mathit{u}_{{\mathrm{2}}}  \approx  \mathit{u}}%
}{
\mathit{u}_{{\mathrm{1}}}  \,,,\,  \mathit{u}_{{\mathrm{2}}}  \approx  \mathit{u}}{%
{\ottdrulename{C\_mergel}}{}%
}}

\newcommand{\ottdruleCXXmerger}[1]{\ottdrule[#1]{%
\ottpremise{\mathit{u}  \approx  \mathit{u}_{{\mathrm{1}}}}%
\ottpremise{\mathit{u}  \approx  \mathit{u}_{{\mathrm{2}}}}%
}{
\mathit{u}  \approx  \mathit{u}_{{\mathrm{1}}}  \,,,\,  \mathit{u}_{{\mathrm{2}}}}{%
{\ottdrulename{C\_merger}}{}%
}}

\newcommand{\ottdefnconsistent}[1]{\begin{ottdefnblock}[#1]{$\mathit{u}_{{\mathrm{1}}}  \approx  \mathit{u}_{{\mathrm{2}}}$}{\ottcom{Consistency}}
\ottusedrule{\ottdruleCXXlit{}}
\ottusedrule{\ottdruleCXXanno{}}
\ottusedrule{\ottdruleCXXdisjoint{}}
\ottusedrule{\ottdruleCXXmergel{}}
\ottusedrule{\ottdruleCXXmerger{}}
\end{ottdefnblock}}

\newcommand{\ottdefnsConsistent}{
\ottdefnconsistent{}}

\newcommand{\ottdruleStepXXpapp}[1]{\ottdrule[#1]{%
\ottpremise{ \mathit{v}  \bullet  \mathit{e} ~ \hookrightarrow ~ \mathit{u} }%
}{
 \mathit{v} \, \mathit{e} \, \hookrightarrow \, \mathit{u} }{%
{\ottdrulename{Step\_papp}}{}%
}}

\newcommand{\ottdruleStepXXpproj}[1]{\ottdrule[#1]{%
\ottpremise{ \mathit{v}  \bullet  \ottsym{\{}  \ottmv{l}  \ottsym{\}} ~ \hookrightarrow ~ \mathit{u} }%
}{
 \mathit{v}  \ottsym{.}  \ottmv{l} \, \hookrightarrow \, \mathit{u} }{%
{\ottdrulename{Step\_pproj}}{}%
}}

\newcommand{\ottdruleStepXXptapp}[1]{\ottdrule[#1]{%
\ottpremise{ \mathit{v}  \bullet  \mathit{A} ~ \hookrightarrow ~ \mathit{u} }%
}{
 \mathit{v} \, \mathit{A} \, \hookrightarrow \, \mathit{u} }{%
{\ottdrulename{Step\_ptapp}}{}%
}}

\newcommand{\ottdruleStepXXfix}[1]{\ottdrule[#1]{%
}{
  \mathbf{fix}~ \ottmv{x} \!\vcentcolon\! \mathit{A} .\, \mathit{e}  \, \hookrightarrow \, \mathit{e}  \ottsym{[}  \ottmv{x}  \mapsto   \mathbf{fix}~ \ottmv{x} \!\vcentcolon\! \mathit{A} .\, \mathit{e}   \ottsym{]}  \!\vcentcolon\!  \mathit{A} }{%
{\ottdrulename{Step\_fix}}{}%
}}

\newcommand{\ottdruleStepXXannov}[1]{\ottdrule[#1]{%
\ottpremise{ \mathsf{pre\text{-}value}~ \mathit{v} }%
\ottpremise{ \mathit{v} \, \hookrightarrow _{ \mathit{A} }\, \mathit{v}' }%
}{
 \mathit{v}  \!\vcentcolon\!  \mathit{A} \, \hookrightarrow \, \mathit{v}' }{%
{\ottdrulename{Step\_annov}}{}%
}}

\newcommand{\ottdruleStepXXmerge}[1]{\ottdrule[#1]{%
\ottpremise{ \mathit{e}_{{\mathrm{1}}} \, \hookrightarrow \, \mathit{e}'_{{\mathrm{1}}} }%
\ottpremise{ \mathit{e}_{{\mathrm{2}}} \, \hookrightarrow \, \mathit{e}'_{{\mathrm{2}}} }%
}{
 \mathit{e}_{{\mathrm{1}}}  \,,,\,  \mathit{e}_{{\mathrm{2}}} \, \hookrightarrow \, \mathit{e}'_{{\mathrm{1}}}  \,,,\,  \mathit{e}'_{{\mathrm{2}}} }{%
{\ottdrulename{Step\_merge}}{}%
}}

\newcommand{\ottdruleStepXXcntx}[1]{\ottdrule[#1]{%
\ottpremise{ \mathit{e} \, \hookrightarrow \, \mathit{e}' }%
}{
 \ottnt{E}  \ottsym{[}  \mathit{e}  \ottsym{]} \, \hookrightarrow \, \ottnt{E}  \ottsym{[}  \mathit{e}'  \ottsym{]} }{%
{\ottdrulename{Step\_cntx}}{}%
}}

\newcommand{\ottdefnstep}[1]{\begin{ottdefnblock}[#1]{$ \mathit{e}_{{\mathrm{1}}} \, \hookrightarrow \, \mathit{e}_{{\mathrm{2}}} $}{\ottcom{Small-Step Semantics}}
\ottusedrule{\ottdruleStepXXpapp{}}
\ottusedrule{\ottdruleStepXXpproj{}}
\ottusedrule{\ottdruleStepXXptapp{}}
\ottusedrule{\ottdruleStepXXfix{}}
\ottusedrule{\ottdruleStepXXannov{}}
\ottusedrule{\ottdruleStepXXmerge{}}
\ottusedrule{\ottdruleStepXXcntx{}}
\end{ottdefnblock}}

\newcommand{\ottdefnsReduction}{
\ottdefnstep{}}

\newcommand{\ottdruleEWXXtop}[1]{\ottdrule[#1]{%
\ottpremise{  \cdot    \vdash  \rceil A^\circ \lceil }%
}{
 \mathit{e} \,\rightsquigarrow_{ A^\circ }\,  [\![  A^\circ  ]\!]  }{%
{\ottdrulename{EW\_top}}{}%
}}

\newcommand{\ottdruleEWXXanno}[1]{\ottdrule[#1]{%
\ottpremise{  \cdot    \vdash  \neg \rceil B^\circ \lceil }%
}{
 \mathit{e} \,\rightsquigarrow_{ B^\circ }\, \mathit{e}  \!\vcentcolon\!  B^\circ }{%
{\ottdrulename{EW\_anno}}{}%
}}

\newcommand{\ottdruleEWXXand}[1]{\ottdrule[#1]{%
\ottpremise{  \mathit{B}_{{\mathrm{1}}}  \mathbin{\lhd}  \mathit{A}  \mathbin{\rhd}  \mathit{B}_{{\mathrm{2}}}   \and     \mathit{e} \,\rightsquigarrow_{ \mathit{B}_{{\mathrm{1}}} }\, \mathit{u}_{{\mathrm{1}}}   \and   \mathit{e} \,\rightsquigarrow_{ \mathit{B}_{{\mathrm{2}}} }\, \mathit{u}_{{\mathrm{2}}}    }%
}{
 \mathit{e} \,\rightsquigarrow_{ \mathit{A} }\, \mathit{u}_{{\mathrm{1}}}  \,,,\,  \mathit{u}_{{\mathrm{2}}} }{%
{\ottdrulename{EW\_and}}{}%
}}

\newcommand{\ottdefnCastExpression}[1]{\begin{ottdefnblock}[#1]{$ \mathit{e} \,\rightsquigarrow_{ \mathit{A} }\, \mathit{u} $}{\ottcom{Expression Wrapping}}
\ottusedrule{\ottdruleEWXXtop{}}
\ottusedrule{\ottdruleEWXXanno{}}
\ottusedrule{\ottdruleEWXXand{}}
\end{ottdefnblock}}

\newcommand{\ottdefnsCastExpression}{
\ottdefnCastExpression{}}

  \renewottcommands[ott]

\newcommand{\lambdai}{\texorpdfstring{$\lambda_{i}$}{Lambda-i}\xspace}
\newcommand{\lambdaip}{\texorpdfstring{$\lambda_{i}^{\!+}$}{Lambda-i-plus}\xspace}
\newcommand{\lambdamerge}{$\lambda^{||}$\xspace}
\newcommand{\fname}{$\mathsf{F}_{i}$\xspace}
\newcommand{\fip}{$\mathsf{F}_{i}^{+}$\xspace}
\newcommand{\fco}{$\mathsf{F}_{co}$\xspace}
\newcommand{\fsub}{$\mathsf{F}_{<:}$\xspace}
\newcommand{\systemf}{System \textsf{F}\xspace}

\title{Direct~Foundations~for Compositional~Programming}

\nolinenumbers

\author{Andong Fan\footnote[1]{The first two authors contributed equally to this work.}}
       {Zhejiang University, Hangzhou, China}{afan2018@zju.edu.cn}{https://orcid.org/0000-0003-2124-9625}{}

\author{Xuejing Huang\footnotemark[1]}{The University of Hong Kong, China}{xjhuang@cs.hku.hk}{https://orcid.org/0000-0002-8496-491X}{}

\author{Han Xu}{Peking University, Beijing, China}{1800012917@pku.edu.cn}{}{}

\author{Yaozhu Sun}{The University of Hong Kong, China}{yzsun@cs.hku.hk}{}{}

\author{Bruno C. d. S. Oliveira}{The University of Hong Kong, China}{bruno@cs.hku.hk}{}{}

\funding{This research was funded by the University of Hong Kong and Hong Kong Research Grants Council
projects number 17209519, 17209520 and 17209821.}

\Copyright{Andong Fan, Xuejing Huang, Han Xu, Yaozhu Sun, and Bruno C. d. S. Oliveira}

\authorrunning{A. Fan, X. Huang, H. Xu, Y. Sun, and B.\,C.\,d.\,S.\,Oliveira}

\acknowledgements{We thank the anonymous reviewers for their helpful comments.}

\ccsdesc[500]{Theory of computation~Type theory}

\keywords{Intersection types, disjoint polymorphism, operational semantics} %mandatory; please add comma-separated list of keywords

\renewcommand{\paragraph}[1]{\vspace{3pt}\noindent {\sffamily \normalsize \bfseries #1.}}

\begin{document}

\maketitle

\begin{abstract}
The recently proposed CP language adopts Compositional Programming:
a new modular programming
style that solves challenging problems such as the Expression
Problem. CP is implemented on top of a polymorphic core language with
disjoint intersection types called \fip.  The semantics of \fip
employs an elaboration to a target language and relies on a
sophisticated proof technique to prove the \emph{coherence} of
the elaboration. Unfortunately, the proof technique is
technically challenging and hard to scale to many common features,
including recursion or impredicative polymorphism. Thus, the
original formulation of \fip does not support the two later features,
which creates a gap between theory and practice, since CP fundamentally
relies on them.

This paper presents a new formulation of \fip based on a
\emph{type-directed operational semantics} (TDOS).  The TDOS approach
was recently proposed to model the semantics of languages with
disjoint intersection types (but without polymorphism).  Our work
shows that the TDOS approach can be extended to languages with
disjoint polymorphism and model the full \fip calculus.  Unlike the
elaboration semantics, which gives the semantics to \fip indirectly
via a target language, the TDOS approach gives a semantics to \fip
directly. With a TDOS, there is no need for a coherence proof. Instead,
we can simply prove that the semantics is \emph{deterministic}. The
proof of determinism only uses simple reasoning techniques, such as
straightforward induction, and is able to handle problematic
features such as recursion and impredicative polymorphism. This
removes the gap between theory and practice and validates the original
proofs of correctness for CP.
We formalized the TDOS variant of the
\fip calculus and all its proofs in the Coq proof assistant.

\end{abstract}

\section{Introduction}\label{sec:intro}

\emph{Compositional Programming}~\cite{zhang2021compositional} is a recently
proposed modular programming paradigm. It offers a natural
solution to the \emph{Expression Problem}~\cite{wadler1998expression} and
novel approaches to modular pattern matching and dependency
injection. The CP language adopts Compositional Programming.
In CP, several new programming language constructs
enable Compositional Programming. Of particular interest for this
paper, CP has a notion of \emph{typed first-class traits}~\cite{bi_et_al:LIPIcs:2018:9214},
which are extended in CP to also enable a form of \emph{family polymorphism}~\cite{Ernst_2001}.

The semantics of CP and its notion of traits is defined via an
elaboration to the core calculus \fip~\cite{xuanbiesop}:
a polymorphic core language with a \emph{merge operator}~\cite{reynolds1988preliminary}
and \emph{disjoint intersection types}~\cite{oliveira2016disjoint}.
The elaboration of traits is inspired by Cook's \emph{denotational semantics of
  inheritance}~\cite{cookthesis}. In the denotational semantics
of inheritance, the key idea is that mechanisms such as classes or traits,
which support \emph{self-references} (a.k.a. the \lstinline{this} keyword in conventional
OOP languages), can be modeled via \emph{open recursion}.
In other words,
the encoding of classes or traits is parametrized by a self-reference. This allows
late binding of self-references at the point of instantiation and
enables the modification and composition of traits before instantiation.
Instantiation happens when \lstinline{new} is used, just as in conventional
OOP languages. When \lstinline{new} is used, it
essentially closes the recursion by binding the self-reference, which then
becomes a recursive reference to the instantiated object. In the denotational semantics of inheritance,
\lstinline{new} is just a fixpoint operator.

The semantics of the original formulation of \fip~\cite{xuanbiesop} itself is
also given by an elaboration into \fco, a \systemf-like language
with products.
Unlike \fip, \fco has no subtyping or intersection types, and it
has a conventional operational semantics.
The main reason for \fip to use elaboration is that \fip has a
\emph{type-dependent} semantics: types may affect the runtime behavior of a
program. The elaboration semantics for \fip seems like a natural choice, since
this is commonly seen in various other type-dependent languages and
calculi.
For instance, the semantics of type-dependent languages with \emph{type
classes}~\cite{wadler1989make}, \emph{Scala-style implicits}~\cite{Oliveira2010}
or \emph{gradual typing}~\cite{siek2006gradual} all usually adopt an
elaboration approach.
In contrast, in the past, more conventional direct formulations
using an operational semantics have been avoided for
languages with a type-dependent semantics.
The appeals of the elaboration semantics are simple type-safety proofs,
and the fact that they directly offer an implementation technique
over conventional languages without a type-dependent semantics.

There are also important drawbacks when using an elaboration
semantics. One of them is simply that more infrastructure is needed for
a target language (such as \fco) and its associated semantics and metatheory.
Moreover, the elaboration semantics is indirect, and to understand
the semantics of a program, we must first translate it to the target language
(which may be significantly different from the source) and then reason
in terms of the target. More importantly, besides type-safety,
another property that is often desirable for an elaboration semantics
is \emph{coherence}~\cite{Reynolds_1991}.
Many elaboration semantics are non-deterministic:
the same source program can elaborate into different target programs.
If those different programs have a different semantics, then this is
problematic, as it would imply that the source language would
have a non-deterministic or ambiguous semantics. Coherence
ensures that even if the same program elaborates to different target
expressions, the different target expressions are semantically equivalent,
eventually evaluating to the same result.

For some languages, including \fip, proving coherence is highly
non-trivial and hard to scale to common programming language
features.  For the original \fip, the proof of coherence comes at the cost of
simple features such as \emph{recursion} and \emph{impredicative polymorphism}.
The proof of coherence for \fip is
based on a logical relation called \emph{canonicity}~\cite{bi_et_al:LIPIcs:2018:9227}.
Together with a notion of contextual equivalence, the two techniques are used to
prove coherence. The use of logical relations is a source
of complexity in the proof and the reason why recursion and
impredicative polymorphism have not been supported. For recursion, in
principle, the use of a more sophisticated \emph{step-indexed} logical
relation~\cite{appel01step} may enable a proof of coherence, at the cost of some
additional complexity. However, due to the extra complexity, this was
left for future work. For impredicative polymorphism, Bi et
al.~\cite{xuanbiesop} identified important technical challenges, and it is not
known if the proof can be extended with such a feature.

The absence of recursion and impredicative polymorphism creates a
gap between theory and practice, since CP fundamentally relies on them.
Moreover, the
proofs of correctness of CP rely on the assumption that \fip with
recursion and impredicative polymorphism would preserve all the
properties of \fip.
Impredicative polymorphism is needed in CP to allow the types of
traits with polymorphic methods to be used as type parameters for
other polymorphic functions. Recursion is needed in CP because
the denotational semantics uses fixpoint operators to instantiate traits.
In addition, the fixpoint operators must be \emph{lazy}; otherwise,
self-references can easily trigger non-termination. Therefore,
a \emph{call-by-name} (CBN) semantics is more natural and also assumed
in the CP encoding of traits. However, the semantics
of the \fco calculus is \emph{call-by-value} (CBV) and, by inheritance, the elaboration
semantics of \fip has a CBV semantics as well.

This paper presents a new formulation of \fip based on a
\emph{type-directed operational semantics} (TDOS)~\cite{huang_et_al:LIPIcs:2020:13183}.
The TDOS approach
has recently been proposed to model the semantics of languages with
disjoint intersection types (but without polymorphism). Although \fip is not a
new calculus, we revise its formulation significantly in this paper. Our new
formulation of \fip is different from the original one in three aspects. Firstly,
the semantics of the original \fip is given by elaborating to \fco,
while our semantics for \fip is a direct operational semantics.
Secondly, our new formulation of \fip supports recursion and impredicative polymorphism.
Finally, we employ a call-by-name evaluation strategy.

Our work shows that the TDOS approach can be extended to languages
with disjoint polymorphism and model the complete \fip calculus with recursion
and impredicative polymorphism.
Moreover, %With a TDOS,
there is no need for a coherence proof.
Instead, we can simply prove that the semantics is
\emph{deterministic}. The proof of determinism uses only simple
reasoning techniques, such as straightforward induction, and is able
to %easily
handle problematic features such as recursion and
impredicative polymorphism.
Thus, this removes the gap between theory
and practice and validates the original proofs of correctness for the
CP language.
Figure~\ref{fig:contrib} contrasts the differences in terms of proofs and implementation
of CP using Zhang et al.'s original work and our own work.
We formalized the TDOS variant of the \fip calculus, together with its
type-soundness and determinism proof in the Coq proof assistant.
Moreover, we have a new implementation of CP based on our new
reformulation of \fip.

In summary, the contributions of this work are:

\begin{figure}[t]
\includegraphics[width=.98\textwidth]{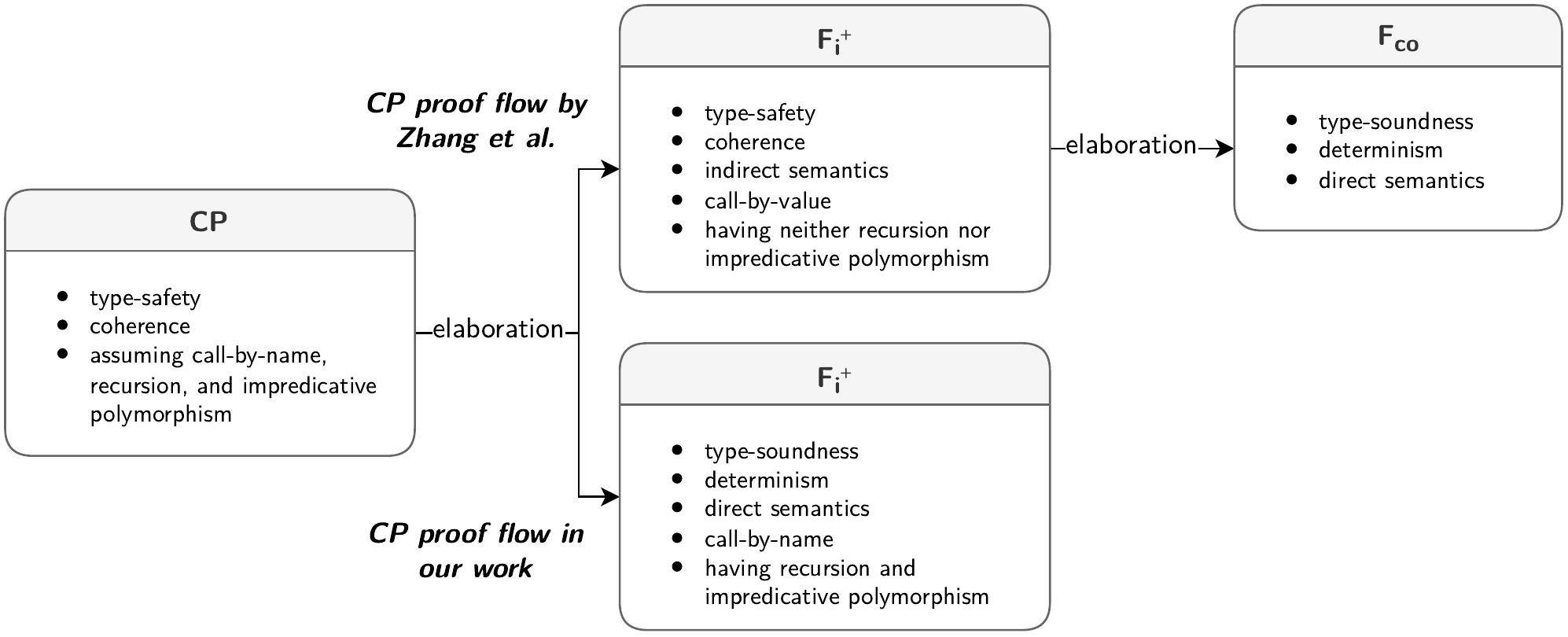}
\caption{Contrasting the flow of results for CP using the original formulation, and our work.}
\label{fig:contrib}
\end{figure}

\begin{itemize}

\item {\bf CBN \fip with recursion and impredicative polymorphism.}
  This paper presents a CBN variant of \fip extended with recursion and
  impredicative polymorphism.

\item {\bf Determinism and type-soundness for \fip using a TDOS.}
  We prove the type-soundness and determinism of \fip using a direct
  TDOS. These proofs validate the
  proofs of correctness previously presented for CP by Zhang et al.~\cite{zhang2021compositional}.

\item {\bf Technical innovations.} Our formulation of \fip has various
  technical innovations over the original one, including a new formulation
  of subtyping using splittable types~\cite{tamingmerge} and more flexible
  term applications.

\begin{comment}
\item {\bf New implementation of CP using our \fip variant.} We have a
new implementation of CP built on top of a TDOS formulation of \fip.
The implementation employs call-by-\emph{need} as an optimization to call-by-name,
and is available at \url{https://plground.org}.
\end{comment}

\item {\bf Implementation and Mechanical formalization.}
We formalized the TDOS variant of the \fip calculus, together with its type-soundness and
determinism proof in the Coq proof assistant. We also have a new implementation of CP built on top of a
TDOS formulation of \fip available at \url{https://plground.org}.
The full Coq formalization is available at:

\begin{center}
  \url{https://github.com/andongfan/CP-Foundations}
\end{center}

\end{itemize}

\section{Motivations and Technical Innovations}

In this section, we introduce Compositional Programming by example and
show how CP traits elaborate to \fip expressions. After that, we
will discuss the practical issues that motivate us to reformulate \fip, as well as
technical challenges and innovations.

\subsection{Compositional Programming by Example}\label{sec:cp}

To demonstrate the capabilities of Compositional Programming, we
show how to solve a variant of the \emph{Expression Problem}~\cite{wadler1998expression} in the CP language.
Our solution is adapted from the original one by Zhang et al.~\cite{zhang2021compositional}.
In this variant, in addition to the usual challenge of extensibility in multiple
directions, we also consider the problem of \emph{context evolution}~\cite{liang1995monad,schrijvers2011monads},
so the interpreter may require different contextual information %(such as environments)
for different features of the interpreter.
The CP language allows a modular solution to both challenges,
which also illustrates some key
features in Compositional Programming, including
\emph{first-class traits}~\cite{bi_et_al:LIPIcs:2018:9214},
\emph{nested composition}~\cite{bi_et_al:LIPIcs:2018:9227}, and
\emph{disjoint polymorphism}~\cite{alpuimdisjoint}.

Examples are based on a simple
expression language, and the goal is to perform various operations over it, such as evaluation and free variable bookkeeping.
The expression language consists of numbers, addition, variables, and
let-bindings.
Besides CP code, we also provide analogous Haskell code in the initial examples
so that readers can connect them with existing concepts in functional languages.

\begin{figure}
\begin{subfigure}[t]{.55\linewidth}
\begin{lstlisting}
type NumSig<Exp> = {
  Lit : Int -> Exp;
  Add : Exp -> Exp -> Exp;
};

type Eval Ctx = { eval : Ctx -> Int };
evalNum Ctx = trait implements NumSig<Eval Ctx> => {
  (Lit     n).eval _   = n;
  (Add e1 e2).eval ctx =
     e1.eval ctx + e2.eval ctx;
};
\end{lstlisting}
\caption{CP code.} \label{fig:eval-cp}
\end{subfigure}
\begin{subfigure}[t]{.45\linewidth}
\begin{lstlisting}[language=Haskell,deletekeywords=Eval]
data Exp where
  Lit :: Int -> Exp
  Add :: Exp -> Exp -> Exp


type Eval ctx = ctx -> Int

eval :: Exp -> Eval ctx
eval (Lit     n) _   = n
eval (Add e1 e2) ctx =
   eval e1 ctx + eval e2 ctx
\end{lstlisting}
\bigskip
\caption{Haskell counterpart.} \label{fig:eval-hs}
\end{subfigure}
\caption{Initial expression language: numbers and addition.}
\end{figure}

\paragraph{Compositional interfaces} First, we define the compositional
interface for numeric literals and addition. The compositional interface at the top of \Cref{fig:eval-cp}
is similar to Haskell's algebraic data type at the top of \Cref{fig:eval-hs}.
\lstinline{Exp} is a special kind of type parameter in CP called a \emph{sort},
which serves as the return type of both constructors \lstinline{Lit} and
\lstinline{Add}. Sorts will be instantiated with concrete representations
later. Internally, sorts are
handled differently from normal type parameters~\cite{zhang2021compositional}.
In accordance with the compositional interface, we can then define how to
evaluate the expression language.

\paragraph{Polymorphic contexts} As shown in the middle of \Cref{fig:eval-cp},
the type \lstinline{Eval} declares a method \lstinline{eval} that
takes a context and returns an integer.
\lstinline{Ctx} is a type parameter that can be instantiated later, % in various ways,
enabling particular traits to assume particular
contextual information for the needs of various features.
The technique is called \emph{polymorphic contexts}~\cite{zhang2021compositional}
in Compositional Programming.

\paragraph{Compositional traits} The trait \lstinline{evalNum} in
\Cref{fig:eval-cp} is parametrized by a type parameter \lstinline{Ctx}.
Note that, in CP, type parameters always start with a capital letter, while
regular parameters are lowercase. The trait \lstinline{evalNum}
implements the compositional interface \lstinline{NumSig} by instantiating it
with the sort \lstinline{Eval Ctx}.
\emph{Traits} are the basic reusable unit in CP,
which are usually type-checked against compositional interfaces. In this trait, we use
a lightweight syntax called \emph{method patterns} to define how to evaluate
different expressions. Such a definition is analogous to pattern matching in
\Cref{fig:eval-hs}. Since
\lstinline{Lit} and \lstinline{Add} do not need to be conscious of any information
in the context, the type parameter \lstinline{Ctx} is unconstrained. The
only thing that we can do to the polymorphic context is either to ignore it (like
in \lstinline{Lit}) or pass it to recursive calls
(like in \lstinline{Add}).

\begin{figure}[t]
\begin{lstlisting}
type VarSig<Exp> = {
  Let : String -> Exp -> Exp -> Exp;
  Var : String -> Exp;
};

type Env = { env : String -> Int };
evalVar (Ctx*Env) = trait implements VarSig<Eval (Env&Ctx)> => {
  (Let s e1 e2).eval ctx = e2.eval
    { ctx with env = insert s (e1.eval ctx) ctx.env };
  (Var       s).eval ctx = lookup s ctx.env;
};
\end{lstlisting}
\caption{Adding more expressions: variables and let-bindings.} \label{fig:var}
\end{figure}

\paragraph{More expressions} Adding more constructs to the expression
language is awkward in Haskell because algebraic data types are \emph{closed}.
However, language components can be modularly declared in CP. Two new constructors, \lstinline{Let} and \lstinline{Var}, are
declared in the second compositional interface \lstinline{VarSig}, as shown in \Cref{fig:var}. Then the two
traits implement \lstinline{VarSig} using method patterns for the new constructors. Since the two new
expressions need to inspect or update some information in the context, we expose
the appropriate \lstinline{Env} part to \lstinline{evalVar}, while the remaining
context is kept polymorphic. This is achieved with the  \emph{disjointness constraint}~\cite{alpuimdisjoint} \lstinline{Ctx*Env}
in \lstinline{evalVar}. A disjointness constraint denotes that the type parameter \lstinline{Ctx} is disjoint to the type
\lstinline{Env}. In other words, types that instantiate \lstinline{Ctx} cannot
overlap with the type \lstinline{Env}.
Also note that the notation \lstinline/{ ctx with env = ... }/ denotes a
\emph{polymorphic record update}~\cite{cardelli1989operations}.
In the code for let-expressions, we need to update the environment
in the recursive calls to extend it with a new entry for the
let-variable.

\paragraph{Intersection types} Independently defined interfaces can be
composed using \emph{intersection types}.
For example, \lstinline{ExpSig} below
is an intersection of \lstinline{NumSig} and \lstinline{VarSig}, containing all
of the four constructors:

\begin{lstlisting}
type ExpSig<Exp> = NumSig<Exp> & VarSig<Exp>;
--               = { Lit : ...; Add : ...; Let : ...; Var : ... };
\end{lstlisting}

\begin{figure}
\begin{lstlisting}
type FV = { fv : [String] };
fv = trait implements ExpSig<FV> => {
  (Lit       n).fv = [];
  (Add   e1 e2).fv = union e1.fv e2.fv;
  (Let s e1 e2).fv = union e1.fv (delete s e2.fv);
  (Var       s).fv = [s];
};

evalWithFV (Ctx*Env) = trait implements ExpSig<FV => Eval (Env&Ctx)> => {
  (Lit       n).eval _   = n;
  (Add   e1 e2).eval ctx = e1.eval ctx + e2.eval ctx;
  (Let s e1 e2).eval ctx = if elem s e2.fv
    then e2.eval { ctx with env = insert s (e1.eval ctx) ctx.env }
    else e2.eval ctx;
  (Var       s).eval ctx = lookup s ctx.env;
};
\end{lstlisting}
\caption{Adding more operation: free variable bookkeeping and another version of evaluation.} \label{fig:shadow}
\end{figure}

\paragraph{More operations} Not only can expressions be modularly extended,
but we can easily add more operations. In \Cref{fig:shadow}, a new trait
\lstinline{fv} modularly implements a new operation that records free variables
in an expression. Here, \lstinline{union} and \lstinline{delete} are two library
functions for arrays.
The modular definition of \lstinline{fv} is quite natural in functional
programming, but it is hard in traditional object-oriented programming. We have
to modify the existing class definitions and supplement them with a method.
This is typical of the well-known Expression Problem. In summary, we have shown that Compositional
Programming can solve both dimensions of this problem: adding expressions and operations.

\paragraph{Dependency injection} Besides the Expression Problem,
\Cref{fig:shadow} also shows another significant feature of CP: dependency
injection. In \lstinline{evalWithFV}, a new implementation of
evaluation is defined with a dependency on free variables. The method pattern for
\lstinline{Let} will check if \lstinline{s} appears as a free variable in
\lstinline{e2}. If so, it evaluates \lstinline{e1} first as usual; otherwise, we
do not need to do any computation or update the environment since \lstinline{s} is
not used at all. Note that the compositional interface \lstinline{ExpSig} is
instantiated with two types separated by a fat arrow (\lstinline{=>}) (\lstinline{=>} was
originally denoted by \lstinline{%} in Zhang et al.'s implementation of CP).
\lstinline{FV} on the left-hand side is the dependency of \lstinline{evalWithFV}. In
other words, the definition of \lstinline{evalWithFV} depends on another trait that
implements \lstinline{ExpSig<FV>}. The static type checker of CP will check this
fact later at the point of trait instantiation. With such dependency injection,
we can call \lstinline{e2.fv} even if \lstinline{evalWithFV} does not have an
implementation of \lstinline{fv}. In other words, \lstinline{evalWithFV} depends
only on the interface of \lstinline{fv} (the type \lstinline{FV}),
but not any concrete implementation.

\paragraph{Self-type annotations}
Before we show how to perform the new version of the evaluation
over the whole expression language, we want to create a repository of
expressions for later use. We expect that these expressions are unaware of any
concrete operation, so we use a polymorphic \lstinline{Exp} type to denote
some abstract type of expressions. The code
that creates the repository of expressions is\footnote{In Zhang et al.'s
original work~\cite{zhang2021compositional}, the \lstinline{new} operator
must be added before every constructor. However, our new implementation will
implicitly insert \lstinline{new} (see \Cref{sec:encoding} for details).}:

\begin{lstlisting}
repo Exp = trait [self : ExpSig<Exp>] => {
  num = Add (Lit 4) (Lit 8);
  var = Let "x" (Lit 4) (Let "y" (Lit 8) (Add (Var "x") (Var "y")));
};
\end{lstlisting}

\noindent
To make constructors available from the compositional interface, we add a
\emph{self-type annotation} to the trait \lstinline{repo}. The self type annotation
\lstinline{[self : ExpSig<Exp>]} imposes a requirement that the \lstinline{repo}
should finally be merged with some trait implementing \lstinline{ExpSig<Exp>}.
This requirement is also statically enforced by the static type checker of CP. This is
the second mechanism in Compositional Programming to modularly inject dependencies.

\begin{figure}[t]
\includegraphics[width=\textwidth]{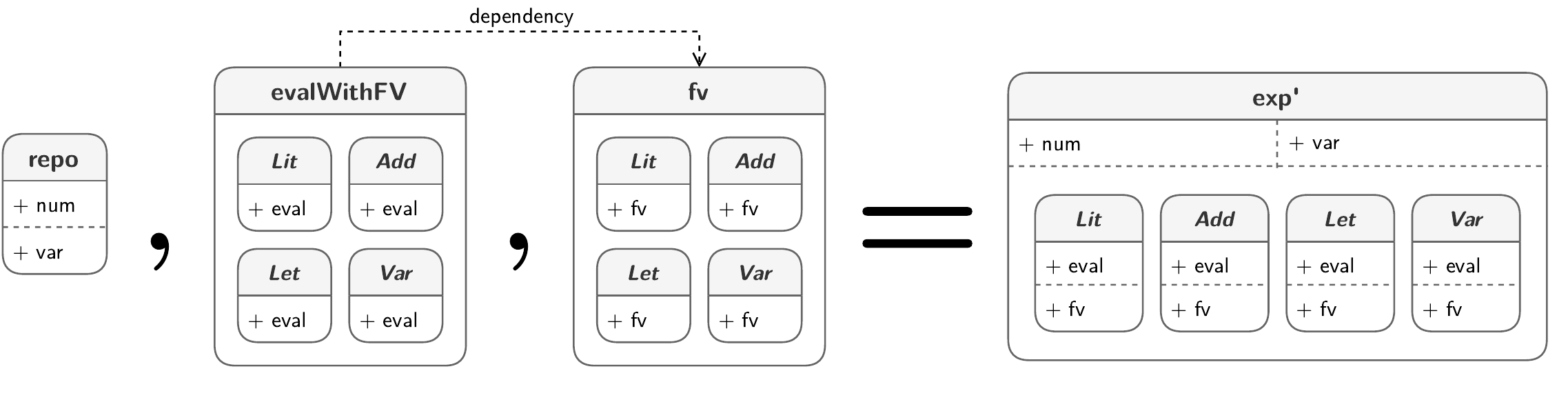}
\caption{Visualization of nested composition.} \label{fig:vis}
\end{figure}

\paragraph{Nested trait composition} With the language components
ready, we can compose them using the merge operator~\cite{dunfield2014elaborating}, which
in the CP language is denoted as a single comma (\lstinline{,}).
First, we show how to compose the old version of the evaluation:
\begin{lstlisting}
exp = new repo @(Eval Env) , evalNum @Env , evalVar @Top;
exp.var.eval { env = empty }  --> 12
\end{lstlisting}
\noindent Since the context has evolved after we add variables, we pass
different type arguments to the two traits to make the final context consistent.
The final context type is \lstinline{Env}, so we pass \lstinline{Env}
to \lstinline{evalNum} and \lstinline{Top} to \lstinline{evalVar}. Type arguments
are prefixed by \lstinline{@} in CP. A more
interesting example is to merge the new version of evaluation with free variable bookkeeping:
\begin{lstlisting}
exp' = new repo @(Eval Env & FV) , evalWithFV @Top , fv;
exp'.var.eval { env = empty }  --> 12
\end{lstlisting}
\noindent After the trait composition, both operations (\lstinline{eval}
and \lstinline{fv}) are available for expressions that
are built with the four constructors (\lstinline{Lit}, \lstinline{Add},
\lstinline{Let}, and \lstinline{Var}). Note that here \lstinline{fv} satisfies
the dependency of \lstinline{evalWithFV}. If no implementation of the type \lstinline{FV}
is present in the composition, there would be a type error, since
the requirement for \lstinline{evalWithFV} would not be satisfied. The
composition of the three traits is \emph{nested} because the two methods
nested in the four constructors are composed, as visualized in \Cref{fig:vis}.
With nested trait composition, the Expression Problem is elegantly solved in Compositional Programming.
Moreover, we allow context evolution using a relatively simple way with
polymorphic contexts.

\paragraph{Impredicative polymorphism}
Another feature of CP is that it allows the creation of objects with polymorphic
methods, similar to most OOP languages with generics where classes can contain polymorphic methods
(like Java).
However, for this to work properly, CP must support \emph{impredicative polymorphism}
(the ability to instantiate type parameters with polymorphic types) as \systemf does.
For example, consider:

\begin{lstlisting}
type Poly = { id : forall A. A -> A };
idTrait = trait implements Poly => { id = /\A. \(x:A) -> x };

(new idTrait).id @Poly  -- impredicative
\end{lstlisting}

\noindent
While accepted by our variant of CP and \fip, such polymorphic
instantiations are forbidden in the original formulation of \fip.

\subsection{Elaborating CP to \fip}\label{sec:encoding}

Under the surface of CP, the
foundation for Compositional Programming is the \fip calculus. We
present the key features in \fip and take a closer look at the connection
between CP and \fip expressions. Here we focus on the elaboration of
traits, which are the most important aspect of this paper.
We refer curious readers to the work by Zhang et
al.~\cite{zhang2021compositional} for the full formulation of the type-directed
elaboration of CP.

\paragraph{Key features of the \fip calculus}
\fip is basically a variant of \systemf~\cite{girard1972interpretation,reynolds1974towards}
extended with intersection types and a merge operator. In the \fip calculus, we denote the
merge operator with a double comma ($,,$) (instead of the single comma notation in CP),
following the original notation proposed by Dunfield~\cite{dunfield2014elaborating}. 
The merge operator allows us to
introduce terms of intersection types. For example,
$ \mathsf{1}   \,,,\,   \mathsf{true} $ is a term of type $ \mathsf{Int}   \, \& \,  \mathsf{Bool}$. Moreover, record
concatenation, which is used to encode multi-field traits, is encoded as merges of records in \fip. Thus, multi-field records
are represented as merges of multiple single-field records. However, to ensure
determinism of operational semantics, not all terms can be merged with each
other. We impose a disjointness check when typing merges:
a merge can only type check when the types of the terms being merged are
disjoint. It ensures that every part in a merge can be distinguished by its type.
For traits, for example, the disjointness restriction ensures that traits cannot
have two fields/methods with the same name \lstinline{m} and overlapping types, which
could otherwise lead to ambiguity when doing method lookup.
Here, we show an example of ambiguity if there is no disjointness check.
With intersection types, both $ \mathit{A}  \, \& \,  \mathit{B}  \leq  \mathit{A} $ and $ \mathit{A}  \, \& \,  \mathit{B}  \leq  \mathit{B} $
are valid. Therefore, a merge $ \mathsf{1}   \,,,\,   \mathsf{2} $ of type $ \mathsf{Int}   \, \& \,   \mathsf{Int} $ can be typed
with $ \mathsf{Int} $, but at runtime, two different values of type $ \mathsf{Int} $ are found.
Thus, an expression such as $\ottsym{(}   \mathsf{1}   \,,,\,   \mathsf{2}   \ottsym{)}  \ottsym{+}   \mathsf{1} $ could evaluate to either $2$ or $3$.
Since we wish for a deterministic semantics, we use disjointness to prevent such
forms of ambiguity. On the other hand, $\ottsym{(}   \mathsf{1}   \,,,\,   \mathsf{true}   \ottsym{)}  \ottsym{+}   \mathsf{1} $ type checks because
$ \mathsf{Int} $ and $\mathsf{Bool}$ are disjoint, and it evaluates to $2$ unambiguously.
A disjointness constraint can also be added to a type variable in
a \systemf-style polymorphic type, such as $ \forall   \ottmv{X} *  \mathsf{Int}  .\, \ottmv{X}   \, \& \,   \mathsf{Int} $.
Moreover, to support unrestricted intersection types like $ \mathsf{Int}   \, \& \,   \mathsf{Int} $, the
disjointness check is relaxed to \emph{consistency} for certain terms, so that
merges with duplications like $ \mathsf{1}   \,,,\,   \mathsf{1} $ are allowed.

\paragraph{Elaborating traits into \fip}
The elaboration of traits is inspired by Cook's denotational semantics of
inheritance~\cite{cookthesis}. To use a concrete example, we
revisit the trait \lstinline{repo} defined in \Cref{sec:cp}.
Both the creation and instantiation of traits are included in
the definition of \lstinline{repo}:

\begin{comment}
-- is desugared to

repo = /\Exp. trait [self : { ... }] =>
  open self in
    { num = new $Add (new $Lit 4) (new $Lit 8) } ,
    { var = ... };
\end{comment}

\begin{lstlisting}
repo Exp = trait [self : ExpSig<Exp>] => {
  num = Add (Lit 4) (Lit 8);
  var = ...
};
\end{lstlisting}

\noindent The CP code above is elaborated to corresponding \fip code of the form:

\begin{lstlisting}
repo = /\Exp. \_(self : ExpSig<Exp>).
  let $Lit = self.Lit in let $Add = self.Add in
  let $Let = self.Let in let $Var = self.Var in
    { num = fix self:Exp. $Add (fix self:Exp. $Lit 4 self)
                               (fix self:Exp. $Lit 8 self) self } ,,
    { var = ... };
\end{lstlisting}

\noindent The type parameter \lstinline{Exp} in the
\lstinline{repo} trait is expressed by a \systemf-style type lambda ($\Lambda A.\,e$).
Note that CP employs a form of syntactic sugar for constructors to allow concise use
of constructors and avoid explicit uses of \lstinline{new}.
The source code
\lstinline{Add (Lit 4) (Lit 8)} is first expanded into
\lstinline{new $Add (new $Lit 4) (new $Lit 8)}, which insert \lstinline{new} operators.
Next we describe the elaboration process of creating and instantiating traits:
\begin{itemize}
\item \textbf{Creation of traits}: A \lstinline{trait} is elaborated to a
\emph{generator} function whose parameter is a self-reference (like \lstinline{self} above) and whose body
is a record of methods;
\item \textbf{Instantiation of traits}: The \lstinline{new} construct is
  used to instantiate a trait. Uses of \lstinline{new} are
  elaborated to a \emph{fixpoint} which applies the elaborated trait function
  to a self-reference. In the definition of the field
  \lstinline{num} there are three elaborations of \lstinline{new}.
  For instance, the CP code \lstinline{new $Lit 4} corresponds to the \fip
  code \lstinline{fix self:Exp. $Lit 4 self}.
\end{itemize}

It is clear now that our trait encoding is heavily dependent on recursion,
due to the self-references employed by the encoding.
However, the original \fip~\cite{xuanbiesop} does not
support recursion, which reveals a gap between theory and practice.

\subsection{The Gap Between Theory and Practice}

Our primary motivation to reformulate \fip is to bridge the gap
between theory and practice. The original formulation of \fip lacks recursion,
impredicative polymorphism
and uses the traditional call-by-value (CBV) evaluation strategy. However, the
recent work of CP assumes a different variant of \fip that is equipped with
fixpoints and the call-by-name (CBN) evaluation. It is worthwhile to probe into
the causes of such differences.

\paragraph{Non-triviality of coherence}
Recursion is essential for general-purpose computation in
programming. More importantly, our encoding of traits requires recursion.
For example, \lstinline{new e} is elaborated to \lstinline{fix self. e self}.
However, adding recursion to the original version of \fip turns out to be
highly non-trivial.
The original \fip is defined using an elaboration semantics.
A fundamental property of \fip is \emph{coherence}~\cite{Reynolds_1991}, which states that the semantics is unambiguous.
Coherence is non-trivial
due to the presence of the merge operator~\cite{dunfield2014elaborating}.
To prove coherence, a logical relation, called \emph{canonicity}~\cite{xuanbiesop}, is
used to reason about contextual equivalence in the original work of \fip.
For example, with contextual equivalence, we can show that
the two possible elaborations for the same \fip source expression into \fco
are contextually equivalent:
\begin{displaymath}
 \mathsf{1}   \!\vcentcolon\!   \mathsf{Int}   \, \& \,   \mathsf{Int}   \!\vcentcolon\!   \mathsf{Int} \; \rightsquigarrow \;\mathsf{fst}\, (1, 1) \\
 \mathsf{1}   \!\vcentcolon\!   \mathsf{Int}   \, \& \,   \mathsf{Int}   \!\vcentcolon\!   \mathsf{Int} \; \rightsquigarrow \;\mathsf{snd}\, (1, 1)
\end{displaymath}
\noindent Two typing derivations lead to two elaborations in this example, which
pick different sides of the merge. However, both elaborated expressions will be
reduced to $1$ eventually. %In other words, they are contextually equivalent.

Unfortunately, the proof technique for coherence based on logical relations
does not immediately scale to recursive programs and programs with impredicative polymorphism.
A possible solution, known
from the research of logical relations, is to move to a more sophisticated
\emph{step-indexed} form of logical relations~\cite{ahmed2006step}. However, this requires a major
reformulation of the proofs and metatheory of the original \fip, and it is not clear
whether additional challenges would be present in such an extension.
Thus, the lack of the two features in the theory of the original \fip remains a
serious limitation since only terminating programs and predicative polymorphism
are considered. In other words, we cannot encode traits
as presented in Section~\ref{sec:encoding} in the original \fip.
To get around this issue and enable the encoding of traits, Zhang et al.~\cite{zhang2021compositional}
simply assumed an extension of \fip with recursion and their proof
of coherence for CP was done under the assumption that the original
\fip with recursion was coherent or deterministic.

Our work rectifies this gap in the theory of Compositional Programming and the CP
language. We reformulate \fip using a direct type-directed operational
semantics~\cite{tamingmerge} that allows recursion and prove that the semantics
is deterministic. Thus, our reformulation of \fip can serve as a target language
to encode traits and validate the proofs of the elaboration of CP
in terms of \fip with recursion. In addition, our approach gives a semantics
to \fip directly, instead of relying on an indirect elaboration semantics to a \systemf-like
language.

\paragraph{Evaluation strategies}
Most mainstream programming languages use CBV,
but CBN is a more natural evaluation strategy for object encodings
such as Cook's denotational semantics of inheritance. As
stated by Bruce et al.\ in their work
on object encodings~\cite{bruce1997comparing}:

\begin{quotation} \noindent \itshape
``Although we shall perform conversion steps in whatever order is convenient for
the sake of examples, we could just as well impose a call-by-name reduction
strategy. (Most of the examples would diverge under a call-by-value strategy.
This can be repaired at the cost of some extra lambda abstractions and
applications to delay evaluation at appropriate points.)''
\end{quotation}

In our elaboration of traits, we adopt a similar approach to object encodings.
For example, consider the following CP expression:
\begin{lstlisting}
  type A = { l1 : Int; l2 : Int };
  new (trait [self : A] => { l1 = 1; l2 = self.l1 })
\end{lstlisting}
which is elaborated to the following (slightly simplified) \fip expression:
\begin{displaymath}
 \mathbf{fix}~ \ottmv{self} \!\vcentcolon\! \mathit{A} .\,  \{  \ottmv{l_{{\mathrm{1}}}} \ottsym{=}  \mathsf{1}   \}    \,,,\,   \{  \ottmv{l_{{\mathrm{2}}}} \ottsym{=} \ottmv{self}  \ottsym{.}  \ottmv{l_{{\mathrm{1}}}}  \} 
\end{displaymath}
\noindent The \lstinline{trait} expression is elaborated to a function, and the
\lstinline{new} expression turns the function into a fixpoint.
Unfortunately, this expression terminates under CBN but diverges under CBV. If
evaluated under CBV, the variable $\ottmv{self}$ will be evaluated repeatedly,
despite the fact that only $\ottmv{self}  \ottsym{.}  \ottmv{l_{{\mathrm{1}}}}$ is used:
\begin{align*}
&  \mathbf{fix}~ \ottmv{self} \!\vcentcolon\! \mathit{A} .\,  \{  \ottmv{l_{{\mathrm{1}}}} \ottsym{=}  \mathsf{1}   \}    \,,,\,   \{  \ottmv{l_{{\mathrm{2}}}} \ottsym{=} \ottmv{self}  \ottsym{.}  \ottmv{l_{{\mathrm{1}}}}  \}  \\
 \hookrightarrow  \quad &  \{  \ottmv{l_{{\mathrm{1}}}} \ottsym{=}  \mathsf{1}   \}   \,,,\,   \{  \ottmv{l_{{\mathrm{2}}}} \ottsym{=} \ottsym{(}   \mathbf{fix}~ \ottmv{self} \!\vcentcolon\! \mathit{A} .\,  \{  \ottmv{l_{{\mathrm{1}}}} \ottsym{=}  \mathsf{1}   \}    \,,,\,   \{  \ottmv{l_{{\mathrm{2}}}} \ottsym{=} \ottmv{self}  \ottsym{.}  \ottmv{l_{{\mathrm{1}}}}  \}   \ottsym{)}  \ottsym{.}  \ottmv{l_{{\mathrm{1}}}}  \}  \\
 \hookrightarrow  \quad &  \{  \ottmv{l_{{\mathrm{1}}}} \ottsym{=}  \mathsf{1}   \}   \,,,\,   \{  \ottmv{l_{{\mathrm{2}}}} \ottsym{=} \ottsym{(}   \{  \ottmv{l_{{\mathrm{1}}}} \ottsym{=}  \mathsf{1}   \}   \,,,\,   \{  \ottmv{l_{{\mathrm{2}}}} \ottsym{=} \ottsym{(}   \mathbf{fix}~ \ottmv{self} \!\vcentcolon\! \mathit{A} .\,  \{  \ottmv{l_{{\mathrm{1}}}} \ottsym{=}  \mathsf{1}   \}    \,,,\,   \{  \ottmv{l_{{\mathrm{2}}}} \ottsym{=} \ottmv{self}  \ottsym{.}  \ottmv{l_{{\mathrm{1}}}}  \}   \ottsym{)}  \ottsym{.}  \ottmv{l_{{\mathrm{1}}}}  \}   \ottsym{)}  \ottsym{.}  \ottmv{l_{{\mathrm{1}}}}  \}  \\
 \hookrightarrow  \quad & \cdots
\end{align*}
We may tackle the problem of non-termination by wrapping self-references in \emph{thunks}, but
CBN provides a simpler and more natural way. In our CBN formulation of \fip,
$ \{  \ottmv{l} \ottsym{=} \mathit{e}  \} $ is already a value (instead of $ \{  \ottmv{l} \ottsym{=} \mathit{v}  \} $), so we do not need
to further evaluate $\mathit{e}$:
\begin{align*}
&  \mathbf{fix}~ \ottmv{self} \!\vcentcolon\! \mathit{A} .\,  \{  \ottmv{l_{{\mathrm{1}}}} \ottsym{=}  \mathsf{1}   \}    \,,,\,   \{  \ottmv{l_{{\mathrm{2}}}} \ottsym{=} \ottmv{self}  \ottsym{.}  \ottmv{l_{{\mathrm{1}}}}  \}  \\
 \hookrightarrow  \quad &  \{  \ottmv{l_{{\mathrm{1}}}} \ottsym{=}  \mathsf{1}   \}   \,,,\,   \{  \ottmv{l_{{\mathrm{2}}}} \ottsym{=} \ottsym{(}   \mathbf{fix}~ \ottmv{self} \!\vcentcolon\! \mathit{A} .\,  \{  \ottmv{l_{{\mathrm{1}}}} \ottsym{=}  \mathsf{1}   \}    \,,,\,   \{  \ottmv{l_{{\mathrm{2}}}} \ottsym{=} \ottmv{self}  \ottsym{.}  \ottmv{l_{{\mathrm{1}}}}  \}   \ottsym{)}  \ottsym{.}  \ottmv{l_{{\mathrm{1}}}}  \} 
\end{align*}
The $\ottmv{l_{{\mathrm{2}}}}$ field is further evaluated only when a record projection is performed:
\begin{align*}
& \ottsym{(}   \mathbf{fix}~ \ottmv{self} \!\vcentcolon\! \mathit{A} .\,  \{  \ottmv{l_{{\mathrm{1}}}} \ottsym{=}  \mathsf{1}   \}    \,,,\,   \{  \ottmv{l_{{\mathrm{2}}}} \ottsym{=} \ottmv{self}  \ottsym{.}  \ottmv{l_{{\mathrm{1}}}}  \}   \ottsym{)}  \ottsym{.}  \ottmv{l_{{\mathrm{2}}}} \\
 \hookrightarrow  \quad & \ottsym{(}   \{  \ottmv{l_{{\mathrm{1}}}} \ottsym{=}  \mathsf{1}   \}   \,,,\,   \{  \ottmv{l_{{\mathrm{2}}}} \ottsym{=} \ottsym{(}   \mathbf{fix}~ \ottmv{self} \!\vcentcolon\! \mathit{A} .\,  \{  \ottmv{l_{{\mathrm{1}}}} \ottsym{=}  \mathsf{1}   \}    \,,,\,   \{  \ottmv{l_{{\mathrm{2}}}} \ottsym{=} \ottmv{self}  \ottsym{.}  \ottmv{l_{{\mathrm{1}}}}  \}   \ottsym{)}  \ottsym{.}  \ottmv{l_{{\mathrm{1}}}}  \}   \ottsym{)}  \ottsym{.}  \ottmv{l_{{\mathrm{2}}}} \\
 \hookrightarrow  \quad & \ottsym{(}   \mathbf{fix}~ \ottmv{self} \!\vcentcolon\! \mathit{A} .\,  \{  \ottmv{l_{{\mathrm{1}}}} \ottsym{=}  \mathsf{1}   \}    \,,,\,   \{  \ottmv{l_{{\mathrm{2}}}} \ottsym{=} \ottmv{self}  \ottsym{.}  \ottmv{l_{{\mathrm{1}}}}  \}   \ottsym{)}  \ottsym{.}  \ottmv{l_{{\mathrm{1}}}} \\
 \hookrightarrow  \quad & \ottsym{(}   \{  \ottmv{l_{{\mathrm{1}}}} \ottsym{=}  \mathsf{1}   \}   \,,,\,   \{  \ottmv{l_{{\mathrm{2}}}} \ottsym{=} \ottsym{(}   \mathbf{fix}~ \ottmv{self} \!\vcentcolon\! \mathit{A} .\,  \{  \ottmv{l_{{\mathrm{1}}}} \ottsym{=}  \mathsf{1}   \}    \,,,\,   \{  \ottmv{l_{{\mathrm{2}}}} \ottsym{=} \ottmv{self}  \ottsym{.}  \ottmv{l_{{\mathrm{1}}}}  \}   \ottsym{)}  \ottsym{.}  \ottmv{l_{{\mathrm{1}}}}  \}   \ottsym{)}  \ottsym{.}  \ottmv{l_{{\mathrm{1}}}} \\
 \hookrightarrow  \quad &  \mathsf{1} 
\end{align*}
This example illustrates how our new CBN formulation of \fip avoids
non-termination of trait instantiation.

%%%%%%%%%%%%%%%%%%%%%%%%%%%%%%%%%%%%%%%%%%%%%%%%%%%%%%%%%%%%%%%%%%%%%%%%%%%%%%%%
\subsection{Technical Challenges and Innovations}
The main novelty of our reformulation of \fip is the use of a
type-directed operational semantics~\cite{huang_et_al:LIPIcs:2020:13183}
instead of an elaboration semantics.
With a TDOS, adding recursion and impredicative polymorphism to our proof of
determinism is trivial.
Our work is an extension of the
\lambdaip calculus~\cite{tamingmerge} which adapts the TDOS approach.
We also follow the subtyping algorithm design in \lambdaip.
While \lambdaip supports BCD-style distributive subtyping~\cite{Barendregt_1983},
the addition of disjoint polymorphism does bring some technical challenges.
Moreover, there are some smaller changes to \fip that enable us to
type-check more programs and improve the design of the original \fip.
We will give an overview of the technical challenges and innovations next.

\paragraph{The role of casting}
A merge like $ \mathsf{1}   \,,,\,   \mathsf{true} $ has multiple meanings under different types
(e.g. $ \mathsf{Int} $ or $\mathsf{Bool}$).
Eventually, we have to extract some components via the
elimination of merges, which is a key issue when designing a direct operational
semantics for a calculus with the merge operator.
A non-deterministic semantics could allow $ \mathit{e}_{{\mathrm{1}}}  \,,,\,  \mathit{e}_{{\mathrm{2}}} \, \hookrightarrow \, \mathit{e}_{{\mathrm{1}}} $ and $ \mathit{e}_{{\mathrm{1}}}  \,,,\,  \mathit{e}_{{\mathrm{2}}} \, \hookrightarrow \, \mathit{e}_{{\mathrm{2}}} $
without any constraints, at the cost of losing both type preservation and
determinism~\cite{dunfield2014elaborating}.
To obtain a non-ambiguous and type-safe semantics, we follow the TDOS
approach~\cite{huang_et_al:LIPIcs:2020:13183}:
which uses (up)casts to ensure that values have the right form during reduction.
In a TDOS, there is a casting relation, which is used in the reduction rule
for annotated values:
\begin{mathpar}
  \inferrule* [Right=Step-annov] {
    {  \mathit{v} \, \hookrightarrow _{ \mathit{A} }\, \mathit{v}' } }
    {  \mathit{v}  \!\vcentcolon\!  \mathit{A} \, \hookrightarrow \, \mathit{v}'  }
\end{mathpar}
Casting % plays a key role in the operational semantics, as it
enables us to drop certain parts from a term (e.g., $  \mathsf{1}   \,,,\,   \mathsf{true}  \, \hookrightarrow _{  \mathsf{Int}  }\,  \mathsf{1}  $).
Very often, it is necessary for us to do so to satisfy the disjointness
constraint. % and obtain a well-typed merge.
Consider a function $ \lambda  \ottmv{x} \!\vcentcolon\!  \mathsf{Int}  .\, \ottmv{x}   \,,,\,   \mathsf{false} $. For its body to be well-typed,
$\ottmv{x}$ cannot contain a boolean.
Hence, when the function is applied to $ \mathsf{1}   \,,,\,   \mathsf{true} $, we cannot directly
substitute the argument in.
Instead, it is wrapped by (and later cast to) $ \mathsf{Int} $ to resolve the potential
conflict.
\[\begin{aligned}
& \ottsym{(}  \ottsym{(}   \lambda  \ottmv{x} \!\vcentcolon\!  \mathsf{Int}  .\, \ottmv{x}   \,,,\,   \mathsf{false}   \ottsym{)}  \!\vcentcolon\!   \mathsf{Int}   \, \& \,  \mathsf{Bool}  \rightarrow   \mathsf{Int}   \, \& \,  \mathsf{Bool}  \ottsym{)} \, \ottsym{(}   \mathsf{1}   \,,,\,   \mathsf{true}   \ottsym{)} \\
 \hookrightarrow  \quad & \ottsym{(}  \ottsym{(}   \mathsf{1}   \,,,\,   \mathsf{true}   \ottsym{)}  \!\vcentcolon\!   \mathsf{Int}   \,,,\,   \mathsf{false}   \ottsym{)}  \!\vcentcolon\!   \mathsf{Int}   \, \& \,  \mathsf{Bool} \\
 \hookrightarrow  \quad & \ottsym{(}   \mathsf{1}   \,,,\,   \mathsf{false}   \ottsym{)}  \!\vcentcolon\!   \mathsf{Int}   \, \& \,  \mathsf{Bool} \\
 \hookrightarrow  \quad &  \mathsf{1}   \,,,\,   \mathsf{false}  \\
\end{aligned}\]

\paragraph{TDOS and function annotations}
In casting, values in a merge are selected based on type information.
In the absence of runtime type-checking, we need to know the type of input value
syntactically to match it with the target type.
Thus, functions must be accompanied by type annotations.
The previous work \lambdaip~\cite{tamingmerge} defines the syntax of functions
like $ \lambda  \ottmv{x} .\, \mathit{e} \!\vcentcolon\! \mathit{A} \rightarrow \mathit{C} $. While the original argument type $\mathit{A}$ is always kept
during reduction, \lambdaip's casting relation may generate a value that has a proper subtype
of the requested type: $  \lambda  \ottmv{x} .\, \mathit{e} \!\vcentcolon\! \mathit{A} \rightarrow \mathit{C}  \, \hookrightarrow _{ \mathit{B}_{{\mathrm{1}}}  \rightarrow  \mathit{B}_{{\mathrm{2}}} }\,  \lambda  \ottmv{x} .\, \mathit{e} \!\vcentcolon\! \mathit{A} \rightarrow \mathit{B}_{{\mathrm{2}}}  $.
We make casting more precise with a more liberal syntax in \fip.
We allow bare abstractions $ \lambda  \ottmv{x} \!\vcentcolon\! \mathit{A} .\, \mathit{e} $ while \lambdaip does not.
Our casting relation requires lambdas to be annotated $\ottsym{(}   \lambda  \ottmv{x} \!\vcentcolon\! \mathit{A} .\, \mathit{e}   \ottsym{)}  \!\vcentcolon\!  \mathit{B}$,
but the full annotation $\mathit{B}$ does not have to be a function type.
For example, $\ottsym{(}   \lambda  \ottmv{x} \!\vcentcolon\!  \mathsf{Int}  .\, \ottmv{x}   \,,,\,   \mathsf{true}   \ottsym{)}  \!\vcentcolon\!  \ottsym{(}   \mathsf{Int}   \rightarrow   \mathsf{Int}   \ottsym{)}  \, \& \,  \ottsym{(}   \mathsf{Int}   \rightarrow  \mathsf{Bool}  \ottsym{)}$ still acts as a
function,
and is equivalent to
$\ottsym{(}   \lambda  \ottmv{x} \!\vcentcolon\!  \mathsf{Int}  .\, \ottmv{x}   \,,,\,   \mathsf{true}   \ottsym{)}  \!\vcentcolon\!   \mathsf{Int}   \rightarrow   \mathsf{Int}   \, \& \,  \mathsf{Bool}$.

\paragraph{Algorithmic subtyping with disjoint polymorphism}
\fip extends BCD-style distributive subtyping~\cite{Barendregt_1983}
to disjoint polymorphism.
$ \forall   \ottmv{X} *  \mathsf{Int}  .\, \ottmv{X}   \, \& \,   \mathsf{Int} $ represents the intersection
of some type $\ottmv{X}$ and $ \mathsf{Int} $ assuming $\ottmv{X}$ is disjoint to $ \mathsf{Int} $.
Like arrows or records, such universal types distribute over intersections.
Hence, $\ottsym{(}   \forall   \ottmv{X} *  \mathsf{Int}  .\, \ottmv{X}   \ottsym{)}  \, \& \,  \ottsym{(}   \forall   \ottmv{X} *  \mathsf{Int}  .\,  \mathsf{Int}    \ottsym{)}$ is a subtype of
$ \forall   \ottmv{X} *  \mathsf{Int}  .\, \ottmv{X}   \, \& \,   \mathsf{Int} $.
A well-known challenge in supporting distributivity in
the BCD-style subtyping is to obtain an algorithmic formulation
of subtyping.
There have been many efforts to eliminate
the explicit transitivity rule to obtain an algorithmic
formulation~\cite{pierce1989decision,muehlboeck2018empowering,siek2019transitivity}.
Compared with the original \fip~\cite{xuanbiesop}, we employ
a different subtyping algorithm design, using
\emph{splittable types}~\cite{huang2021distributing}.
This approach employs a type-splitting operation ($ \mathit{B}  \mathbin{\lhd}  \mathit{A}  \mathbin{\rhd}  \mathit{C} $) that converts a given
type $\mathit{A}$ to an equivalent intersection type $\mathit{B}  \, \& \,  \mathit{C}$,
for example, $\mathit{A}  \rightarrow  \mathit{B}_{{\mathrm{1}}}  \, \& \,  \mathit{B}_{{\mathrm{2}}}$ is split to $\mathit{A}  \rightarrow  \mathit{B}_{{\mathrm{1}}}$ and $\mathit{A}  \rightarrow  \mathit{B}_{{\mathrm{2}}}$.
The subtyping algorithm uses type splitting whenever an intersection
type is expected in the conventional algorithm for subtyping without
distributivity, and therefore handles distributivity smoothly and modularly.

\paragraph{Enhanced subtyping and disjointness with more top-like types}
Unlike previous systems with disjoint polymorphism~\cite{alpuimdisjoint,xuanbiesop},
we add a context in subtyping judgments to track the disjointness assumption $ \ottmv{X}   *   \mathit{A} $ % of type variables
whenever we open a universal type $ \forall   \ottmv{X} * \mathit{A} .\, \mathit{B} $,
similar to the subtyping with F-bounded quantification.
The extra information enhances our subtyping:
we know a type must be a supertype of $ \mathsf{Top} $, if it is disjoint with $ \mathsf{Bot} $.
This also fixes the following broken property in the original \fip, as we now have more types
that are \emph{top-like}.

\begin{definition}[Disjointness specification]\label{def:disjoint-spec}
  If $\mathit{A}$ is disjoint with $\mathit{B}$, any common supertypes they have must be
  equivalent to $ \mathsf{Top} $.
\end{definition}

Keeping this property is necessary for us to obtain a deterministic operational
semantics.
Meanwhile, we prove our subtyping and disjointness relations are decidable
in Coq.
Note that in the original \fip, the decidability of the two relations was proved
manually, although the rest of the proof was mechanized.

\section{The \fip Calculus and Its Operational Semantics}\label{sec:calculus}
This section introduces the \fip calculus, including its static and dynamic semantics.

%%%%%%%%%%%%%%%%%%%%%%%%%%%%%%%%%%%%%%%%%%%%%%%%%%%%%%%%%%%%%%%%%%%%%%%%%%%%%%%%
\subsection{Syntax}

The syntax of \fip is as follows: % shown in Figure~\ref{fig:syntax}
\begin{small}
\begin{align*}
    &\text{Types} &A, B, C::=&~ \ottmv{X} ~|~  \mathsf{Int}  ~|~  \mathsf{Top}  ~|~  \mathsf{Bot}  ~|~ \mathit{A}  \, \& \,  \mathit{B} ~|~ \mathit{A}  \rightarrow  \mathit{B} ~|~  \forall   \ottmv{X} * \mathit{A} .\, \mathit{B}  ~|~ \ottsym{\{}  \ottmv{l}  \!\vcentcolon\!  \mathit{A}  \ottsym{\}}  \\
    &\text{Checkable terms} &p ::=&~  \lambda  \ottmv{x} \!\vcentcolon\! \mathit{A} .\, \mathit{e}  ~|~  \Lambda  \ottmv{X} .\, \mathit{e}  ~|~  \{  \ottmv{l} \ottsym{=} \mathit{e}  \}  \\
    &\text{Expressions} &e    ::=&~ \mathit{p} ~|~ x ~|~ i ~|~ \top ~|~ e:A ~|~ \mathit{e}_{{\mathrm{1}}}  \,,,\,  \mathit{e}_{{\mathrm{2}}} ~|~  \mathbf{fix}~ \ottmv{x} \!\vcentcolon\! \mathit{A} .\, \mathit{e}  ~|~ \mathit{e}_{{\mathrm{1}}} \, \mathit{e}_{{\mathrm{2}}} ~|~ \mathit{e} \, \mathit{A} ~|~ \mathit{e}  \ottsym{.}  \ottmv{l} \\
  &\text{Values} &v   ::=&~ \mathit{p} ~|~ \mathit{p}  \!\vcentcolon\!  \mathit{A} ~|~ \ottmv{i} ~|~ \top ~|~ \mathit{v}_{{\mathrm{1}}}  \,,,\,  \mathit{v}_{{\mathrm{2}}} \\
    &\text{Term contexts} &\Gamma ::=&~   \cdot  ~|~ \Gamma  ,  \ottmv{x}  \!\vcentcolon\!  \mathit{A} \\
    &\text{Type contexts} &\Delta ::=&~   \cdot  ~|~ \Delta  ,  \ottmv{X}  *  \mathit{A}\\
\end{align*}
\end{small}
\vspace{-10pt}

\paragraph{Types} % Metavariables $\mathit{A}$, $\mathit{B}$, and $\mathit{C}$ range over types.
Types include the $ \mathsf{Top} $ type and the uninhabited
type $ \mathsf{Bot} $.  Intersection types are created with $\mathit{A}  \, \& \,  \mathit{B}$.
Disjoint polymorphism, a key feature of \fip, is based on universal
types with a disjointness quantifier $ \forall   \ottmv{X} * \mathit{A} .\, \mathit{B} $, expressing
that the type variable $\ottmv{X}$ is bound inside $\mathit{B}$ and disjoint to
type $\mathit{A}$.  $\ottsym{\{}  \ottmv{l}  \!\vcentcolon\!  \mathit{A}  \ottsym{\}}$ denotes single-field record types, where
$\ottmv{l}$ is the record label.  Multi-field record types are desugared
to intersections of single-field ones~\cite{reynolds1997design}:
\begin{align*}
  &\{l_1:A_1;\;\dots;\;l_n:A_n\} \triangleq \{l_1 : A_1\}  \, \& \,  \cdots  \, \& \,  \{l_n : A_n\}
\end{align*}

\paragraph{Expressions} %% Metavariable $\mathit{e}$ ranges over expressions,
As we will explain later with the typing rules, some expressions do
not have an inferred type (or principal type), including lambda abstractions,
type abstractions, and single-field records. We use metavariable
$\mathit{p}$ to represent these expressions, which with optional
annotations, are values.
Also, note that expressions inside record values do not have to be
a value since our calculus employs call-by-name.
The merge operator $ \,,,\, $ composes two expressions to make a term of
an intersection type.
The top value $\top$ can be viewed as a merge of zero elements.
Fixpoint expressions $ \mathbf{fix}~ \ottmv{x} \!\vcentcolon\! \mathit{A} .\, \mathit{e} $ construct recursive programs.
The type annotation $\mathit{A}$ denotes the type of $\ottmv{x}$ as well as the whole
expression.
Like record types, multi-field records are desugared to merges of single-field ones:
\begin{align*}
&\{l_1=e_1;\;\dots;\;l_n=e_n\} \triangleq \{l_1 = e_1\}  \,,,\,  \dots  \,,,\,  \{l_n = e_n\}
\end{align*}

\paragraph{Contexts} We have two contexts:
$\Gamma$ tracks the types of term variables;
$\Delta$ tracks the disjointness information of type variables,
which follows the original design of \fip.
We use $\Delta  \vdash  \mathit{A}$, $\vdash  \Delta$, and $\Delta  \vdash  \Gamma$ judgments for
the type well-formedness and the context well-formedness
(defined in \cref{appendix:wfness}).
For multiple type well-formedness judgments, we combine them into one, i.e.,
$\Delta  \vdash  \mathit{A}  ,  \mathit{B}$ is equivalent to $\Delta  \vdash  \mathit{A} \wedge \Delta  \vdash  \mathit{B}$.

%%%%%%%%%%%%%%%%%%%%%%%%%%%%%%%%%%%%%%%%%%%%%%%%%%%%%%%%%%%%%%%%%%%%%%%%%%%%%%%%
\subsection{Subtyping} \label{sec:subtyping}

\begin{figure}[t]
\begin{small}
    \drules[DS]{$ \Delta   \vdash   \mathit{A}  <:  \mathit{B} $}{Declarative Subtyping}{refl, trans, top, bot,
      and, andl, andr, arrow, distArrow, topArrow, rcd, distRcd, topRcd, all, topAll, distAll,
      topVar}
\end{small}
    \caption{Declarative subtyping rules.}\label{fig:subtyping-declarative}
\end{figure}

\Cref{fig:subtyping-declarative} shows our subtyping relation, which
extends BCD-style subtyping~\cite{Barendregt_1983} with disjoint
polymorphism, records, and the bottom type.
Compared with the original \fip, we add a context to track type variables
and their disjointness information.
The context not only ensures the well-formedness of types but is also important
to our new \rref{DS-topVar}.
An equivalence relation (\cref{def:type-equivalence}) is defined on
types that are subtype of each other. These equivalent types can be
converted back and forth without loss of information.

\begin{definition}[Type equivalence]\label{def:type-equivalence}
  $ \Delta \vdash  \mathit{A} \sim  \mathit{B} $ $\triangleq$ $ \Delta   \vdash   \mathit{A}  <:  \mathit{B} $ and $ \Delta   \vdash   \mathit{B}  <:  \mathit{A} $.
\end{definition}

For functions (\rref{DS-arrow}) and disjoint quantifications
(\rref{DS-all}), subtyping is covariant in positive positions and
contravariant in negative positions. The intuition is that type
abstractions of the more specific type (subtype) should have a
\emph{looser} disjointness constraint for the parameter type.
$ \forall   \ottmv{X} *  \mathsf{Top}  .\, \mathit{A} $ denotes that there is no constraint on $\ottmv{X}$, since
$ \mathsf{Top} $ is disjoint to all types.
On the contrary, $ \mathsf{Bot} $ is the strictest constraint.
It is useful in types like $ \forall   \ottmv{X} * \ottsym{\{}  \ottmv{l}  \!\vcentcolon\!   \mathsf{Bot}   \ottsym{\}} .\, \mathit{A} $ which expresses that
$\ottmv{X}$ does not contain any informative field of label $\ottmv{l}$~\cite{xie2020row}.
For intersection types, \rref{DS-andl,DS-andr,DS-and} axiomatize that
$\mathit{A}  \, \& \,  \mathit{B}$ is the greatest lower bound of $\mathit{A}$ and $\mathit{B}$.
As a typical characteristic of BCD-style subtyping, type constructors
distribute over intersections, including arrows (\rref{DS-distArrow}),
records (\rref{DS-distRcd}) and disjoint quantifications (\rref{DS-distAll}).

Another feature of BCD subtyping, which is often overlooked, is the
generalization of \emph{top-like types}, i.e. supertypes of $ \mathsf{Top} $.

\begin{definition}[Specification of top-like types]\label{def:toplikespec}
  $ \Delta   \vdash  \rceil \mathit{A} \lceil $ $\triangleq$ $ \Delta \vdash  \mathit{A} \sim   \mathsf{Top}  $.
\end{definition}

Initially, top-like types include $ \mathsf{Top} $ and intersections like $ \mathsf{Top}   \, \& \,   \mathsf{Top} $.
But the BCD subtyping adds $ \mathsf{Top}   \rightarrow   \mathsf{Top} $ to it via \rref{DS-topArrow},
as well as $\mathit{A}  \rightarrow   \mathsf{Top} $ for any type $\mathit{A}$ due to the contravariance of
function parameters.
\Rref{DS-topArrow} can be viewed as a special case of \rref{DS-distArrow} where
intersections are replaced by $ \mathsf{Top} $ (one can consider it as an
intersection of zero components).
Like the original \fip, we extend this idea to universal types and record types
(\rref{DS-topAll,DS-topRcd}).

The most important change is the \rref{DS-topVar}.
This rule means a type variable is top-like if it is disjoint with the bottom type.
Every type $\mathit{B}$ is a common supertype of $\mathit{B}$ itself and $ \mathsf{Bot} $.
If $\mathit{B}$ is disjoint with $ \mathsf{Bot} $, then it must be top-like.
We proved that subtyping is decidable via an equivalent algorithmic formulation.

The discussion about algorithmic subtyping is
in \cref{appendix:algo-sub}.

\begin{lemma}[Decidability of subtyping]
    $ \Delta   \vdash   \mathit{A}  <:  \mathit{B} $ is decidable.
\end{lemma}

\paragraph{Disjointness} The notion of disjointness (\Cref{def:disjoint-spec}),
defined via subtyping, is used in the original \fip, as well as calculi with
disjoint intersection types~\cite{oliveira2016disjoint}.
We proved that our algorithmic definition of disjointness (written as $\Delta  \vdash  \mathit{A}  *  \mathit{B}$,
in \cref{appendix:disjointness}) is sound to a specification in terms of top-like types.

\begin{lemma}[Disjointness soundness]\label{lemma:disjoint-soundness}
  If $\Delta  \vdash  \mathit{A}  *  \mathit{B}$ then $\forall \mathit{C}$ that $ \Delta   \vdash   \mathit{A}  <:  \mathit{C} $ and $ \Delta   \vdash   \mathit{B}  <:  \mathit{C} $
  we have $ \Delta   \vdash  \rceil \mathit{C} \lceil $.
\end{lemma}

\noindent
Informally, two disjoint types do not have common supertypes, except for top-like types.
This definition is motivated by the desire to prevent
ambiguous upcasts on merges.  That is, we wish to avoid casts that can
extract \emph{different} values of the same type from a merge. Thus in
\fip and other calculi with disjoint intersection types, we only
allow merges of expressions whose only common supertypes are types
that are (equivalent to) the top type.  For instance, consider the
merge $\ottsym{(}   \mathsf{1}   \,,,\,   \mathsf{true}   \ottsym{)}  \,,,\,  \ottsym{(}   \mathsf{2}   \,,,\,   \text{'c'}   \ottsym{)}$. The first component of the merge
($ \mathsf{1}   \,,,\,   \mathsf{true} $) has type $ \mathsf{Int}   \, \& \,  \mathsf{Bool}$, while the second component
($ \mathsf{2}   \,,,\,   \text{'c'} $) has type $ \mathsf{Int}   \, \& \,  \mathsf{Char}$. This merge is problematic
because $ \mathsf{Int} $ is a supertype of the type of the merge
$\ottsym{(}   \mathsf{Int}   \, \& \,  \mathsf{Bool}  \ottsym{)}  \, \& \,  \ottsym{(}   \mathsf{Int}   \, \& \,  \mathsf{Char}  \ottsym{)}$, allowing us to extract two different
integers by casting the two terms to $ \mathsf{Int} $. Fortunately, our disjointness
restriction rejects such merges, since the supertype $ \mathsf{Int} $ is not
top-like.

%%%%%%%%%%%%%%%%%%%%%%%%%%%%%%%%%%%%%%%%%%%%%%%%%%%%%%%%%%%%%%%%%%%%%%%%%%%%%%%%
\subsection{Bidirectional Typing}

The type system of \fip is bidirectional~\cite{bidirectional2021},
where the subsumption rule is triggered by type annotations.
Calculi with a merge operator are incompatible with a general
subsumption rule because it cancels disjointness checking.
For example, with a general subsumption rule, we can directly use
$ \mathsf{1}   \,,,\,   \mathsf{true} $ as a term of type $ \mathsf{Int} $ since $  \mathsf{Int}   \, \& \,  \mathsf{Bool}  <:   \mathsf{Int}  $.
Then, merging $ \mathsf{1}   \,,,\,   \mathsf{true} $ with the term $ \mathsf{false} $ would
type-check since disjointness simply checks whether the static types
of merging terms are disjoint, and $ \mathsf{Int} $ is disjoint with
$\mathsf{Bool}$. But now, the merge contains two booleans, which would
lead to ambiguity if later we wish to extract a boolean value from the
merge.
The key issue is that a general subsumption rule loses static
type information that is necessary to reject ambiguous merges. A
bidirectional type system solves this problem by having a more
restricted form of subsumption that only works in the checking
mode where the type is provided.
A more detailed description of the problem for calculi with the merge operator
can be found in Huang et al.'s work~\cite{tamingmerge}.
We should also remark that this issue of incompatibility with a general
subsumption rule is not unique to calculi with a merge operator. It shows
up, for instance, in calculi with \emph{gradual typing}~\cite{siek07objects}
and calculi with \emph{record concatenation} and subtyping~\cite{cardelli1989operations}.

\begin{figure}[t!]
\begin{small}
\begin{align*}
  &\text{Typing modes} &\Leftrightarrow  &::= ~ \Leftarrow  ~|~   \Rightarrow  \\
  &\text{Pre-values} &\mathit{u} &::= \ottmv{i} ~|~ \top ~|~\mathit{e}  \!\vcentcolon\!  \mathit{A}~|~ \mathit{u}_{{\mathrm{1}}}  \,,,\,  \mathit{u}_{{\mathrm{2}}}\\
\end{align*}
\drules[Typ]{$\Delta  ;  \Gamma  \vdash  \mathit{e} \, \Leftrightarrow \, \mathit{A}$}{Bidirectional Typing}{top, lit, var,
  abs, tabs, rcd, app, tapp, proj, merge, mergev, inter, fix, anno, sub}
\end{small}
    \caption{Bidirectional typing rules for \fip.}\label{fig:typing-rules}
\end{figure}

\paragraph{Typing} As presented in \Cref{fig:typing-rules},
there are two modes of typing: synthesis ($ \Rightarrow $) and checking ($ \Leftarrow $).
We use $ \Leftrightarrow $ as a metavariable for typing modes.
$\Delta  ;  \Gamma  \vdash  \mathit{e} \, \Leftrightarrow \, \mathit{A}$ indicates that under type context $\Delta$ and term context $\Gamma$,
the expression $\mathit{e}$ has type $\mathit{A}$ in mode $ \Leftrightarrow $.
A bidirectional type system directly provides a type-checking algorithm.
$\Delta$, $\Gamma$, $\mathit{e}$ are all inputs in both modes.
Type synthesis generates a \emph{unique} type as the output (also called the
inferred type), while type checking takes a type as an input and
examines the term.

\begin{lemma}[Uniqueness of type synthesis]\label{lemma:inferred-uniq}
   If $\Delta  ;  \Gamma  \vdash  \mathit{e} \, \Rightarrow \, \mathit{A}_{{\mathrm{1}}}$ and $\Delta  ;  \Gamma  \vdash  \mathit{e} \, \Rightarrow \, \mathit{A}_{{\mathrm{2}}}$ then $ \mathit{A}_{{\mathrm{1}}}  =  \mathit{A}_{{\mathrm{2}}} $.
\end{lemma}

Conversion of typing modes happens in \rref{Typ-sub}.
With it, a term with inferred type $\mathit{A}$ can be checked against any $\mathit{B}$
that is a supertype of $\mathit{A}$.
Compared to the original \fip, fixpoints are new.
They model recursion with a self-reference ($\ottmv{x}$ in $ \mathbf{fix}~ \ottmv{x} \!\vcentcolon\! \mathit{A} .\, \mathit{e} $).
Other than this, \rref{Typ-fix} is almost the same as \rref{Typ-anno}.
It checks the expression $\mathit{e}$ by the annotated type $\mathit{A}$, with
assumption that $\ottmv{x}$ has type $\mathit{A}$ in $\mathit{e}$.

\paragraph{Checking abstractions, type abstractions, and records}
To check a function $ \lambda  \ottmv{x} \!\vcentcolon\! \mathit{A} .\, \mathit{e} $ against $\mathit{B}_{{\mathrm{1}}}  \rightarrow  \mathit{B}_{{\mathrm{2}}}$ by \rref{Typ-abs},
we track the type of the term variable as \emph{the precise parameter type} $\mathit{A}$,
and check if $\mathit{e}$ can be checked against $\mathit{B}_{{\mathrm{2}}}$.
$\mathit{B}_{{\mathrm{1}}}$ must be a subtype of $\mathit{A}$ to guarantee the safety of the function application.
The type-checking of type abstractions $ \Lambda  \ottmv{X} .\, \mathit{e} $ works by
tracking the disjointness relation of the type variable with the context
and checking $\mathit{e}$ against the quantified type $\mathit{B}$.
Typing of records works similarly.
Additionally, there is a \rref{Typ-inter}, which checks an expression against an intersection
type by seperately checking the expression against the composing two types. With this design,
we allow $ \lambda  \ottmv{x} \!\vcentcolon\!  \mathsf{Int}  .\, \ottmv{x}   \,,,\,   \mathsf{true} $ to be checked against $\ottsym{(}   \mathsf{Int}   \rightarrow   \mathsf{Int}   \ottsym{)}  \, \& \,  \ottsym{(}   \mathsf{Int}   \rightarrow  \mathsf{Bool}  \ottsym{)}$.

\begin{figure}[t]
\begin{small}
    \ottdefnsApplicativeDistribution
\end{small}
    \caption{Applicative distribution rules.}\label{fig:typing-appdist}
\end{figure}

\paragraph{Application, record projection, and conversion of applicable types}
It is not surprising that a merge can act as a function.
But in the original \fip, this requires annotations since the expression
being applied in an application must have an inferred arrow type.
Our design, following the \lambdaip calculus~\cite{tamingmerge},
allows a term of an intersection type to directly apply, as long as the
intersection type can be converted into an applicable form.
For example, $\ottsym{(}   \mathsf{Int}   \rightarrow   \mathsf{Int}   \ottsym{)}  \, \& \,  \ottsym{(}   \mathsf{Int}   \rightarrow  \mathsf{Bool}  \ottsym{)}$ is converted into
$\ottsym{(}   \mathsf{Int}   \, \& \,   \mathsf{Int}   \ottsym{)}  \rightarrow  \ottsym{(}   \mathsf{Int}   \, \& \,  \mathsf{Bool}  \ottsym{)}$, which is a supertype of the former.
When inferring the type of the application $\mathit{e}_{{\mathrm{1}}} \, \mathit{e}_{{\mathrm{2}}}$, \rref{Typ-app} first
\emph{converts} the inferred type of $\mathit{e}_{{\mathrm{1}}}$ into an arrow form $\mathit{B}  \rightarrow  \mathit{C}$
and then checks the argument $\mathit{e}_{{\mathrm{2}}}$ against $\mathit{B}$.
If the check succeeds, the whole expression has inferred type $\mathit{C}$.

In \fip, we have three applicable forms: \emph{arrow types},
\emph{record types}, \emph{universal types}.
Like \rref{Typ-app}, the typing of type application and record projection
also allows the applied term to have an intersection type,
and relies on \emph{applicative distribution} to convert the type.

Applicative distribution $ \mathit{A}  \rhd  \mathit{B} $ (defined in \cref{fig:typing-appdist})
takes type $\mathit{A}$ and generates a supertype $\mathit{B}$ that has an applicable form.
The first three rules bring all parts of the input intersection type together.
For example, assuming that we apply several merged functions whose types are
$\mathit{A}_{{\mathrm{1}}}  \rightarrow  \mathit{B}_{{\mathrm{1}}}$, $\mathit{A}_{{\mathrm{2}}}  \rightarrow  \mathit{B}_{{\mathrm{2}}}$, ..., $\mathit{A}_{\ottmv{n}}  \rightarrow  \mathit{B}_{\ottmv{n}}$, the combined function type is
$\ottsym{(}  \mathit{A}_{{\mathrm{1}}}  \, \& \, \, ... \, \, \& \,  \mathit{A}_{\ottmv{n}}  \ottsym{)}  \rightarrow  \ottsym{(}  \mathit{B}_{{\mathrm{1}}}  \, \& \, \, ... \, \, \& \,  \mathit{B}_{\ottmv{n}}  \ottsym{)}$.
It is equivalent to the input type only when $\mathit{A}_{{\mathrm{1}}}$, $\mathit{A}_{{\mathrm{2}}}$, ...,
and $\mathit{A}_{\ottmv{n}}$ are all equivalent.
Essentially, applicative distribution ($ \mathit{A}  \rhd  \mathit{B} $) is a subset of
subtyping ($ \mathit{A}  <:  \mathit{B} $). The supertype is selected to ensure that
when a merge is applied to an argument, every component in the merge
are satisfied.
Although each one of the three first rules overlaps with the reflexivity rule,
for any given type, at most one result has an applicable form.

Since merges are treated as a whole applicable term,
programmers can extend functions via a \emph{compositional} approach
without modifying the original implementation.
It also enables the modular extension of type abstractions
and especially records, which play a core role in the trait encoding
used in Compositional Programming.

Davies and Pfenning also employ a similar design in their bidirectional type
system for refinement intersections~\cite{Davies_2000}. Their type conversion
procedure respects subtyping as well. Instead of combining function types,
it makes use of $ \mathit{A}  \, \& \,  \mathit{B}  <:  \mathit{A} $ and $ \mathit{A}  \, \& \,  \mathit{B}  <:  \mathit{B} $ to enumerate components in
intersections and uncover arrows.

%% TYP-MERGE & TYP-MERGEV
\paragraph{Typing merges with disjointness and consistency}
Well-typed merges always have inferred types.
There are two type synthesis rules for merges, both combining the
inferred types of the two parts into an intersection.
\Rref*{Typ-merge} requires the two subterms to have \emph{disjoint} inferred
types, like $ \mathsf{1}   \,,,\,   \mathsf{true} $.
\Rref*{Typ-mergev} relaxes the disjointness constraint to \emph{consistency}
checking (written as $\mathit{u}_{{\mathrm{1}}}  \approx  \mathit{u}_{{\mathrm{2}}}$) to accept overlapping terms like $ \mathsf{1}   \,,,\,   \mathsf{1} $.
Such duplication is meaningless to users but may appear during evaluation.
In fact, \rref{Typ-mergev} is designed for metatheory properties, and not to
allow more user-written programs~\cite{tamingmerge}.
We will state the formal specification of consistency in \cref{sec:determinism}
and show how it is involved in the proofs of determinism and type soundness.
Informally, consistent merges cause no ambiguity in the runtime.
For practical reasons, we only consider \emph{pre-values} (defined at
the top of \Cref{fig:typing-rules}) in consistency checking,
for which the inferred type can be told directly.
The algorithms for disjointness and consistency are presented
in \cref{appendix:algo}.
In general, disjointness and consistency avoid introducing ambiguity of merges,
and enable a deterministic semantics for \fip.

%%%%%%%%%%%%%%%%%%%%%%%%%%%%%%%%%%%%%%%%%%%%%%%%%%%%%%%%%%%%%%%%%%%%%%%%%%%%%%%%
\subsection{Small-Step Operational Semantics}\label{sec:reduction}

\begin{figure}[t]
\begin{small}
\begin{align*}
    &\text{Arguments} &\ottnt{arg}  &::= ~\mathit{e} ~|~ \mathit{A} ~|~ \ottsym{\{}  \ottmv{l}  \ottsym{\}} \\
    &\text{Evaluation contexts} &\ottnt{E}  &::= [~]~e ~|~ [~]~A ~|~ [~]~.l
    ~|~ [~]~,,~v ~|~ v~,,~[~] ~|~ [~]:A
\end{align*}
    \ottdefnsParallelApplication
    \ottdefnsReduction
\end{small}
    \caption{Small-step semantics rules.}\label{fig:semantics-parallel-smallstep}
\end{figure}

We specify the \emph{call-by-name} reduction of \fip using a small-step operational
semantics in \cref{fig:semantics-parallel-smallstep}.
\Rref*{Step-papp}, \rref*{Step-pproj}, and \rref*{Step-ptapp} are reduction rules
for application and record projection. They trigger \emph{parallel application}
(defined in the middle of \cref{fig:semantics-parallel-smallstep})
of merged values to the argument.
\Rref{Step-fix} substitutes the fixpoint term variable with the fixpoint
expression itself. Note that the result is annotated with $\mathit{A}$. With the
explicit type annotation, the result of reduction preserves the type of the
original fixpoint expression.
Through \rref{Step-annov}, values are \emph{cast} to their annotated
type. Such values must also be pre-values.
This is to filter out checkable terms $\mathit{p}$ including bare abstractions
or records without annotations, as $\mathit{p}  \!\vcentcolon\!  \mathit{A}$ is a form of value itself
and thus should not step.

A merge of multiple terms may reduce in parallel, as shown in
\rref{Step-merge}.  Only when one side cannot step, the other side
steps alone, as suggested by the evaluation context $\mathit{E~,,~v}$ and
$\mathit{v~,,~E}$.  \Rref{Step-cntx} is the reduction rule of
expressions within an \emph{evaluation context}. Since the rule can be
applied repeatedly, we only need evaluation contexts of depth one
(shown at the top of \cref{fig:semantics-parallel-smallstep}).
Our operational semantics substitutes arguments \emph{wrapped} by type
annotations into function bodies, while it forbids the reduction of
records since records are values.

%%%%%%%%%%%%%%%%%%%%%%%%%%%%%%%%%%%%%%%%%%%%%%%%%%%%%%%%%%%%%%%%%%%%%%%%%%%%%%%%

\paragraph{Parallel application}
Parallel application is at the heart of what we call \emph{nested composition} in CP.
It provides the runtime behavior that is necessary to implement nested composition, and
it reflects the subtyping distributivity rules at the term level.
A merge of functions is treated as one function.
The beta reduction of all functions in a merge happens in \emph{parallel}
to keep the consistency of merged terms.
For type abstractions or records, things are similar.
The parallel application handles these applicable merges uniformly via \rref{PApp-merge}.
To align record projection with the other two kinds of application,
we define \emph{arguments} which abstract expressions, types, and record labels
(at the top of \cref{fig:semantics-parallel-smallstep}).
In \rref{PApp-abs}, the argument expression is \emph{wrapped} by the function
argument type before we substitute it into the function body.
Parallel application of type abstractions substitutes the type
argument into the body and annotates the body with the substituted
disjoint quantified type.  \Rref{PApp-proj} projects record fields.
Note these three rules have types to annotate the result, since in
\rref*{Typ-abs}, \rref*{Typ-tabs}, and \rref*{Typ-rcd} we only type
the expression $\mathit{e}$ inside in \emph{checking} mode. With an
explicit type annotation, the application preserves types.

%%%%%%%%%%%%%%%%%%%%%%%%%%%%%%%%%%%%%%%%%%%%%%%%%%%%%%%%%%%%%%%%%%%%%%%%%%%%%%%%
\paragraph{Splittable types}
Before explaining wrapping or casting, we first introduce
\emph{splittable types}~\cite{tamingmerge}, which are a key component
of our algorithmic formulations of various relations. Ordinary
types are the basic units, values of ordinary types can be constructed without the merge
operator. As defined at the top of
\Cref{fig:split-cast}, ordinary types do not have intersection types
in positive positions.
By contrast, splittable types are isomorphic to
appropriate intersections. Recall that in BCD-style distributive
rules, arrows distribute over intersection, making $ \mathsf{Int}   \rightarrow   \mathsf{Int}   \, \& \,  \mathsf{Bool}$
equivalent to the intersection $\ottsym{(}   \mathsf{Int}   \rightarrow   \mathsf{Int}   \ottsym{)}  \, \& \,  \ottsym{(}   \mathsf{Int}   \rightarrow  \mathsf{Bool}  \ottsym{)}$.
Therefore we say that the former type \emph{splits} into the latter two arrow types.
In \Cref{fig:split-cast}, we extend the type splitting algorithm of
\lambdaip to universal types in correspondence to the distributive
subtyping rules (\rref{DS-distArrow}, \rref{DS-distRcd}, and
\rref{DS-distAll}). It gives a decision procedure to check whether a
type is splittable or ordinary.

\begin{lemma}[Type splitting loses no information]\label{def:splittable}
    $ \mathit{B}  \mathbin{\lhd}  \mathit{A}  \mathbin{\rhd}  \mathit{C} $ only if $  \cdot  \vdash  \mathit{A} \sim  \mathit{B}  \, \& \,  \mathit{C} $.
\end{lemma}

\begin{figure}[t]
  \begin{small}
    \begin{align*}
      &\text{Ordinary types} &A^\circ, B^\circ, C^\circ::=&~ \ottmv{X} ~|~  \mathsf{Int}  ~|~  \mathsf{Top}  ~|~  \mathsf{Bot}  ~|~ \mathit{A}  \rightarrow  B^\circ ~|~  \forall   \ottmv{X} * \mathit{A} .\, B^\circ  ~|~ \ottsym{\{}  \ottmv{l}  \!\vcentcolon\!  A^\circ  \ottsym{\}}
    \end{align*}
  \drules[Sp]{$ \mathit{B}  \mathbin{\lhd}  \mathit{A}  \mathbin{\rhd}  \mathit{C} $}{Splittable Types}{arrow, rcd, all, and}
  \ottdefnsCastExpression
  \drules[Cast]{$ \mathit{v}_{{\mathrm{1}}} \, \hookrightarrow _{ \mathit{A} }\, \mathit{v}_{{\mathrm{2}}} $}{Casting}{int, top, mergel, merger, anno, and}
\end{small}
  \caption{Type splitting, expression wrapping and value casting rules.}\label{fig:split-cast}
\end{figure}

\paragraph{Expression wrapping}
Rules for expression wrapping ($ \mathit{e} \,\rightsquigarrow_{ \mathit{A} }\, \mathit{u} $)
are listed in the middle of \cref{fig:split-cast}.
Basically, it splits the type $\mathit{A}$ when possible, annotates a duplication of
$\mathit{e}$ by each ordinary part of $\mathit{A}$, and then composes all of them.
The only exception is that it never uses top-like types to annotate terms,
to avoid ill-typed results like $ \{  \ottmv{l} \ottsym{=}  \mathsf{1}   \}   \!\vcentcolon\!   \mathsf{Int}   \rightarrow   \mathsf{Top} $,
but rather generates a normal value whose inferred type is that top-like type,
like $\ottsym{(}   \lambda  \ottmv{x} \!\vcentcolon\!  \mathsf{Int}  .\, \top   \ottsym{)}  \!\vcentcolon\!   \mathsf{Int}   \rightarrow   \mathsf{Top} $
(via the \emph{top-like value generating} function $ [\![  A^\circ  ]\!] $, defined
in \cref{appendix:generator}).

\paragraph{Casting}
Casting (shown in \cref{fig:split-cast}) is the core of the TDOS, and
is triggered by the \rref*{Step-annov} rule.
Recalling that only values that are also pre-values will be cast,
we can always tell the inferred type of the input value
and cast it by any supertype of that inferred type.
The definition of casting uses the notion of splittable types.
In \rref{Cast-and}, the value is cast under two parts of a splittable type
separately, and the results are put together by the merge operator.
The following example shows that a merge retains its form when cast
under equivalent types.
\[\begin{aligned}
  && \ottsym{(}  \ottsym{(}   \lambda  \ottmv{x} \!\vcentcolon\!  \mathsf{Int}  .\, \ottmv{x}   \ottsym{)}  \!\vcentcolon\!   \mathsf{Int}   \rightarrow   \mathsf{Int}   \ottsym{)}  \,,,\,  \ottsym{(}  \ottsym{(}   \lambda  \ottmv{x} \!\vcentcolon\!  \mathsf{Int}  .\,  \mathsf{true}    \ottsym{)}  \!\vcentcolon\!   \mathsf{Int}   \rightarrow  \mathsf{Bool}  \ottsym{)} \\
  &  \hookrightarrow _{\ottsym{(}   \mathsf{Int}   \rightarrow   \mathsf{Int}   \ottsym{)}  \, \& \,  \ottsym{(}   \mathsf{Int}   \rightarrow  \mathsf{Bool}  \ottsym{)}} ~& \ottsym{(}  \ottsym{(}   \lambda  \ottmv{x} \!\vcentcolon\!  \mathsf{Int}  .\, \ottmv{x}   \ottsym{)}  \!\vcentcolon\!   \mathsf{Int}   \rightarrow   \mathsf{Int}   \ottsym{)}  \,,,\,  \ottsym{(}  \ottsym{(}   \lambda  \ottmv{x} \!\vcentcolon\!  \mathsf{Int}  .\,  \mathsf{true}    \ottsym{)}  \!\vcentcolon\!   \mathsf{Int}   \rightarrow  \mathsf{Bool}  \ottsym{)} \\
  &  \hookrightarrow _{ \mathsf{Int}   \rightarrow   \mathsf{Int}   \, \& \,  \mathsf{Bool}} ~& \ottsym{(}  \ottsym{(}   \lambda  \ottmv{x} \!\vcentcolon\!  \mathsf{Int}  .\, \ottmv{x}   \ottsym{)}  \!\vcentcolon\!   \mathsf{Int}   \rightarrow   \mathsf{Int}   \ottsym{)}  \,,,\,  \ottsym{(}  \ottsym{(}   \lambda  \ottmv{x} \!\vcentcolon\!  \mathsf{Int}  .\,  \mathsf{true}    \ottsym{)}  \!\vcentcolon\!   \mathsf{Int}   \rightarrow  \mathsf{Bool}  \ottsym{)}
\end{aligned}\]
In the latter case,
the requested type is a \emph{function} type,
but the result has an intersection type.
This change of type causes a major challenge for type preservation.

For ordinary types, \rref{Cast-int} casts an integer to itself under type $ \mathsf{Int} $.
Under any ordinary top-like type, the cast result is the output of the
\emph{top-like value generator}.
The casting of values with annotations
works by changing the type annotation to the casting (not top-like) supertype.
\Rref{Cast-mergel} and \rref{Cast-merger} make a selection between two merged values.
The two rules overlap, but for a well-typed value, the casting result
is unique.

%%%%%%%%%%%%%%%%%%%%%%%%%%%%%%%%%%%%%%%%%%%%%%%%%%%%%%%%%%%%%%%%%%%%%%%%%%%%%%%%
\paragraph{Example} We show an example to illustrate the behavior of our semantics:

\vspace{8pt}
\resizebox{.92\textwidth}{!}{
\[\begin{aligned}
  & \text{Let}~\mathit{f}:=~ \lambda  \ottmv{x} \!\vcentcolon\!  \mathsf{Int}   \, \& \,   \mathsf{Top}  .\, \ottmv{x}   \,,,\,   \mathsf{false} ~\text{in} \\
  & \ottsym{(}  \ottsym{(}  \mathit{f}  \!\vcentcolon\!  \ottsym{(}   \mathsf{Int}   \, \& \,   \mathsf{Top}   \rightarrow   \mathsf{Int}   \ottsym{)}  \, \& \,  \ottsym{(}   \mathsf{Int}   \, \& \,   \mathsf{Top}   \rightarrow  \mathsf{Bool}  \ottsym{)}  \ottsym{)}  \!\vcentcolon\!   \mathsf{Int}   \, \& \,  \mathsf{Bool}  \rightarrow   \mathsf{Int}   \, \& \,  \mathsf{Bool}  \ottsym{)} \, \ottsym{(}   \mathsf{1}   \,,,\,   \mathsf{true}   \ottsym{)} \\
   \hookrightarrow ^{*} \quad & \text{\{by \rref{Step-annov,Cast-and,Cast-anno}\}}  \\
  & \ottsym{(}  \mathit{f}  \!\vcentcolon\!   \mathsf{Int}   \, \& \,  \mathsf{Bool}  \rightarrow   \mathsf{Int}   \ottsym{)}  \,,,\,  \ottsym{(}  \mathit{f}  \!\vcentcolon\!   \mathsf{Int}   \, \& \,  \mathsf{Bool}  \rightarrow  \mathsf{Bool}  \ottsym{)} \, \ottsym{(}   \mathsf{1}   \,,,\,   \mathsf{true}   \ottsym{)} \\
   \hookrightarrow ^{*} \quad & \text{\{by \rref{Step-papp,EW-and,EW-anno,EW-top}\}}\\
  & \ottsym{(}  \ottsym{(}  \ottsym{(}   \mathsf{1}   \,,,\,   \mathsf{true}   \ottsym{)}  \!\vcentcolon\!   \mathsf{Int}   \,,,\,  \top  \ottsym{)}  \,,,\,   \mathsf{false}   \ottsym{)}  \!\vcentcolon\!   \mathsf{Int}   \,,,\,  \ottsym{(}  \ottsym{(}  \ottsym{(}   \mathsf{1}   \,,,\,   \mathsf{true}   \ottsym{)}  \!\vcentcolon\!   \mathsf{Int}   \,,,\,  \top  \ottsym{)}  \,,,\,   \mathsf{false}   \ottsym{)}  \!\vcentcolon\!  \mathsf{Bool} \\
  %  \hookrightarrow ^{*} \quad & \text{by \rref{Step-annov,Cast-int}}\\
  % & \ottsym{(}  \ottsym{(}   \mathsf{1}   \,,,\,  \top  \ottsym{)}  \,,,\,   \mathsf{false}   \ottsym{)}  \!\vcentcolon\!   \mathsf{Int}   \,,,\,  \ottsym{(}  \ottsym{(}   \mathsf{1}   \,,,\,  \top  \ottsym{)}  \,,,\,   \mathsf{false}   \ottsym{)}  \!\vcentcolon\!  \mathsf{Bool} \\
   \hookrightarrow ^{*} \quad & \text{\{by \rref{Step-merge,Step-annov,Cast-int,Cast-mergel,Cast-merger}\}}\\
  &  \mathsf{1}   \,,,\,   \mathsf{false}  \\
  \end{aligned}\]
}
\vspace{8pt}

\noindent This example shows that
a function with a splittable type will be cast to a merge of two copies of itself
with different type annotations, i.e., two split results. The application of
a merge of functions works by distributing the argument to both functions.
Finally, casting selects one side of the merge under the annotated type.
From this example, we can see that without the precise parameter annotation of
a lambda function (here $ \mathsf{Int}   \, \& \,   \mathsf{Top} $), there is no way to filter the argument $ \mathsf{1}   \,,,\,   \mathsf{true} $,
causing a conflict.

\section{Algorithmics}\label{appendix:algo}
In \cref{sec:calculus} we have presented several relations
in a declarative form, including subtyping, disjointness, and consistency.
For the purposes of implementation,
it is important to formulate the corresponding algorithmic versions.

\subsection{Algorithmic Subtyping}\label{appendix:algo-sub}
To obtain an equivalent algorithmic formulation of subtyping, we firstly define
algorithms for \emph{top-like types} and \emph{bottom-like types} inductively in
\Cref{fig:botlike-toplike},
then we introduce our algorithmic subtyping and argue that it is equivalent to
the declarative subtyping.

\begin{figure}[t]
\begin{small}
    \drules[BL]{$ \rfloor \mathit{A} \lfloor $}{Bottom-like Types}{bot, andl, andr}
    \drules[TL]{$ \Delta   \vdash  \rceil \mathit{A} \lceil $}{Top-like Types}{top, and, arrow, rcd, all, var}
\end{small}
    \caption{Bottom-like type rules and top-like type rules.}\label{fig:botlike-toplike}
\end{figure}

\paragraph{Top-like and bottom-like types}
Every top-like type is a supertype of all types (see \cref{def:toplikespec}),
which is equivalent to $ \mathsf{Top} $. Compared to the definition of top-like types
in the original \fip~\cite{xuanbiesop},
we add a type context in the subtyping judgment to keep track of type variables
that are disjoint to the bottom type. With type contexts,
we can derive $  \cdot    \vdash  \rceil  \forall   \ottmv{X} *  \mathsf{Bot}  .\, \ottmv{X}  \lceil $
by \rref*{TL-var} since only top-like types are disjoint to $ \mathsf{Bot} $.
The corresponding declarative subtyping rule is the novel \rref*{DS-topVar}.
To eliminate the dependence of our top-like type algorithm on subtyping,
we define \emph{bottom-like types} as a separate relation and use $ \rfloor \mathit{A} \lfloor $
when $ \Delta   \vdash   \mathit{A}  <:   \mathsf{Bot}  $ is needed.

\begin{lemma}[Equivalence of bottom-like types]\label{lemma:subtyping-botlike}
  If $\vdash  \Delta$ and $\Delta  \vdash  \mathit{A}$ and
  $ \rfloor \mathit{A} \lfloor $ if and only if $ \Delta   \vdash   \mathit{A}  <:   \mathsf{Bot}  $.
\end{lemma}
Then we obtain an algorithmic definition of top-like
types (\cref{fig:botlike-toplike}) that
is equivalent to \Cref{def:toplikespec}.

\begin{lemma}[Top-like equivalence]\label{lemma:subtyping-toplike}
  $ \Delta   \vdash  \rceil \mathit{A} \lceil $ if and only if $ \Delta   \vdash    \mathsf{Top}   <:  \mathit{A} $.
\end{lemma}

\begin{figure}[t]
\begin{small}
  \ottdefnsAlgorithmicSubtyping
\end{small}
    \caption{Algorithmic subtyping rules.}\label{fig:algosub}
\end{figure}

\paragraph{Algorithmic subtyping}
Our subtyping algorithm is shown in \Cref{fig:algosub}. This algorithm
is an extension of the algorithm used in \lambdaip with splittable types~\cite{tamingmerge}.
The novel additions are the rules involving disjoint polymorphism, which \lambdaip
does not have.
While \rref{S-and} requires the supertype $\mathit{B}$ to be splittable,
the remaining rules only apply to ordinary $\mathit{B}$.
The basic idea is to split the intersection-like right-hand-side type by
\rref{S-and} until types are in more atomic forms, i.e., ordinary,
and then apply the remaining rules to decide whether
the left-hand-side type is a subtype of each ordinary part.
For a subtyping $ \mathit{A}  \leq  \mathit{B} $ checking to succeed where $\mathit{B}$ is splittable,
we need every sub-checking of split types to succeed,
as described by the inversion lemma:

\begin{lemma}[Inversion of the supertype in algorithmic subtyping]\label{lemma:subtyping-right-inv}
    If $ \mathit{B}_{{\mathrm{1}}}  \mathbin{\lhd}  \mathit{B}  \mathbin{\rhd}  \mathit{B}_{{\mathrm{2}}} $ then $ \Delta   \vdash   \mathit{A}  \leq  \mathit{B} $ if and only if $ \Delta   \vdash   \mathit{A}  \leq  \mathit{B}_{{\mathrm{1}}} $ and $ \Delta   \vdash   \mathit{A}  \leq  \mathit{B}_{{\mathrm{2}}} $.
\end{lemma}

The key for the distributive subtyping rules to work is encoded in how we split types.
For example, only after we split $ \mathsf{Int}   \rightarrow   \mathsf{Int}   \, \& \,  \mathsf{Bool}$ into $\ottsym{(}   \mathsf{Int}   \rightarrow   \mathsf{Int}   \ottsym{)}$ and
$\ottsym{(}   \mathsf{Int}   \rightarrow  \mathsf{Bool}  \ottsym{)}$, it becomes straightforward to tell it is a
supertype of $\ottsym{(}   \mathsf{Int}   \rightarrow   \mathsf{Int}   \ottsym{)}  \, \& \,  \ottsym{(}   \mathsf{Int}   \rightarrow  \mathsf{Bool}  \ottsym{)}$.
Assuming that we only have \rref{Sp-and} (see Figure~\ref{fig:split-cast})
that splits intersection types,
the system degenerates to the conventional intersection subtyping rules
with disjoint polymorphism (like \fname).
In other words, the ordinary-type rules (all rules except for \rref{S-and})
are standard and the algorithm design is modular.
They are mostly the same as the declarative formulation with the additional
ordinary-type condition,
except that \rref{S-andl,S-andr} are embedded with transitivity.
For two universal
types $ \forall   \ottmv{X} * \mathit{A}_{{\mathrm{1}}} .\, \mathit{A}_{{\mathrm{2}}} $ and $ \forall   \ottmv{X} * \mathit{B}_{{\mathrm{1}}} .\, \mathit{B}_{{\mathrm{2}}} $, the subtyping
of $\mathit{A}_{{\mathrm{1}}}$ and $\mathit{B}_{{\mathrm{1}}}$ is contravariant, and the subtyping of $\mathit{A}_{{\mathrm{2}}}$ and $\mathit{B}_{{\mathrm{2}}}$
is covariant.
When deciding the subtyping
of $\mathit{A}_{{\mathrm{2}}}$ and $\mathit{B}_{{\mathrm{2}}}$, we add $ \ottmv{X}   *   \mathit{B}_{{\mathrm{1}}} $ into the context to track the disjointness
of type variable. For the special case of $\ottmv{X}$ disjoint to bottom-like types as
\rref{DS-topVar}, we have the context to decide $\ottmv{X}$ to be top-like
by our top-like type algorithm in \rref{S-top}.
Our algorithmic subtyping is equivalent to the declarative subtyping
and decidable:

\begin{lemma}[Equivalence of subtyping]\label{lemma:subtyping-equiv} % Equivalence of declarative and algorithmic subtyping
    $ \Delta   \vdash   \mathit{A}  \leq  \mathit{B} $ if and only if $ \Delta   \vdash   \mathit{A}  <:  \mathit{B} $.
\end{lemma}

\begin{lemma}[Decidability of algorithmic subtyping]\label{lemma:subtyping-decidable}
    $ \Delta   \vdash   \mathit{A}  \leq  \mathit{B} $ is decidable.
\end{lemma}

%%%%%%%%%%%%%%%%%%%%%%%%%%%%%%%%%%%%%%%%%%%%%%%%%%%%%%%%%%%%%%%%%%%%%%%%%%%%%%%%
\subsection{Disjointness}\label{appendix:disjointness}

\begin{figure}[t]
\begin{small}
  \ottdefnsConsistent
  \ottdefnsTypeDisjointness
\end{small}
    \caption{Algorithmic type disjointness and pre-value consistency rules.}\label{fig:disjointness-consistency}
\end{figure}

As shown in \cref{fig:disjointness-consistency},
the disjointness definition
is almost the same as the original \fip~\cite{xuanbiesop}. The disjointness judgment
$\Delta  \vdash  \mathit{A}  *  \mathit{B}$ ensures that under the context $\Delta$, any common supertype of
$\mathit{A}$ and $\mathit{B}$ is a top-like type. Disjointness helps ensuring the determinism of \fip by
forbidding merging terms of types that are not disjoint with each other (\rref{Typ-merge}).
Basic disjointness axioms $\mathit{A}  *_{ax}  \mathit{B}$, such as $ \mathsf{Int} $ is disjoint with functions, are shown in
\cref{appendix:disjoint-ax}. Our algorithm of disjointness is
sound with respect to our specification (\cref{lemma:disjoint-soundness}). %in \cref{def:disjoint-spec-formal}.

Compared to the original \fip, there are two main novelties.
\begin{comment}
First, we modify the uncontexted
top-like in (\rref{D-topl,D-topr}) into a contexted one. Top-like types are disjoint
to any types, because top-like types are supertypes of themselves and any other types.
\end{comment}
Firstly, now the context may tell us if a type variable is top-like,
i.e. is disjoint to bottom-like types, and a top-like type variable should be disjoint to
any types.
Secondly, in \rref{D-andl,D-andr}, we split types instead of allowing only
intersection types for the convenience of the proof.
We have an alternative disjointness definition in the original style
proved to be equivalent in our Coq formalization.
We can prove the following properties:

\begin{theorem}[Covariance of disjointness]\label{lemma:disjoint-covariance}
    If $\Delta  \vdash  \mathit{A}  *  \mathit{B}$ and $ \Delta   \vdash   \mathit{B}  \leq  \mathit{C} $ then $\Delta  \vdash  \mathit{A}  *  \mathit{C}$.
\end{theorem}

\begin{theorem}[Substitution of disjointness]\label{lemma:disjoint-subst}
    If $\Delta  ,   \ottmv{X} * \mathit{C}   ,  \Delta'  \vdash  \mathit{A}  *  \mathit{B}$ and $\Delta  \vdash  \mathit{C}'  *  \mathit{C}$ then $\Delta  ,  \Delta'  \ottsym{[}  \ottmv{X}  \mapsto  \mathit{C}'  \ottsym{]}  \vdash  \mathit{A}  \ottsym{[}  \ottmv{X}  \mapsto  \mathit{C}'  \ottsym{]}  *  \mathit{B}  \ottsym{[}  \ottmv{X}  \mapsto  \mathit{C}'  \ottsym{]}$.
\end{theorem}

\noindent \Cref{lemma:disjoint-covariance} is a generalization of \rref{D-varl,D-varr}.
In short disjointness is covariant: supertypes of disjoint types are still disjoint.
\Cref{lemma:disjoint-subst} ensures the correct behavior of the type
instantiation of type application. Disjoint types are still disjoint with each
other after instantiation.

\subsection{Consistency}\label{appendix:consistency}
\Rref{Typ-mergev} is aimed to type merges
of \emph{pre-values} produced by \emph{type casting} and \emph{expression wrapping}.
The intention is not to accept more programs, so this rule is not supposed to be
exposed to users.
The first question is:
% WHY PREVALUES
why are we considering the consistency of pre-values rather than values or expressions?
The answer is that our system allows
an expression (or a value) to be duplicated, and merged with type annotations
in our dynamic semantics by \rref{Cast-and} and \rref{EW-and}.
For example,
\[\begin{aligned}
  & \ottsym{(}  \ottsym{(}   \lambda  \ottmv{x} \!\vcentcolon\!  \mathsf{Int}  .\, \ottmv{x}   \,,,\,   \mathsf{false}   \!\vcentcolon\!   \mathsf{Int}   \rightarrow   \mathsf{Int}   \, \& \,  \mathsf{Bool}  \ottsym{)}  \!\vcentcolon\!   \mathsf{Int}   \rightarrow   \mathsf{Int}   \, \& \,  \mathsf{Bool}  \ottsym{)} \,  \mathsf{1}  \\
   \hookrightarrow  \quad & \ottsym{(}   \lambda  \ottmv{x} \!\vcentcolon\!  \mathsf{Int}  .\, \ottmv{x}   \,,,\,   \mathsf{false}   \!\vcentcolon\!   \mathsf{Int}   \rightarrow   \mathsf{Int}   \,,,\,   \lambda  \ottmv{x} \!\vcentcolon\!  \mathsf{Int}  .\, \ottmv{x}   \,,,\,   \mathsf{false}   \!\vcentcolon\!   \mathsf{Int}   \rightarrow  \mathsf{Bool}  \ottsym{)} \,  \mathsf{1}  \\
   \hookrightarrow  \quad & \ottsym{(}   \mathsf{1}   \,,,\,   \mathsf{false}   \ottsym{)}  \!\vcentcolon\!   \mathsf{Int}   \,,,\,  \ottsym{(}   \mathsf{1}   \,,,\,   \mathsf{false}   \ottsym{)}  \!\vcentcolon\!  \mathsf{Bool} \\
   \hookrightarrow  \quad &  \mathsf{1}   \,,,\,   \mathsf{false} 
  \end{aligned}\]
Moreover, it is also allowed to produce a result of duplicated terms with the same type
annotation, like $ \mathsf{1}   \!\vcentcolon\!   \mathsf{Int}   \,,,\,   \mathsf{1}   \!\vcentcolon\!   \mathsf{Int} $. Such values cannot pass disjointness checking
but are harmless at runtime.
To keep type preservation for the application results like the merge after
the first and second steps, consistency judgments on annotated terms are necessary.
% RULES
In consistency checking (\cref{fig:disjointness-consistency}),
we take two pre-values and analyze them structurally.
A merged pre-value $\mathit{u}_{{\mathrm{1}}}  \,,,\,  \mathit{u}_{{\mathrm{2}}}$ is consistent to another pre-value $\mathit{u}$
if pre-values composing the merge are consistent with $\mathit{u}$.
Any two basic components from each term must be either disjoint (\rref{C-disjoint})
or only differ in the annotation (except for the argument annotation of lambda abstractions
in \rref{C-anno}).
For instance, functions with different annotations but the same body
are consistent (as the one after the first step in the example above),
and $ \mathsf{1}   \,,,\,   \mathsf{true} $ is consistent with $ \mathsf{1}   \,,,\,   \text{'a'} $.
Note that we use $\mathit{u}  \!\vcentcolon\!  \mathit{A}$ to represent the principal type of $\mathit{u}$ is $\mathit{A}$,
which is a syntactical approach to compute the types from pre-values. Principal
type rules are listed \cref{appendix:principal}.
Also note that any disjoint pre-values are also
consistent. So \rref{Typ-mergev} is a strict relaxation of \rref{Typ-merge}
on pre-values.

\section{Type Soundness and Determinism}\label{sec:type-soundness}
In this section, we show that the operational semantics of \fip is type-sound and
deterministic.
In \fip, determinism also plays a key role in the proof of type soundness.

%%%%%%%%%%%%%%%%%%%%%%%%%%%%%%%%%%%%%%%%%%%%%%%%%%%%%%%%%%%%%%%%%%%%%%%%%%%%%%%%
\subsection{Determinism}\label{sec:determinism}

A common problem of determinism for calculi with a merge operator is
the ambiguity of selection between merged values. In our system,
ambiguity is removed by employing disjointness and consistency
constraints on merges via typing.

\begin{definition}[Consistency specification]\label{definition:consistency-spec}
  $ \mathit{v}_{{\mathrm{1}}} \approx_{spec} \mathit{v}_{{\mathrm{2}}} $ $\triangleq$ $\forall$ $\mathit{A}$ that $ \mathit{v}_{{\mathrm{1}}} \, \hookrightarrow _{ \mathit{A} }\, \mathit{v}'_{{\mathrm{1}}} $ and $ \mathit{v}_{{\mathrm{2}}} \, \hookrightarrow _{ \mathit{A} }\, \mathit{v}'_{{\mathrm{2}}} $
  then $ \mathit{v}'_{{\mathrm{1}}}  =  \mathit{v}'_{{\mathrm{2}}} $.
\end{definition}

Two values in a merge have no conflicts as long as casting both values under any type
leads to the same result.
This specification allows $\mathit{v}_{{\mathrm{1}}}$ and $\mathit{v}_{{\mathrm{2}}}$ to contain identical expressions
(may differ in annotations), and terms with disjoint types
as such terms can only be cast under top-like types, and the cast
result is only decided by that top-like type.

\begin{lemma}[Top-like casting is term irrelevant]\label{lemma:casting-toplike}
  If $  \cdot    \vdash  \rceil \mathit{A} \lceil $ and $ \mathit{v}_{{\mathrm{1}}} \, \hookrightarrow _{ \mathit{A} }\, \mathit{v}'_{{\mathrm{1}}} $ and $ \mathit{v}_{{\mathrm{2}}} \, \hookrightarrow _{ \mathit{A} }\, \mathit{v}'_{{\mathrm{2}}} $ then $ \mathit{v}'_{{\mathrm{1}}}  =  \mathit{v}'_{{\mathrm{2}}} $.
\end{lemma}

\noindent This is because casting only happens when the given type is a supertype
of the cast value's type, and disjoint types only share top-like types as common
supertypes (\Cref{lemma:disjoint-soundness}).

\begin{lemma}[Upcast only]\label{lemma:upcast-only}
  If $ \cdot   ;   \cdot   \vdash  \mathit{v} \, \Rightarrow \, \mathit{B}$ and $ \mathit{v} \, \hookrightarrow _{ \mathit{A} }\, \mathit{v}' $ then $  \cdot    \vdash   \mathit{B}  <:  \mathit{A} $.
\end{lemma}

\noindent With consistency, casting all well-typed values leads to a unique result.
The remaining reduction rules, including expression wrapping
and parallel application, are trivially deterministic.

\begin{lemma}[Determinism of casting]\label{lemma:casting-unique}
  If $ \cdot   ;   \cdot   \vdash  \mathit{v} \, \Rightarrow \, \mathit{B}$ and
  $ \mathit{v} \, \hookrightarrow _{ \mathit{A} }\, \mathit{v}_{{\mathrm{1}}} $ and $ \mathit{v} \, \hookrightarrow _{ \mathit{A} }\, \mathit{v}_{{\mathrm{2}}} $,
  then $ \mathit{v}_{{\mathrm{1}}}  =  \mathit{v}_{{\mathrm{2}}} $.
\end{lemma}

\begin{theorem}[Determinism of reduction]\label{theorem:step-unique}
  If $ \cdot   ;   \cdot   \vdash  \mathit{e} \, \Rightarrow \, \mathit{A}$ and $ \mathit{e} \, \hookrightarrow \, \mathit{e}_{{\mathrm{1}}} $ and $ \mathit{e} \, \hookrightarrow \, \mathit{e}_{{\mathrm{2}}} $ then $ \mathit{e}_{{\mathrm{1}}}  =  \mathit{e}_{{\mathrm{2}}} $.
\end{theorem}

%%%%%%%%%%%%%%%%%%%%%%%%%%%%%%%%%%%%%%%%%%%%%%%%%%%%%%%%%%%%%%%%%%%%%%%%%%%%%%%%
\subsection{Progress}
Annotated values trigger casting, for which the progress lemma can be directly
proved, as we know $\mathit{v}$ must have an inferred type that is a subtype of $\mathit{B}$.

\begin{lemma}[Progress of casting]\label{lemma:cast-progress}
  If $ \cdot   ;   \cdot   \vdash  \mathit{v} \, \Rightarrow \, \mathit{A}$ and $ \cdot   ;   \cdot   \vdash  \mathit{v} \, \Leftarrow \, \mathit{B}$
  then there exists a $\mathit{v}'$ such that $ \mathit{v} \, \hookrightarrow _{ \mathit{B} }\, \mathit{v}' $.
\end{lemma}

The progress lemma for expression wrapping is more relaxed.
It does not enforce that $\mathit{e}$ is checked against the wrapping type $\mathit{A}$
because that is the typical situation where we need to use the relation.

\begin{lemma}[Progress of expression wrapping]
    If $ \cdot   ;   \cdot   \vdash  \mathit{e} \, \Leftarrow \, \mathit{A}$ and $ \cdot   \vdash  \mathit{B}$ then
    there exists an $\mathit{e}'$ that $ \mathit{e} \,\rightsquigarrow_{ \mathit{B} }\, \mathit{e}' $.
\end{lemma}

Parallel applications deal with function application, type application,
and record projection. We use the term \emph{general application} of a value $\mathit{v}$
to an argument $\ottnt{arg}$ next to denote all of these for simplicity.

\begin{lemma}[Progress of parallel application]
    If $ \cdot   ;   \cdot   \vdash  \mathit{v} \,  \bullet\text{arg}  \, \Rightarrow \, \mathit{A}$ then
    there exists a $\mathit{u}$ such that $ \mathit{v}  \bullet  \ottnt{arg} ~ \hookrightarrow ~ \mathit{u} $.
\end{lemma}

Finally, the progress property of reduction can be proved.

\begin{theorem}[Progress of reduction]\label{theorem:progress}
    If $ \cdot   ;   \cdot   \vdash  \mathit{e} \, \Leftrightarrow \, \mathit{A}$ then either $\mathit{e}$ is a value
    or there exists a $\mathit{e}'$ such that $ \mathit{e} \, \hookrightarrow \, \mathit{e}' $.
\end{theorem}

%%%%%%%%%%%%%%%%%%%%%%%%%%%%%%%%%%%%%%%%%%%%%%%%%%%%%%%%%%%%%%%%%%%%%%%%%%%%%%%%
\subsection{Preservation}
Retaining preservation is challenging.
When typing merges, we need to satisfy the extra side conditions in
\rref{Typ-merge,Typ-mergev}: disjointness and consistency.
While the former only depends on types, the latter needs special care.

\paragraph{Consistency}
As discussed in \cref{sec:reduction}, casting may duplicate terms.
For example, $  \mathsf{1}  \, \hookrightarrow _{  \mathsf{Int}   \, \& \,   \mathsf{Int}  }\,  \mathsf{1}   \,,,\,   \mathsf{1}  $ by \rref{Cast-and}. \Rref{Typ-mergev}
is a relaxation of \rref{Typ-merge} to type such merges.
We have to ensure any two merged casting results are consistent:

\begin{lemma}[Value consistency after casting]\label{lemma:consaftercast}
  If $ \cdot   ;   \cdot   \vdash  \mathit{v} \, \Rightarrow \, \mathit{C}$ and $ \mathit{v} \, \hookrightarrow _{ \mathit{A} }\, \mathit{v}_{{\mathrm{1}}} $ and $ \mathit{v} \, \hookrightarrow _{ \mathit{B} }\, \mathit{v}_{{\mathrm{2}}} $ then
  $\mathit{v}_{{\mathrm{1}}}  \approx  \mathit{v}_{{\mathrm{2}}}$.
\end{lemma}

Then we need to make sure that consistency is preserved during reduction.
\ifdefined\undefined
Recall that consistency is defined on pre-values.
Consider any three components from consistent merges,
since merges are reduced in parallel (\rref{Step-merge}), the sub-expression $\mathit{e}$
in merges like $\mathit{e}  \!\vcentcolon\!  \mathit{A}  \,,,\,  \mathit{e}  \!\vcentcolon\!  \mathit{B}  \,,,\,  \mathit{e}  \!\vcentcolon\!  \mathit{C}$ remains the same
until it is cast by $\mathit{A}$, $\mathit{B}$, and $\mathit{C}$ respectively.
Guarded by the determinism theorem of reduction (\cref{theorem:step-unique})
and the consistency lemma of casting (\cref{lemma:consaftercast}),
we prove consistent pre-values are always consistent after possible reduction.
\fi

\begin{lemma}[Reduction keeps consistency]\label{lemma:consistent-steps}
  If $ \cdot   ;   \cdot   \vdash  \mathit{u}_{{\mathrm{1}}} \, \Rightarrow \, \mathit{A}$ and $ \cdot   ;   \cdot   \vdash  \mathit{u}_{{\mathrm{2}}} \, \Rightarrow \, \mathit{B}$ and
  $\mathit{u}_{{\mathrm{1}}}  \approx  \mathit{u}_{{\mathrm{2}}}$ then
  \begin{itemize}
  \item if $\mathit{u}_{{\mathrm{1}}}$ is a value and $ \mathit{u}_{{\mathrm{2}}} \, \hookrightarrow \, \mathit{u}'_{{\mathrm{2}}} $ then $\mathit{u}_{{\mathrm{1}}}  \approx  \mathit{u}'_{{\mathrm{2}}}$;
  \item if $\mathit{u}_{{\mathrm{2}}}$ is a value and $ \mathit{u}_{{\mathrm{1}}} \, \hookrightarrow \, \mathit{u}'_{{\mathrm{1}}} $ then $\mathit{u}'_{{\mathrm{1}}}  \approx  \mathit{u}_{{\mathrm{2}}}$;
  \item if $ \mathit{u}_{{\mathrm{1}}} \, \hookrightarrow \, \mathit{u}'_{{\mathrm{1}}} $ and $ \mathit{u}_{{\mathrm{2}}} \, \hookrightarrow \, \mathit{u}'_{{\mathrm{2}}} $ then $\mathit{u}'_{{\mathrm{1}}}  \approx  \mathit{u}'_{{\mathrm{2}}}$.
  \end{itemize}
\end{lemma}

%% PAPP KEEPS CONSISTENCY
Besides, when parallel application substitutes arguments into merges of
applicable terms or projects the wished field, consistency is preserved as well.
This requirement enforces us to define consistency not only on values but also
on pre-values since the application transforms a value merge into a pre-value merge.

\begin{lemma}[Parallel application keeps consistency]\label{lemma:papp-consistent}
  If $ \cdot   ;   \cdot   \vdash  \mathit{v}_{{\mathrm{1}}} \, \Rightarrow \, \mathit{A}$ and
  $ \cdot   ;   \cdot   \vdash  \mathit{v}_{{\mathrm{2}}} \, \Rightarrow \, \mathit{B}$ and
  $\mathit{v}_{{\mathrm{1}}}  \approx  \mathit{v}_{{\mathrm{2}}}$ and
  $ \mathit{v}_{{\mathrm{1}}}  \bullet  \ottnt{arg} ~ \hookrightarrow ~ \mathit{u}_{{\mathrm{1}}} $ and
  $ \mathit{v}_{{\mathrm{2}}}  \bullet  \ottnt{arg} ~ \hookrightarrow ~ \mathit{u}_{{\mathrm{2}}} $ then $\mathit{u}_{{\mathrm{1}}}  \approx  \mathit{u}_{{\mathrm{2}}}$
  when % the following conditions are satisfied:
  \begin{itemize}
  \item $\ottnt{arg}$ is a well-typed expression;
  \item or $\ottnt{arg}$ is a label;
  \item or $\ottnt{arg}$ is a type $\mathit{C}$;
    we know $ \mathit{A}  \rhd   \forall   \ottmv{X} * \mathit{A}_{{\mathrm{1}}} .\, \mathit{A}_{{\mathrm{2}}}  $ and $ \mathit{B}  \rhd   \forall   \ottmv{X} * \mathit{B}_{{\mathrm{1}}} .\, \mathit{B}_{{\mathrm{2}}}  $;
    and $ \cdot   \vdash  \mathit{C}  *  \mathit{A}_{{\mathrm{1}}}  \, \& \,  \mathit{B}_{{\mathrm{1}}}$.
  \end{itemize}
\end{lemma}

\begin{figure}[t]
\begin{small}
  \ottdefnsIsomorphicSubtyping
\end{small}
\caption{Isomorphic subtyping.}\label{fig:subtyping-runtime}
\end{figure}

\paragraph{Isomorphic subtyping}
In \fip, types are not always precisely preserved by all reduction steps.
Specifically, when we cast a value $ \mathit{v} \, \hookrightarrow _{ \mathit{A} }\, \mathit{v}' $ (in \rref{Step-annov})
or wrap a term $ \mathit{e} \,\rightsquigarrow_{ \mathit{A} }\, \mathit{u} $ (in \rref{PApp-abs}),
the context expects $\mathit{v}'$ or $\mathit{u}$ to have type $\mathit{A}$,
but this is not always true.
In our casting rules shown at the bottom of \Cref{fig:split-cast},
most values will be reduced to results with
the exact type we want, except for \rref{Cast-and}.
The inferred type of the result is always an intersection,
which may differ from the original splittable type.
To describe the change of types during reduction accurately,
we define \emph{isomorphic subtyping} (\Cref{fig:subtyping-runtime}).
If $ \mathit{A}  \lesssim  \mathit{B} $, we say $\mathit{A}$ is an isomorphic subtype of $\mathit{B}$.
The following lemma shows that while the two types in an isomorphic subtyping
relation may be syntactically different, they are equivalent under an empty type context
(i.e. $  \cdot    \vdash   \mathit{A}  <:  \mathit{B} $ and $  \cdot    \vdash   \mathit{B}  <:  \mathit{A} $).

\begin{theorem}[Isomorphic subtypes are equivalent]\label{theorem:isomorphic-subsume}
    If $ \mathit{A}  \lesssim  \mathit{B} $ then $  \cdot  \vdash  \mathit{A} \sim  \mathit{B} $.
\end{theorem}

With isomorphic subtyping, we define the preservation property of casting,
expression wrapping, and parallel application as follows.

\begin{lemma}[Casting preserves typing]\label{lemma:casting-preservation}
  If $ \cdot   ;   \cdot   \vdash  \mathit{v} \, \Rightarrow \, \mathit{A}$ and $ \mathit{v} \, \hookrightarrow _{ \mathit{B} }\, \mathit{v}' $
  then there exists a $\mathit{C}$ such that
  $ \cdot   ;   \cdot   \vdash  \mathit{v}' \, \Rightarrow \, \mathit{C}$ and $  \mathit{C}  \lesssim  \mathit{B} $.
\end{lemma}

\begin{lemma}[Expression wrapping preserves typing]\label{lemma:wrapping-preservation}
  If $ \cdot   ;   \cdot   \vdash  \mathit{e} \, \Leftarrow \, \mathit{B}$ and $  \cdot    \vdash   \mathit{B}  <:  \mathit{A} $ and
  $ \mathit{e} \,\rightsquigarrow_{ \mathit{A} }\, \mathit{u} $ then there exists a $\mathit{C}$ such that
  $ \cdot   ;   \cdot   \vdash  \mathit{u} \, \Rightarrow \, \mathit{C}$ and $  \mathit{C}  \lesssim  \mathit{A} $.
\end{lemma}

\begin{lemma}[Parallel application preserves typing]
  If $ \cdot   ;   \cdot   \vdash  \mathit{v} \,  \bullet\text{arg}  \, \Rightarrow \, \mathit{A}$
  and $ \mathit{v}  \bullet  \ottnt{arg} ~ \hookrightarrow ~ \mathit{u} $ then there exists a $\mathit{B}$ such that
  $ \cdot   ;   \cdot   \vdash  \mathit{u} \, \Rightarrow \, \mathit{B}$ and $  \mathit{B}  \lesssim  \mathit{A} $.
\end{lemma}

Of course, we can prove that the result of casting always has a subtype (or
an equivalent type) of the requested type instead of an isomorphic subtype.
But it would be insufficient for type preservation of reduction.
In summary, if casting or wrapping generates a term of type $\mathit{B}$ when
the requested type is $\mathit{A}$, we need $\mathit{B}$ to satisfy:
% COPIED FROM DISCUSSION
\begin{itemize}
\item $\mathit{B}$ is a subtype of $\mathit{A}$
  because we want a preservation theorem that respects subtyping.

\item For any type $\mathit{C}$, $ \mathit{A}   *   \mathit{C} $ implies $ \mathit{B}   *   \mathit{C} $.
  This is for the disjointness and consistency checking in \rref{Typ-merge,Typ-mergev}.
  Note that $ \mathit{B}  <:  \mathit{A} $ is not enough for this property.

\item If $\mathit{A}$ converts into an applicable type $\mathit{C}$, then
  $\mathit{B}$ converts into an applicable type too.
\end{itemize}

Although the type equivalence satisfies the first two conditions, it breaks the last one.
For example, $  \mathsf{Top}  $ is equivalent to $  \mathsf{Top}   \rightarrow   \mathsf{Top}  $, but one may not
convert $ \mathsf{Top} $ to an applicable type by the applicative distribution.
So it is infeasible to replace the isomorphic subtyping with the type equivalence.

\paragraph{Narrowing and Substitution Lemmas}
Before proving type preservation, we have to prove narrowing and substitution lemmas,
including both type and term substitutions.
The narrowing lemma for typing is mainly to deal with proof cases regarding
universal types. To check an expression against a supertype, we need this property to
tighten the disjointness constraint of the type variable in the context of
a typing judgment, since a supertype of a universal type
has a more tight disjointness constraint.

\begin{lemma}[Typing narrowing]\label{lemma:typing-narrowing}
    If $\Delta  ,  \ottmv{X}  *  \mathit{A}  ,  \Delta'  ;  \Gamma  \vdash  \mathit{e} \, \Leftrightarrow \, \mathit{C}$ and $ \Delta   \vdash   \mathit{B}  <:  \mathit{A} $
    then $\Delta  ,  \ottmv{X}  *  \mathit{B}  ,  \Delta'  ;  \Gamma  \vdash  \mathit{e} \, \Leftrightarrow \, \mathit{C}$.
\end{lemma}

The type substitution lemma for typing is necessary for the instantiation of
type variables. We always use types satisfying the disjointness
constraint of a universal type to substitute.
We need to make sure that the typing judgment
still holds after type substitution:

\begin{lemma}[Type substitution preserves typing]\label{lemma:typsubst-type}
    If $\Delta  ,  \ottmv{X}  *  \mathit{A}  ,  \Delta'  ;  \Gamma  \vdash  \mathit{e} \, \Leftrightarrow \, \mathit{C}$ and $\Delta  \vdash  \mathit{A}  *  \mathit{B}$
    then $\Delta  ,  \Delta'  \ottsym{[}  \ottmv{X}  \mapsto  \mathit{B}  \ottsym{]}  ;  \Gamma  \ottsym{[}  \ottmv{X}  \mapsto  \mathit{B}  \ottsym{]}  \vdash  \mathit{e}  \ottsym{[}  \ottmv{X}  \mapsto  \mathit{B}  \ottsym{]} \, \Leftrightarrow \, \mathit{C}  \ottsym{[}  \ottmv{X}  \mapsto  \mathit{B}  \ottsym{]}$.
\end{lemma}

In beta reduction or the reduction of fixpoints,
we want the term substitution of the parameter inside the function body to
always preserve types when the argument has an \emph{isomorphic subtype} of
the function input type. This is different from traditional
term substitution lemmas. Our expression wrapping does not
necessarily produce a wrapped result with the identical type to the wrapping
type (\Cref{lemma:wrapping-preservation}).
In addition, if we substitute a term with an isomorphic subtype into an expression,
the substituted result does not \emph{infer} the original type: it should infer an
isomorphic subtype, though it can still be \emph{checked} by the original checking type.

\begin{lemma}[Term substitution preserves type synthesis]\label{lemma:typsubst-term-inf}
  If $\Delta  ;  \Gamma  ,  \ottmv{x}  \!\vcentcolon\!  \mathit{A}  ,  \Gamma'  \vdash  \mathit{e} \, \Rightarrow \, \mathit{C}$ and $\Delta  ;  \Gamma  \vdash  \mathit{e}' \, \Rightarrow \, \mathit{B}$ and $ \mathit{B}  \lesssim  \mathit{A} $
  then $\exists$ $\mathit{C}'$ that $\Delta  ;  \Gamma  ,  \Gamma'  \vdash  \mathit{e}  \ottsym{[}  \ottmv{x}  \mapsto  \mathit{e}'  \ottsym{]} \, \Rightarrow \, \mathit{C}'$ and $ \mathit{C}'  \lesssim  \mathit{C} $.
\end{lemma}

\begin{lemma}[Term substitution preserves type checking]\label{lemma:typsubst-term-chk}
  If $\Delta  ;  \Gamma  ,  \ottmv{x}  \!\vcentcolon\!  \mathit{A}  ,  \Gamma'  \vdash  \mathit{e} \, \Leftarrow \, \mathit{C}$ and $\Delta  ;  \Gamma  \vdash  \mathit{e}' \, \Rightarrow \, \mathit{B}$ and $ \mathit{B}  \lesssim  \mathit{A} $,
  then $\Delta  ;  \Gamma  ,  \Gamma'  \vdash  \mathit{e}  \ottsym{[}  \ottmv{x}  \mapsto  \mathit{e}'  \ottsym{]} \, \Leftarrow \, \mathit{C}$.
\end{lemma}

Finally, with the lemmas above and isomorphic subtyping,
we have the type preservation property of \fip.
That is, after one or multiple steps of reduction,
the inferred type of the reduced expression is an isomorphic subtype.
Therefore, for checked expressions, the initial type-checking always succeeds.

\begin{theorem}[Type preservation with isomorphic subtyping]\label{theorem:preservation-subsub}
    If $ \cdot   ;   \cdot   \vdash  \mathit{e} \, \Leftrightarrow \, \mathit{A}$ and $ \mathit{e} \hookrightarrow ^* \mathit{e}' $
    then there exists a $\mathit{B}$ such that $ \cdot   ;   \cdot   \vdash  \mathit{e}' \, \Leftrightarrow \, \mathit{B}$ and $ \mathit{B}  \lesssim  \mathit{A} $.
\end{theorem}

\begin{corollary}[Type preservation]\label{corollary:preservation-check}
    If $ \cdot   ;   \cdot   \vdash  \mathit{e} \, \Leftrightarrow \, \mathit{A}$ and $ \mathit{e} \hookrightarrow ^* \mathit{e}' $
    then $ \cdot   ;   \cdot   \vdash  \mathit{e}' \, \Leftarrow \, \mathit{A}$.
\end{corollary}

\section{Related Work}\label{sec:related}

In the following discussion,
sometimes we attach the publication year to
its calculus name for easy distinction.
For instance, \fip'19 means the original formulation of \fip by Bi et al.~\cite{xuanbiesop}.

\paragraph{The merge operator, disjoint intersection types and TDOS}
The merge operator for calculi with intersection types
was proposed by Reynolds~\cite{reynolds1988preliminary}. His original formulation
came with significant restrictions to ensure
that the semantics is not ambiguous. Castagna~\cite{Castagna_1992} showed that
a merge operator restricted to functions could model overloading.
Dunfield~\cite{dunfield2014elaborating} proposed a calculus,
which we refer to as $\lambda_{,,}$,
with an unrestricted merge operator. While powerful, $\lambda_{,,}$
lacked both determinism and subject reduction, though type safety
was proved via a \emph{type-directed} elaboration semantics.

To address the ambiguity problems in Dunfield's calculus,
Oliveira et al.~\cite{oliveira2016disjoint}
proposed \lambdai'16, which only allows intersections of disjoint types.
With that restriction and the use of an elaboration
semantics, it was then possible to prove the coherence of \lambdai'16, showing that the
semantics was not ambiguous. Bi et al.~\cite{bi_et_al:LIPIcs:2018:9227}
relaxed the disjointness restriction,
requiring it only on merges, in a new calculus called \lambdaip'18 (or NeColus). This
enabled the use of unrestricted intersections in \lambdaip'18.
In addition, they added a more powerful subtyping relation based on the well-known
BCD subtyping~\cite{Barendregt_1983} relation. The new subtyping relation, in turn,
enabled nested
composition, which is a fundamental feature of Compositional Programming.
Unfortunately, both unrestricted intersections and BCD subtyping greatly
complicated the coherence proof of \lambdaip'18. To address those issues,
Bi et al. turned to an approach based on logical relations and a notion
of contextual equivalence.

To address the increasing complexities arising from the elaboration
semantics and the coherence proofs, Huang et al.~\cite{huang_et_al:LIPIcs:2020:13183,tamingmerge}
proposed a new approach to model the type-directed semantics of
calculi with a merge operator.  The type-directed elaboration in
\lambdai'16 and \lambdaip'18 is replaced by a direct
\emph{type-directed operational semantics} (TDOS). In the new TDOS
formulations of \lambdai and \lambdaip, coercive subtyping is removed
since subtyping no longer needs to generate explicit coercion for
the elaboration to a target calculus. Instead, runtime implicit
(up)casting is used. This is implemented by the casting
relation, which was originally called \emph{typed reduction}.
Our work adopts TDOS and adds disjoint polymorphism. Disjoint polymorphism
is used in Compositional Programming to enable techniques such as polymorphic contexts.
We also change the evaluation strategy from call-by-value (CBV) to
call-by-name (CBN), motivated by the elaboration of trait instantiation
in Compositional Programming. Otherwise, with a CBV semantics, many uses of trait
instantiation would diverge.

\paragraph{Calculi with disjoint polymorphism}
Disjoint polymorphism was originally introduced in a calculus called \fname by Alpuim et
al.~\cite{alpuimdisjoint}. A disjointness constraint is added to universal
quantification in order to allow merging components whose type contains type
variables. Later, Bi et al.~\cite{xuanbiesop} augment it with distributive
subtyping in the \fip'19 calculus. In addition, the bottom type is added and unrestricted
intersection types are also allowed to fully encode row and bounded polymorphism~\cite{xie2020row}.
Compared to \fip'19, our new formulation of \fip adopts a direct semantics, based on
a TDOS approach,
where simpler proofs of determinism supersede the original proofs of coherence.
As a result, recursion and impredicative polymorphism can be easily added.
Both features are important to fully support the trait
encoding in Compositional Programming. A detailed comparison of calculi with a merge operator, which summarizes
our discussion on related work, can be found in \cref{fig:compare}.

% circles are defined in macro.tex
\begin{figure}[t]

\begin{small}
  \centering
  \setlength\tabcolsep{2pt}
  \newcolumntype{L}{>{\raggedright\arraybackslash}p{3.8cm}}
  \newcolumntype{C}{>{\centering\arraybackslash}p{1.2cm}}

\resizebox{.98\textwidth}{!}{% <------ Don't forget this %
\begin{tabular}{LCCCCCCCC}
& $\lambda_{,,}$ & \lambdai'16 & \fname & \lambdaip'18 &\fip'19  & \lambdai & \lambdaip & \fip \\ %\midrule

Disjointness &  \emptycircle &  \fullcircle &   \fullcircle   & \fullcircle   &   \fullcircle  &   \fullcircle & \fullcircle & \fullcircle  \\

Unrestricted Intersections &  \fullcircle &  \emptycircle &  \emptycircle &    \fullcircle  &   \fullcircle   &   \fullcircle & \fullcircle & \fullcircle  \\

Determinism / Coherence &  No  &  Coh.  &   Coh.   &   Coh. &   Coh. &   Det. &   Det. & Det. \\

Recursion  &  \fullcircle &  \emptycircle  &  \emptycircle  &  \emptycircle &  \emptycircle  &  \fullcircle  & \fullcircle & \fullcircle \\

Direct Semantics &  \fullcircle & \emptycircle  &  \emptycircle  &  \emptycircle   &  \emptycircle  & \fullcircle & \fullcircle  & \fullcircle  \\

Subject Reduction &  \emptycircle &  -  &   -  &   -  &   -  & \fullcircle & \fullcircle & \fullcircle  \\

Distributive Subtyping & \emptycircle & \emptycircle & \emptycircle & \fullcircle & \fullcircle & \emptycircle & \fullcircle & \fullcircle \\

Disjoint Polymorphism & \emptycircle & \emptycircle & \fullcircle & \emptycircle & \fullcircle & \emptycircle & \emptycircle & \fullcircle \\

Evaluation Strategy & CBV & CBV & CBV & CBV & CBV & CBV & CBV & CBN
\end{tabular}% <------ Don't forget this %
}

\end{small}
\caption{Summary of intersection calculi with the merge operator. \\
  (\fullcircle $=$ yes, \emptycircle $=$ no, - $=$ not applicable % , Coh. $=$ Coherence, Det. $=$ Determinism
)}\label{fig:compare}
\end{figure}

\paragraph{\fip versus \fsub}
There are quite a few typed object encodings in the
literature~\cite{bruce1997comparing}, most of which are based on \fsub~\cite{cardelli1985understanding}.
As it is not our goal in this paper to encode full OOP in \fip, we will not compare our trait
encoding with other object encodings. However, it is still interesting to
compare \fip with \fsub. Some disadvantages of \fsub have been studied in the literature.
It has been shown that, with bounded quantification, the subtyping of \fsub is
undecidable~\cite{pierce1994bounded}, and some useful operations like polymorphic
record updates~\cite{cardelli1989operations} are not directly supported. \fip
does not have these drawbacks. \fip has decidable subtyping.
For \fsub the most common decidable fragment is the so-called kernel \fsub variant~\cite{cardelli1985understanding}.
Xie et al.~\cite{xie2020row} have shown that kernel \fsub is encodable in \fip.
Therefore the bounded quantification that is present in kernel \fsub
can be expressed in \fip as well.
In addition, polymorphic record updates can be easily encoded without extra
language constructs. For example, concerning a polymorphic record that contains
an \lstinline{x} field among others (\lstinline|rcd : { x: Int } & R|), the
record update \lstinline/{ rcd with x = 1 }/ can be encoded in \fip as
\lstinline|{ x = 1 } ,, (rcd : R)|. In other words, we can rewrite whichever
fields we want and then merge the remaining polymorphic part back.

\paragraph{\fip versus row-polymorphic calculi}
Row polymorphism provides an alternative way to model extensible record types in
\systemf-like calculi. There are many variants of row-polymorphic calculi in
the literature~\cite{mark2001,leijen2005extensible,cardelli1989operations,Harper:1991:RCB:99583.99603}.
Among them, the most relevant one with respect to our work is \lambdamerge by Harper and
Pierce~\cite{Harper:1991:RCB:99583.99603}. Disjoint quantification has a striking
similarity to \emph{constrained quantification} in \lambdamerge. Their
\emph{compatibility} constraint plays a similar role to \emph{disjointness} in
our system. Furthermore, their merge operator (\lstinline{||}) can concatenate
either two records like our merge operator ($ \,,,\, $) or two record types
like our intersection type operator ($ \, \& \, $). However, their
compatibility constraint and merge operator are only applicable to record types,
while we generalize them to arbitrary types. \lambdamerge has no subtyping and
does not allow for distributivity and nested composition either.
Disjoint polymorphism also subsumes the form of row
polymorphism present in \lambdamerge as demonstrated by Xie et al.\cite{xie2020row}.
We refer to Xie et al.'s work for an extended discussion of the relationship between \fip and
various other row polymorphic calculi.

\paragraph{Semantics for type-dependent languages}
The elaboration semantics approach is commonly used to model the semantics
of type-dependent languages and calculi.
The appeals of the elaboration semantics are simple
type-safety proofs, and the fact that they directly offer an
implementation technique over conventional languages without a
type-dependent semantics. For instance, the semantics
of type-dependent languages with \emph{type
  classes}~\cite{wadler1989make,hall96typeclasses}, \emph{Scala-style
  implicits}~\cite{Oliveira2010,odersky17simplicity} or \emph{gradual
  typing}~\cite{siek2006gradual} all use an elaboration
semantics.  In contrast, in the past, more conventional direct
formulations using an operational semantics have been avoided for
languages with a type-dependent semantics. A problem is that the
type-dependent semantics introduces complexity in the formulation of an
operational semantics, since enough type information should be present
at runtime and type information needs to be properly propagated.
Early work on the semantics of type
classes~\cite{kaes88parametric,odersky95overloading}, for instance,
attempted to employ an operational semantics. However, those approaches had significant
practical restrictions in comparison to conventional type classes.
The TDOS approach has shown how to overcome important issues
when modeling the direct semantics of type-dependent languages.
An important advantage of the TDOS approach
is that it removes the need for non-trivial coherence proofs.
The TDOS approach has also been recently shown to work for modeling
the semantics of gradually typed languages directly~\cite{ye21tdos}.

\section{Conclusion}\label{sec:conclusion}

In this paper, we presented a new formulation of the \fip calculus and showed
how it serves as a direct foundation for Compositional Programming. In contrast
to the original \fip, we adopt a direct semantics based on the TDOS approach
and embrace call-by-name evaluation. As a result, the metatheory of \fip is
significantly simplified, especially due to the fact that a coherence proof
based on logical relations and contextual equivalence is not needed.
In addition, our formulation of \fip enables recursion and impredicative polymorphism,
validating the original trait encoding by Zhang et al.~\cite{zhang2021compositional}.
We proved the type-soundness and determinism of \fip using the Coq proof assistant.
Our research explores further
possibilities of the TDOS approach and shows some novel notions that could
inspire the design of other calculi with similar features.

Although \fip is already expressive enough to work as a core calculus of the CP
language, some useful constructs like type operators are missing. We leave the
extension of type-level operations for future work. Another interesting design
choice that we want to explore is to lazily evaluate both sides of merges, just
like what we have done for record fields, which can help avoid some redundant
computation on the unused side of a merge.
This design choice may be implemented
by considering merges of pre-values as values, but the parallel application of
pre-value merges needs special care.

\bibliography{miscellaneous/reference}

\begin{thebibliography}{10}

\bibitem{ahmed2006step}
Amal Ahmed.
\newblock Step-indexed syntactic logical relations for recursive and quantified
  types.
\newblock In {\em European Symposium on Programming (ESOP)}, 2006.

\bibitem{alpuimdisjoint}
Jo{\~a}o Alpuim, Bruno C. d.~S. Oliveira, and Zhiyuan Shi.
\newblock Disjoint polymorphism.
\newblock In {\em European Symposium on Programming (ESOP)}, 2017.

\bibitem{appel01step}
Andrew~W. Appel and David McAllester.
\newblock An indexed model of recursive types for foundational proof-carrying
  code.
\newblock {\em ACM Trans. Program. Lang. Syst.}, 23(5):657–683, sep 2001.

\bibitem{Barendregt_1983}
Henk Barendregt, Mario Coppo, and Mariangiola Dezani-Ciancaglini.
\newblock A filter lambda model and the completeness of type assignment.
\newblock {\em The journal of symbolic logic}, 48(04):931--940, 1983.

\bibitem{bi_et_al:LIPIcs:2018:9214}
Xuan Bi and Bruno C. d.~S. Oliveira.
\newblock {Typed First-Class Traits}.
\newblock In {\em European Conference on Object-Oriented Programming (ECOOP)},
  2018.

\bibitem{bi_et_al:LIPIcs:2018:9227}
Xuan Bi, Bruno C. d.~S. Oliveira, and Tom Schrijvers.
\newblock {The Essence of Nested Composition}.
\newblock In {\em European Conference on Object-Oriented Programming (ECOOP)},
  2018.

\bibitem{xuanbiesop}
Xuan Bi, Ningning Xie, Bruno C. d.~S. Oliveira, and Tom Schrijvers.
\newblock Distributive disjoint polymorphism for compositional programming.
\newblock In {\em European Symposium on Programming (ESOP)}, 2019.

\bibitem{bruce1997comparing}
Kim~B Bruce, Luca Cardelli, and Benjamin~C Pierce.
\newblock Comparing object encodings.
\newblock In {\em International Symposium on Theoretical Aspects of Computer
  Software}, pages 415--438. Springer, 1997.

\bibitem{cardelli1989operations}
Luca Cardelli and John~C Mitchell.
\newblock Operations on records.
\newblock In {\em International Conference on Mathematical Foundations of
  Programming Semantics}, 1989.

\bibitem{cardelli1985understanding}
Luca Cardelli and Peter Wegner.
\newblock On understanding types, data abstraction, and polymorphism.
\newblock {\em {ACM} Computing Surveys}, 17(4):471--523, 1985.

\bibitem{Castagna_1992}
Giuseppe Castagna, Giorgio Ghelli, and Giuseppe Longo.
\newblock A calculus for overloaded functions with subtyping.
\newblock In {\em Conference on LISP and Functional Programming}, 1992.

\bibitem{cookthesis}
William~R. Cook.
\newblock {\em A Denotational Semantics of Inheritance}.
\newblock PhD thesis, Brown University, 1989.

\bibitem{Davies_2000}
Rowan Davies and Frank Pfenning.
\newblock Intersection types and computational effects.
\newblock In {\em International Conference on Functional Programming (ICFP)},
  2000.

\bibitem{dunfield2014elaborating}
Jana Dunfield.
\newblock Elaborating intersection and union types.
\newblock {\em Journal of Functional Programming (JFP)}, 24(2-3):133--165,
  2014.

\bibitem{bidirectional2021}
Jana Dunfield and Neel Krishnaswami.
\newblock Bidirectional typing.
\newblock {\em ACM Comput. Surv.}, 54(5), May 2021.
\newblock \href {https://doi.org/10.1145/3450952} {\path{doi:10.1145/3450952}}.

\bibitem{Ernst_2001}
Erik Ernst.
\newblock Family polymorphism.
\newblock In {\em European Conference on Object-Oriented Programming (ECOOP)},
  2001.

\bibitem{girard1972interpretation}
Jean-Yves Girard.
\newblock {\em Interpr{\'e}tation fonctionnelle et {\'e}limination des coupures
  de l'arithm{\'e}tique d'ordre sup{\'e}rieur}.
\newblock PhD thesis, Universit{\'e} Paris 7, 1972.

\bibitem{hall96typeclasses}
Cordelia~V. Hall, Kevin Hammond, Simon~L. Peyton~Jones, and Philip~L. Wadler.
\newblock Type classes in haskell.
\newblock {\em ACM Trans. Program. Lang. Syst.}, 18(2):109–138, mar 1996.

\bibitem{Harper:1991:RCB:99583.99603}
Robert Harper and Benjamin Pierce.
\newblock A record calculus based on symmetric concatenation.
\newblock In {\em Principles of Programming Languages (POPL)}, 1991.

\bibitem{huang_et_al:LIPIcs:2020:13183}
Xuejing Huang and Bruno~C. d.~S.~Oliveira.
\newblock A type-directed operational semantics for a calculus with a merge
  operator.
\newblock In Robert Hirschfeld and Tobias Pape, editors, {\em 34th European
  Conference on Object-Oriented Programming (ECOOP 2020)}, volume 166 of {\em
  Leibniz International Proceedings in Informatics (LIPIcs)}, pages
  26:1--26:32, Dagstuhl, Germany, 2020. Schloss Dagstuhl--Leibniz-Zentrum
  f{\"u}r Informatik.
\newblock URL: \url{https://drops.dagstuhl.de/opus/volltexte/2020/13183}, \href
  {https://doi.org/10.4230/LIPIcs.ECOOP.2020.26}
  {\path{doi:10.4230/LIPIcs.ECOOP.2020.26}}.

\bibitem{huang2021distributing}
Xuejing Huang and Bruno C d~S Oliveira.
\newblock Distributing intersection and union types with splits and duality
  (functional pearl).
\newblock {\em Proceedings of the ACM on Programming Languages}, 5(ICFP):1--24,
  2021.

\bibitem{tamingmerge}
Xuejing Huang, Jinxu Zhao, and Bruno~C. d.~S.~Oliveira.
\newblock Taming the merge operator.
\newblock {\em Journal of Functional Programming}, 31:e28, 2021.
\newblock \href {https://doi.org/10.1017/S0956796821000186}
  {\path{doi:10.1017/S0956796821000186}}.

\bibitem{kaes88parametric}
Stefan Kaes.
\newblock Parametric overloading in polymorphic programming languages.
\newblock In H.~Ganzinger, editor, {\em ESOP '88}, Berlin, Heidelberg, 1988.
  Springer Berlin Heidelberg.

\bibitem{leijen2005extensible}
Daan Leijen.
\newblock Extensible records with scoped labels.
\newblock {\em Trends in Functional Programming}, 5:297--312, 2005.

\bibitem{liang1995monad}
Sheng Liang, Paul Hudak, and Mark Jones.
\newblock Monad transformers and modular interpreters.
\newblock In {\em Proceedings of the 22nd ACM SIGPLAN-SIGACT symposium on
  principles of programming languages}, pages 333--343, 1995.

\bibitem{muehlboeck2018empowering}
Fabian Muehlboeck and Ross Tate.
\newblock Empowering union and intersection types with integrated subtyping.
\newblock In {\em OOPSLA}, 2018.

\bibitem{odersky17simplicity}
Martin Odersky, Olivier Blanvillain, Fengyun Liu, Aggelos Biboudis, Heather
  Miller, and Sandro Stucki.
\newblock Simplicitly: Foundations and applications of implicit function types.
\newblock {\em Proc. ACM Program. Lang.}, 2(POPL), dec 2017.

\bibitem{odersky95overloading}
Martin Odersky, Philip Wadler, and Martin Wehr.
\newblock A second look at overloading.
\newblock In {\em Proceedings of the Seventh International Conference on
  Functional Programming Languages and Computer Architecture}, FPCA '95, page
  135–146, New York, NY, USA, 1995. Association for Computing Machinery.

\bibitem{Oliveira2010}
Bruno C. d.~S. Oliveira, Adriaan Moors, and Martin Odersky.
\newblock Type classes as objects and implicits.
\newblock In William~R. Cook, Siobh{\'{a}}n Clarke, and Martin~C. Rinard,
  editors, {\em Proceedings of the 25th Annual {ACM} {SIGPLAN} Conference on
  Object-Oriented Programming, Systems, Languages, and Applications, {OOPSLA}
  2010, October 17-21, 2010, Reno/Tahoe, Nevada, {USA}}, pages 341--360. {ACM},
  2010.
\newblock \href {https://doi.org/10.1145/1869459.1869489}
  {\path{doi:10.1145/1869459.1869489}}.

\bibitem{oliveira2016disjoint}
Bruno C. d.~S. Oliveira, Zhiyuan Shi, and Jo{\~a}o Alpuim.
\newblock Disjoint intersection types.
\newblock In {\em International Conference on Functional Programming (ICFP)},
  2016.

\bibitem{pierce1989decision}
Benjamin~C Pierce.
\newblock A decision procedure for the subtype relation on intersection types
  with bounded variables.
\newblock Technical report, Carnegie Mellon University, 1989.

\bibitem{pierce1994bounded}
Benjamin~C Pierce.
\newblock Bounded quantification is undecidable.
\newblock {\em Information and Computation}, 112(1):131--165, 1994.

\bibitem{reynolds1974towards}
John~C Reynolds.
\newblock Towards a theory of type structure.
\newblock In {\em Programming Symposium}, pages 408--425. Springer, 1974.

\bibitem{reynolds1988preliminary}
John~C Reynolds.
\newblock Preliminary design of the programming language forsythe.
\newblock Technical report, Carnegie Mellon University, 1988.

\bibitem{Reynolds_1991}
John~C. Reynolds.
\newblock The coherence of languages with intersection types.
\newblock In {\em Lecture Notes in Computer Science (LNCS)}, pages 675--700.
  Springer Berlin Heidelberg, 1991.

\bibitem{reynolds1997design}
John~C Reynolds.
\newblock Design of the programming language forsythe.
\newblock In {\em ALGOL-like languages}, pages 173--233. 1997.

\bibitem{schrijvers2011monads}
Tom Schrijvers and Bruno C. d.~S. Oliveira.
\newblock Monads, zippers and views: virtualizing the monad stack.
\newblock In {\em Proceedings of the 16th ACM SIGPLAN international conference
  on functional programming}, pages 32--44, 2011.

\bibitem{mark2001}
Mark Shields and Erik Meijer.
\newblock Type-indexed rows.
\newblock In {\em Proceedings of the 28th ACM SIGPLAN-SIGACT Symposium on
  Principles of Programming Languages}, POPL '01, page 261–275, New York, NY,
  USA, 2001. Association for Computing Machinery.
\newblock \href {https://doi.org/10.1145/360204.360230}
  {\path{doi:10.1145/360204.360230}}.

\bibitem{siek2019transitivity}
Jeremy~G. Siek.
\newblock Transitivity of subtyping for intersection types.
\newblock {\em CoRR}, abs / 1906.09709, 2019.
\newblock URL: \url{http://arxiv.org/abs/1906.09709}, \href
  {http://arxiv.org/abs/1906.09709} {\path{arXiv:1906.09709}}.

\bibitem{siek2006gradual}
Jeremy~G Siek and Walid Taha.
\newblock Gradual typing for functional languages.
\newblock In {\em Scheme and Functional Programming Workshop}, 2006.

\bibitem{siek07objects}
Jeremy~G. Siek and Walid Taha.
\newblock Gradual typing for objects.
\newblock In Erik Ernst, editor, {\em {ECOOP} 2007 - Object-Oriented
  Programming, 21st European Conference, Berlin, Germany, July 30 - August 3,
  2007, Proceedings}, volume 4609 of {\em Lecture Notes in Computer Science},
  pages 2--27. Springer, 2007.

\bibitem{wadler1998expression}
Philip Wadler.
\newblock The expression problem.
\newblock {\em Java-genericity mailing list}, 1998.

\bibitem{wadler1989make}
Philip Wadler and Stephen Blott.
\newblock How to make ad-hoc polymorphism less ad-hoc.
\newblock In {\em Conference Record of the Sixteenth Annual {ACM} Symposium on
  Principles of Programming Languages, Austin, Texas, USA, January 11-13,
  1989}, pages 60--76. {ACM} Press, 1989.
\newblock \href {https://doi.org/10.1145/75277.75283}
  {\path{doi:10.1145/75277.75283}}.

\bibitem{xie2020row}
Ningning Xie, Bruno C d~S Oliveira, Xuan Bi, and Tom Schrijvers.
\newblock Row and bounded polymorphism via disjoint polymorphism.
\newblock In {\em 34th European Conference on Object-Oriented Programming
  (ECOOP 2020)}. Schloss Dagstuhl-Leibniz-Zentrum f{\"u}r Informatik, 2020.

\bibitem{ye21tdos}
Wenjia Ye, Bruno~C. d.~S.~Oliveira, and Xuejing Huang.
\newblock Type-directed operational semantics for gradual typing.
\newblock In Anders M{\o}ller and Manu Sridharan, editors, {\em 35th European
  Conference on Object-Oriented Programming, {ECOOP} 2021, July 11-17, 2021,
  Aarhus, Denmark (Virtual Conference)}, volume 194 of {\em LIPIcs}, pages
  12:1--12:30. Schloss Dagstuhl - Leibniz-Zentrum f{\"{u}}r Informatik, 2021.

\bibitem{zhang2021compositional}
Weixin Zhang, Yaozhu Sun, and Bruno C. d.~S. Oliveira.
\newblock Compositional programming.
\newblock {\em ACM Transactions on Programming Languages and Systems (TOPLAS)},
  43(3):1--61, 2021.

\end{thebibliography}

\clearpage
\appendix
\section{Some Definitions}
\subsection{Type and Context Well-Formedness}\label{appendix:wfness}

\begin{figure}[h!]
  \drules[TW]{$\Delta  \vdash  \mathit{A}$}{Type Well-formedness}{top, bot, int, var, rcd, arrow, and, all}
  \drules[TCW]{$\vdash  \Delta$}{Type Context Well-formedness}{empty, cons}
  \drules[CW]{$\Delta  \vdash  \Gamma$}{Term Context Well-formedness}{empty, cons}
  \caption{Well-formedness rules.}\label{fig:wfness-rules}
\end{figure}

Well-formedness relations are defined in \cref{fig:wfness-rules}.
$\Delta  \vdash  \mathit{A}$ means type $\mathit{A}$ is well-formed under type context $\Delta$.
It checks that all type variables that appear in $\mathit{A}$ are included in $\Delta$.
$\vdash  \Delta$ defines well-formed type contexts: any type in the context can only
use type variables that are in the context in front of itself.
$\Delta  \vdash  \Gamma$ makes sure types in $\Gamma$ are all well-formed to $\Delta$.

\begin{lemma}[Typing Regular Properties]\label{lemma:typing-regular}
     \[ \Delta  ;  \Gamma  \vdash  \mathit{e} \, \Leftrightarrow \, \mathit{A}~then~\vdash  \Delta~and~\Delta  \vdash  \Gamma~and~\Delta  \vdash  \mathit{A}.\]
 \end{lemma}

\begin{figure}[h!]
  \begin{small}
    {
      \setlength{\fboxrule}{0pt}\setlength{\fboxsep}{0pt}
      \framebox[\linewidth]{\setlength{\fboxrule}{0.4pt}\setlength{\fboxsep}{3pt}\framebox{$ [\![  \mathit{A}  ]\!] $} \hfill \textit{(Value Generator)}}
    }
    \begin{align*}
       [\![   \mathsf{Top}   ]\!]  &=  \top &
       [\![  \mathit{A}_{{\mathrm{1}}}  \rightarrow  A^\circ_{{\mathrm{2}}}  ]\!]  &=  \ottsym{(}   \lambda  \ottmv{x} \!\vcentcolon\!  \mathsf{Top}  .\, \top   \ottsym{)}  \!\vcentcolon\!  \mathit{A}_{{\mathrm{1}}}  \rightarrow  A^\circ_{{\mathrm{2}}} \\
       [\![  \ottsym{\{}  \ottmv{l}  \!\vcentcolon\!  A^\circ  \ottsym{\}}  ]\!]  &=   \{  \ottmv{l} \ottsym{=} \top  \}   \!\vcentcolon\!  \ottsym{\{}  \ottmv{l}  \!\vcentcolon\!  A^\circ  \ottsym{\}} &
       [\![   \forall   \ottmv{X} * \mathit{A}_{{\mathrm{1}}} .\, A^\circ_{{\mathrm{2}}}   ]\!]  &=  \ottsym{(}   \Lambda  \ottmv{X} .\, \top   \ottsym{)}  \!\vcentcolon\!   \forall   \ottmv{X} * \mathit{A}_{{\mathrm{1}}} .\, A^\circ_{{\mathrm{2}}}   \\
    \end{align*}

  \end{small}
    \caption{Top-like value generator.}\label{fig:top-like-gen}
  \end{figure}

\subsection{Top-like Value Generator}\label{appendix:generator}

The top-like value generator function is shown as \cref{fig:top-like-gen}, which
generates a value inferring the input top-like type.

\subsection{Type Disjointness Axioms}\label{appendix:disjoint-ax}

\begin{figure}[h!]
    \ottdefnsDisjointnessAxiom
     \caption{Algorithmic type disjointness axioms.}\label{fig:disjointness-axioms}
\end{figure}

The type disjointness axioms are shown in \cref{fig:disjointness-axioms}.

\subsection{Principal Types}\label{appendix:principal}

\begin{figure}[h!]
  \small
\begin{align*}
\end{align*}
    \ottdefnsPrincipalType
    \caption{Principal types.}\label{fig:principaltype}
\end{figure}

Principal types (\cref{fig:principaltype}), compute the types
from pre-values syntatically, using either the annotations available in the pre-values
or, for basic values $\top$ and $i$, just directly returning the corresponding type.
We use $\mathit{u}  \!\vcentcolon\!  \mathit{A}$ to represent the principal type of $\mathit{u}$ is $\mathit{A}$.
This syntactic approach is complete with respect to our type system.
\begin{lemma}[Pre-values have principal types]\label{lemma:prevalue-principal}
  $\exists$ A that $\mathit{u}  \!\vcentcolon\!  \mathit{A}$.
\end{lemma}

\begin{lemma}[Completeness of principal types]\label{lemma:soundness-principal}
  If $\Delta  ;   \cdot   \vdash  \mathit{u} \, \Rightarrow \, \mathit{A}$ then $ \mathit{u}  :  \mathit{A} $.
\end{lemma}

\section{Some Properties}

\subsection{Applicative Distribution}

Merging $ \mathsf{Int}   \rightarrow   \mathsf{Int} $ and $ \mathsf{Int}   \rightarrow  \mathsf{Bool}$ gives us $ \mathsf{Int}   \, \& \,   \mathsf{Int}   \rightarrow   \mathsf{Int}   \, \& \,  \mathsf{Bool}$.
The parameter types are combined and, in this case, duplicated.
Such a design allows any two function types to be merged.
In contrast, we cannot split $ \mathsf{Int}   \, \& \,  \mathsf{Bool}  \rightarrow   \mathsf{Int}   \, \& \,  \mathsf{Bool}$ into $ \mathsf{Int}   \rightarrow   \mathsf{Int} $ and
$\mathsf{Bool}  \rightarrow  \mathsf{Bool}$, because not every subtype of the former is a subtype of the
latter two. This indicates the applicative distribution output is not equivalent
to the input intersection type, unlike type splitting.

An important property of applicative distribution is that it is always deterministic:

\begin{lemma}[Determinism of applicative distribution]\label{lemma:appdist-uniq}
  For any type $\mathit{A}$,
  \begin{itemize}
   \item if $ \mathit{A}  \rhd  \mathit{B}_{{\mathrm{1}}}  \rightarrow  \mathit{B}_{{\mathrm{2}}} $ and $ \mathit{A}  \rhd  \mathit{C}_{{\mathrm{1}}}  \rightarrow  \mathit{C}_{{\mathrm{2}}} $ then $ \mathit{B}_{{\mathrm{1}}}  =  \mathit{C}_{{\mathrm{1}}} $ and $ \mathit{B}_{{\mathrm{2}}}  =  \mathit{C}_{{\mathrm{2}}} $;
   \item if $ \mathit{A}  \rhd   \forall   \ottmv{X} * \mathit{B}_{{\mathrm{1}}} .\, \mathit{B}_{{\mathrm{2}}}  $ and $ \mathit{A}  \rhd   \forall   \ottmv{X} * \mathit{C}_{{\mathrm{1}}} .\, \mathit{C}_{{\mathrm{2}}}  $ then $ \mathit{B}_{{\mathrm{1}}}  =  \mathit{C}_{{\mathrm{1}}} $ and $ \mathit{B}_{{\mathrm{2}}}  =  \mathit{C}_{{\mathrm{2}}} $;
   \item if $ \mathit{A}  \rhd  \ottsym{\{}  \ottmv{l}  \!\vcentcolon\!  \mathit{B}  \ottsym{\}} $ and $ \mathit{A}  \rhd  \ottsym{\{}  \ottmv{l}  \!\vcentcolon\!  \mathit{C}  \ottsym{\}} $ then $ \mathit{B}  =  \mathit{C} $.
  \end{itemize}
\end{lemma}

\subsection{Type-Checking Subsumption}

The type-checking subsumption of our bi-directional type system is restricted
within ordinary and \emph{not}-top-like types, since we may not check arbitrary
expressions by all top-like types, such as checking a lambda abstraction against
a top-like universal type.
Therefore when wrapping an expression by a type annotation, we cannot directly
output $\mathit{e}  \!\vcentcolon\!  \mathit{A}$ when $\mathit{A}$ is top-like.

\begin{lemma}[Restricted check subsumption]\label{lemma:typing-chksub}
  If $\Delta  ;  \Gamma  \vdash  \mathit{e} \, \Leftarrow \, \mathit{A}$ and $ \Delta   \vdash   \mathit{A}  \leq  B^\circ $ and $ \neg   (  \Delta   \vdash  \rceil B^\circ \lceil  \!)  $
  then $\Delta  ;  \Gamma  \vdash  \mathit{e} \, \Leftarrow \, B^\circ$.
\end{lemma}

\end{document}